\documentclass[%
 reprint,
superscriptaddress,
 amsmath,amssymb,
 aps,
prb,
]{revtex4-2}

\usepackage{graphicx}
\usepackage{dcolumn}
\usepackage{bm}
\usepackage{braket}
\usepackage{comment} 
\usepackage{hyperref}
\usepackage{siunitx}

\usepackage{xcolor}

\usepackage{nicematrix}
\NiceMatrixOptions{cell-space-limits = 1pt}

\usepackage{mathtools}

\usepackage{amsthm}
\newtheorem{theorem}{Theorem}
\newtheorem*{theorem*}{Theorem}

\theoremstyle{definition}

\theoremstyle{remark}

\newtheorem*{example*}{Example}
\newtheorem*{algorithm*}{Algorithms}

\usepackage{etoolbox}

\makeatletter
\patchcmd{\frontmatter@abstract@produce}
  {\vskip200\p@\@plus1fil
   \penalty-200\relax
   \vskip-200\p@\@plus-1fil}
  {}
  {}
  {}
\makeatother

\begin{document}

\title{Joint control of coherent transmission, reflection, and absorption}


\author{Shiyu Li}
\affiliation{Department of Electrical and Computer Engineering and Microelectronics Research Center, The University of Texas at Austin, Austin, Texas 78712, USA}

\author{Dongha Kim}
\affiliation{Ginzton Laboratory and Department of Electrical Engineering, Stanford University, Stanford, California 94305, USA}
\affiliation{Department of Physics, Korea University, Seoul, 02483, Republic of Korea}

\author{Shanhui Fan}
\email{shanhui@stanford.edu}
\affiliation{Ginzton Laboratory and Department of Electrical Engineering, Stanford University, Stanford, California 94305, USA}%

\author{Cheng Guo}
\email{chengguo@utexas.edu}
\affiliation{Department of Electrical and Computer Engineering and Microelectronics Research Center, The University of Texas at Austin, Austin, Texas 78712, USA}

\date{\today}

\begin{abstract}
Controlling multiple wave properties simultaneously poses a key challenge in coherent control of wave transport. We present a theory for joint coherent control of transmission, reflection, and absorption in linear systems. We prove that the numerical range provides the mathematical structure governing achievable responses, and reveal non-abelian effects due to non-commutativity between transmission, reflection, and absorption matrices. We provide an algorithm to achieve arbitrary target responses. Our results establish a theoretical foundation for joint coherent control of waves.
\end{abstract}
\maketitle


Controlling wave transport is fundamental to many applications in imaging~\cite{sebbah2001a,ntziachristos2010a,cizmar2012,kang2015,guo2018,guo2018a,yoon2020b,wang2020p,long2021,bertolotti2022,wang2022,long2022}, sensing~\cite{aulbach2011,sarma2015,mounaix2016,jeong2018a,liu2020s}, communications~\cite{miller2013c,miller2019,seyedinnavadeh2024}, and renewable energy~\cite{ottens2011, guha2012, babuty2013, rephaeli2013, raman2014, boriskina2016,raj2017a,fiorino2018a,park2021,zhu2019c,li2019g}. A key development is coherent control~\cite{popoff2014,liew2016,mounaix2016}, where one tailors the wavefront of input waves through techniques such as spatial light modulation~\cite{vellekoop2007,yu2017e} to achieve desired transport characteristics. Recent advances~\cite{guan2012,katz2012,horstmeyer2015,vellekoop2015,yu2015d,pai2021b} have enabled precise control of wave transport in complex media~\cite{rotter2017,cao2022a} including biological tissues~\cite{yaqoob2008,ntziachristos2010a,horstmeyer2015,yu2015d} and multimode optical fibers~\cite{fan2005,cizmar2011,cizmar2012,papadopoulos2012,carpenter2015,xiong2016}, leading to effects such as spatial and temporal focusing~\cite{lerosey2007,vellekoop2008b,katz2011,mccabe2011,xu2011a,papadopoulos2012,park2013a,horstmeyer2015,blochet2017,jeong2018a}, transmission enhancement and suppression~\cite{vellekoop2008,popoff2010,aulbach2011,kim2012a,shi2012,yu2013a,yan2014a,gerardin2014,pena2014a,popoff2014,davy2015,shi2015a,bender2020c,guo2024c}, coherent perfect absorption~\cite{chong2010a,wan2011,sun2014,baranov2017,mullers2018a,pichler2019a,sweeney2019a,chen2020ac,wang2021h,slobodkin2022,guo2024b}, reflectionless scattering modes~\cite{sweeney2020a,stone2021,horodynski2022}, and optical micro-manipulation~\cite{cizmar2010a,gonzalez-ballestero2021,hupfl2023}. Most existing work has focused on controlling a single transport property such as transmission or absorption. However, there is an emerging interest in simultaneously controlling multiple transport characteristics~\cite{shaughnessy2024}.

For coherent control of a single property, the mathematical framework is well-established: one introduces a Hermitian matrix whose eigenvalues and eigenstates determine the achievable range of outcomes and corresponding inputs for an outcome~\cite{rotter2017,cao2022a}. For example, wave transmission through a medium is described by a field transmission matrix $t$. The achievable power transmittance $\tau$ is bounded by the extremal eigenvalues of the Hermitian matrix $T = t^\dagger t$, with the corresponding eigenstates providing the input wavefronts that achieve these bounds~\cite{cao2022a}. The behaviors of other transport properties can be described similarly~\cite{fink1997,xie2014,mounaix2016a,mounaix2016,mullers2018a,katz2019,guo2023b,guo2023g,guo2024a,guo2024e}. However, the mathematical framework for joint coherent control of multiple wave properties remains undeveloped.

Here we present a comprehensive theory for joint coherent control, focusing on the simultaneous manipulation of transmission, reflection, and absorption. Our analysis reveals that for joint control of two quantities, such as transmittance $\tau$ and reflectance $\rho$ with corresponding Hermitian matrices $T$ and $R$, one must consider the composite non-Hermitian matrix $T + iR$. We prove that the achievable range of $(\tau,\rho) \in \mathbb{R}^2$ is determined by the numerical range~\cite{bonsall1971,gustafson1997,wu2021a} of $T + iR$. We derive inner and outer bounds on this achievable range based on the eigenvalues of $T + iR$, and show the inner bound is reached in the abelian case when $T$ and $R$ commute. We also provide a constructive algorithm to find input wavefronts that realize any target $(\tau_0,\rho_0)$ within the achievable range. Our approach can be readily extended to joint control of other wave properties. Our results lay a theoretical foundation for joint coherent control and provide practical guidelines for its implementation.

\begin{figure}[hbtp]
    \centering
    \includegraphics[width=0.45\textwidth]{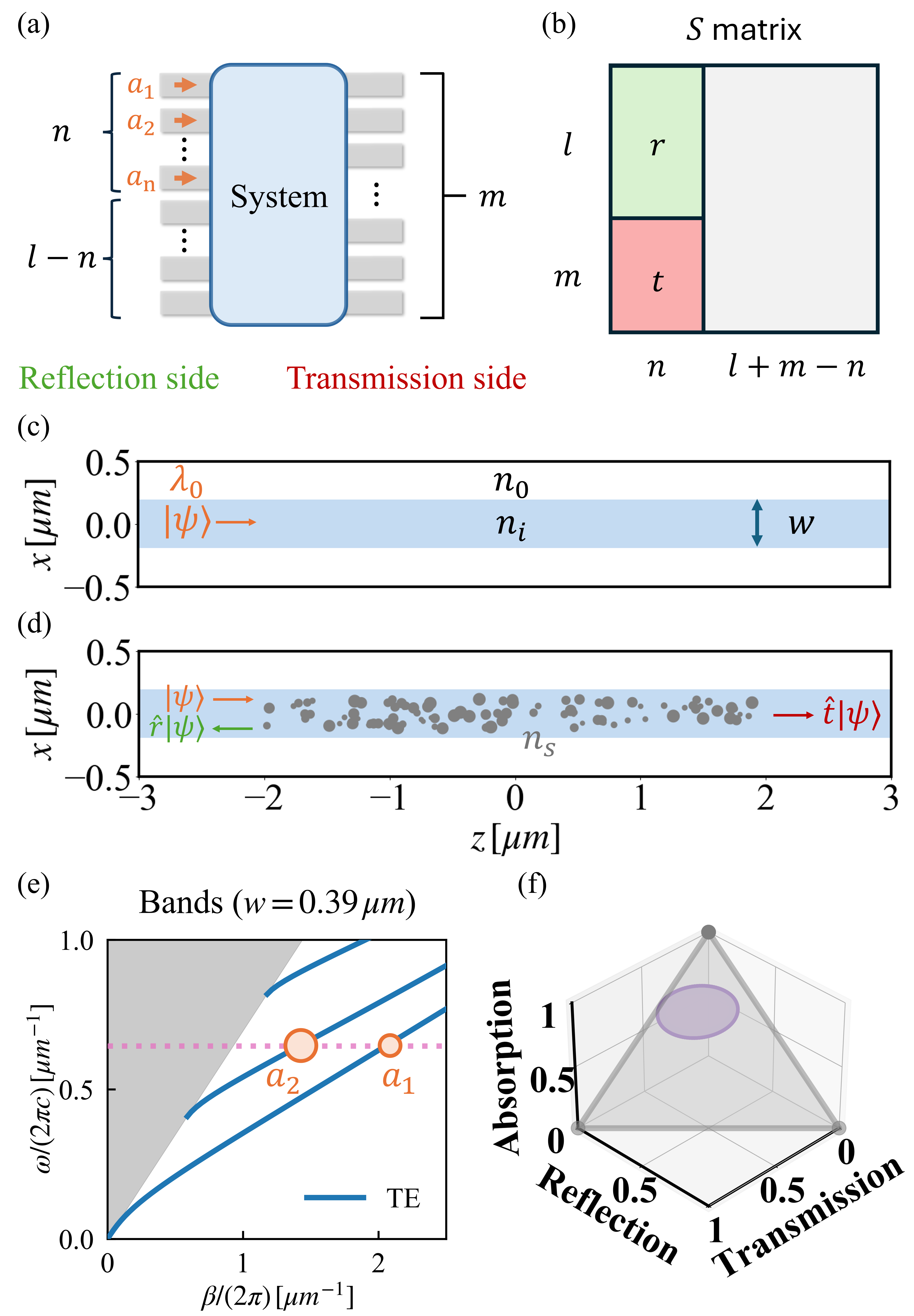}
    \caption{(a) An $(l+m)$-port linear time-invariant system with $l$ ports on the left side and $m$ ports on the right side. A coherent wave $\bm{a}$ input into $n$ left-side ports produces transmitted and reflected waves $\bm{b_t} = t\bm{a}$ and $\bm{b_r} = r\bm{a}$. (b) $r$ and $t$ are block submatrices of the entire $S$-matrix. (c) A silicon waveguide ($n_i=3.48$) embedded in silica ($n_0=1.444$). (d) Modified waveguide section with random lossy silica scatterers ($n_s=1.444+0.100i$). (e) Band dispersion of TE modes in the uniform waveguide with width $w=0.39~\mu\text{m}$, supporting two guided modes at wavelength $\lambda_0=1.55~\mu\text{m}$. (f) Schematic illustration of the set $\Omega$ containing all attainable tuples $(\tau[\bm{a}],\rho[\bm{a}],\alpha[\bm{a}])$ under varying input states $\bm{a}$.}
    \label{fig:system}
\end{figure}

We consider a passive linear time-invariant system with $(l+m)$ ports, having $l$ ports on the left side and $m$ ports on the right side [Fig.~\ref{fig:system}(a)]. A coherent wave characterized by the complex vector
\begin{equation}
\bm{a} = (a_1, a_2, \ldots, a_n)^T    
\end{equation}
is injected into $n\leq l$ input ports on the left side. The case $n < l$ represents scenarios where incident waves are restricted to an accessible $n$ dimensional subspace of the full $l$ dimensional space of left-side waves. The input power $|\bm{a}|^2$ is normalized:
\begin{equation}
    \bm{a}^\dagger \bm{a} = 1
\end{equation}
The output consists of transmitted and reflected waves characterized by
\begin{equation}
    \bm{b_t} = t \bm{a}, \quad \bm{b_r} = r \bm{a},
\end{equation}
where $t$ is the $m \times n$ field transmission matrix and $r$ is the $l \times n$ field reflection matrix, both being block submatrices of the $(l+m)\times (l+m)$ scattering matrix $S$~\cite{haus1984} [Fig.~\ref{fig:system}(b)]. The power transmittance $\tau$, reflectance $\rho$, and absorptance $\alpha$ are defined as
\begin{align}
    \tau[\bm{a}] &\coloneqq \bm{b_t}^\dagger \bm{b_t} = \bm{a}^\dagger t^\dagger t \bm{a} = \bm{a}^\dagger T \bm{a}, \label{eq:def_tau_a}\\
    \rho[\bm{a}] &\coloneqq \bm{b_r}^\dagger \bm{b_r} = \bm{a}^\dagger r^\dagger r \bm{a} = \bm{a}^\dagger R \bm{a}, \label{eq:def_rho_a}\\
    \alpha[\bm{a}] &\coloneqq 1 - \bm{b_t}^\dagger \bm{b_t} - \bm{b_r}^\dagger \bm{b_r} = \bm{a}^\dagger A \bm{a},\label{eq:def_alpha_a}
\end{align}
where we introduce the power matrices~\cite{rotter2017,guo2024j}
\begin{equation}\label{eq:def-T_R_A}
    T = t^\dagger t, \quad R = r^\dagger r, \quad A = I - t^\dagger t - r^\dagger r.
\end{equation}
Here $I$ denotes the identity matrix. $T$, $R$, and $A$ are $n\times n$ positive semidefinite Hermitian matrices. Energy conservation requires that 
\begin{align}
T + R + A = I, \quad 
    \tau[\bm{a}] + \rho[\bm{a}] + \alpha[\bm{a}] = 1.     \label{eq:energy_conservation_condition}
\end{align}

Joint coherent control refers to the method of varying $\bm{a}$ to manipulate the tuple $(\tau[\bm{a}], \rho[\bm{a}], \alpha[\bm{a}])$ simultaneously. Two fundamental questions are immediately raised. For a passive system characterized by $t$ and $r$ matrices:
\begin{enumerate}
    \item What is the set of all attainable tuples:
    \begin{equation}
    \Omega \coloneqq \{(\tau[\bm{a}], \rho[\bm{a}], \alpha[\bm{a}]) \in \mathbb{R}^{3}:  \bm{a} \in \mathbb{C}^n, \bm{a}^\dagger \bm{a} = 1\}?
    \end{equation}    
    \item How to find an input unit vector $\bm{a_0}$ that realizes a given target $(\tau_0, \rho_0, \alpha_0) \in \Omega$: 
    \begin{equation}
     \tau[\bm{a_0}] = \tau_0, \quad \rho[\bm{a_0}] = \rho_0, \quad \alpha[\bm{a_0}] = \alpha_0?   
    \end{equation}
\end{enumerate}
This paper provides complete answers to both questions. 

We illustrate joint coherent control with a concrete example. Consider a silicon slab waveguide (refractive index $n_i=3.48$) embedded in silica cladding ($n_0=1.444$) [Fig.~\ref{fig:system}(c)]. The waveguide has a thickness of $w$ in the $x$ direction and extends along the $y$ and $z$ directions. Light propagates along the $z$ direction with a vacuum wavelength of $\lambda_0 =1.55~\mu\text{m}$, and the electric field is polarized along the $y$ direction (TE polarization). The uniform waveguide supports $n$ eigenmodes at $\lambda_0$ with $n$ depending on $w$ [Fig.~\ref{fig:system}(e)]. Under the eigenmode bases, an input guided wave $\ket{\psi}$ is represented by a complex vector $\bm{a}=(a_1,\ldots,a_n)$. Next, we introduce random cylindrical scatterers made of lossy silica ($n_s = 1.444+0.100i$) into a section of the waveguide [Fig.~\ref{fig:system}(d)]. See Appendix~\ref{SI-sec:geometry_waveguide} for the detailed geometry of the disorders. The input wave $\ket{\psi}$ interacts with the scatterers and undergoes partial transmission, reflection, and absorption. (Some power will be scattered into leaky radiation, which is assumed to be absorbed by an absorbing cladding outside the silica not shown in Fig.~\ref{fig:system}(d).) By joint coherent control, we vary $\ket{\psi}$ to manipulate the distribution of output power among transmission, reflection, and absorption~[Fig.~\ref{fig:system}(f)]. The simulation is performed using Tidy3D~\cite{hughes2021a}, which implements the finite-difference time-domain method.

\begin{figure}[htbp]
    \centering
    \includegraphics[width=0.4\textwidth]{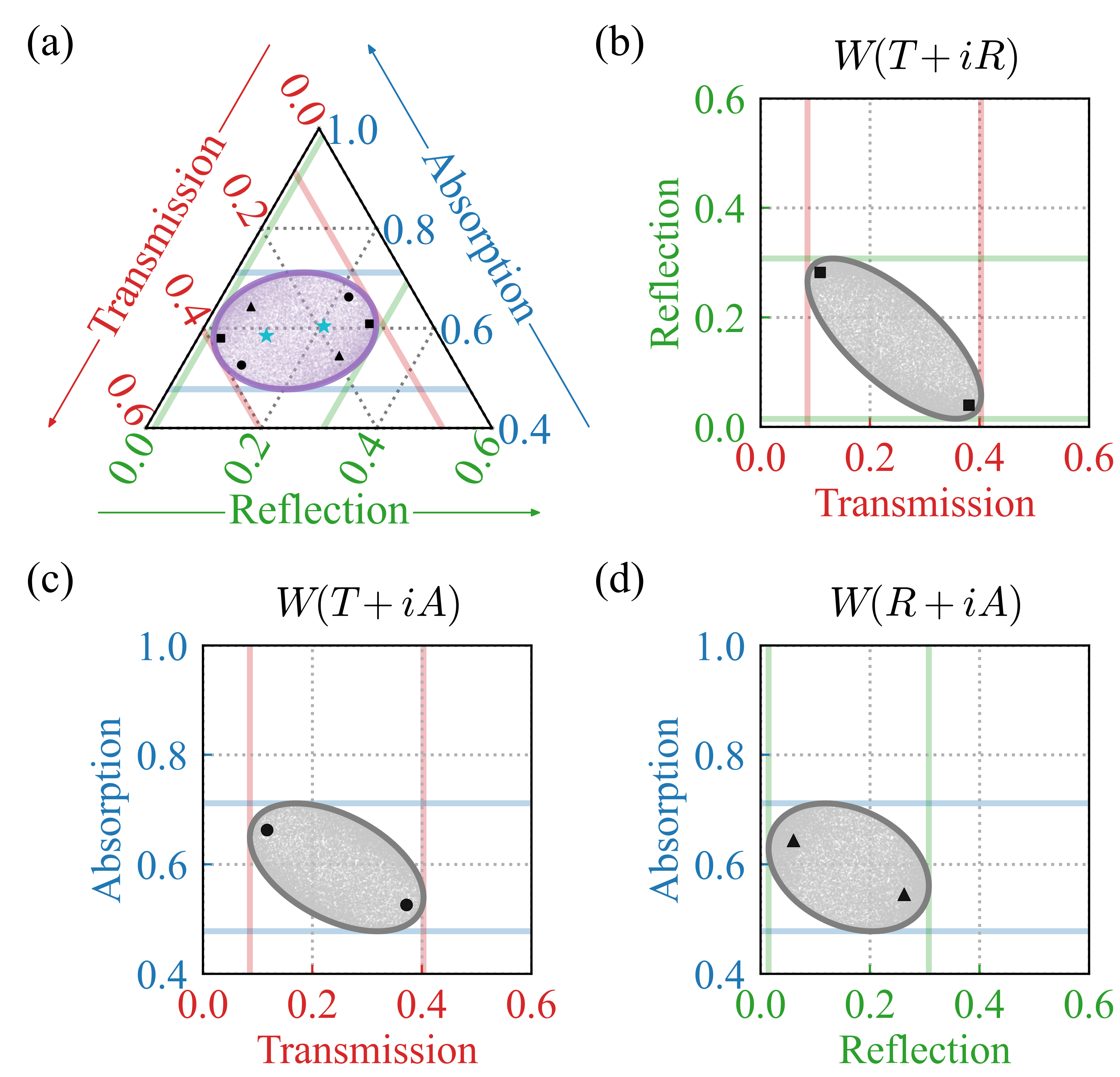}
    \caption{Attainable responses for a two-mode ($n=2$) disordered waveguide. (a) Ternary plot showing the set $\Omega$ of achievable $(\tau,\rho,\alpha)$ tuples, which forms an elliptic disk (purple boundary). Cyan stars mark the foci. Purple dots show numerical results from 30,000 random input states. Red, green, and blue lines indicate bounds on $\tau$, $\rho$, and $\alpha$ from Eq.~(\ref{eq:def-hexagon}). (b-d) Projections of $\Omega$ onto the $(\tau,\rho)$, $(\tau,\alpha)$, and $(\rho,\alpha)$ planes coincide with the numerical ranges $W(T+iR)$, $W(T+iA)$, and $W(R+iA)$. Each projection is an elliptic disk with foci determined by the eigenvalues of the corresponding matrix (marked with distinct symbols). }
    \label{fig:example-2modes}
\end{figure}

In the first example, we set $w=0.39~\mu\text{m}$ so that the waveguide supports $n=2$ modes. We numerically determine the system's $t$ and $r$ matrices and calculate $T$, $R$, and $A$ using Eq.~(\ref{eq:def-T_R_A}). See Appendix~\ref{SI-subsec:T_R_A_matrices-Fig2} for the numerical values of these matrices. Fig.~\ref{fig:example-2modes}(a) shows the scatter plot of $(\tau[\bm{a}], \rho[\bm{a}], \alpha[\bm{a}])$ for $30,000$ random input vectors $\bm{a}$ calculated from Eqs.~(\ref{eq:def_tau_a})-(\ref{eq:def_alpha_a}). The set $\Omega$ forms an elliptic disk. Figs.~\ref{fig:example-2modes}(b-d) show the projections of $\Omega$ onto the $(\tau,\rho)$, $(\tau,\alpha)$, and $(\rho,\alpha)$ planes, respectively. Each projection also forms an elliptic disk, with foci determined by the eigenvalues of $T+iR$, $T+iA$, and $R+iA$, respectively. (We identify the complex plane with $\mathbb{R}^2$.)

\begin{figure}[htbp]
    \centering
    \includegraphics[width=0.4\textwidth]{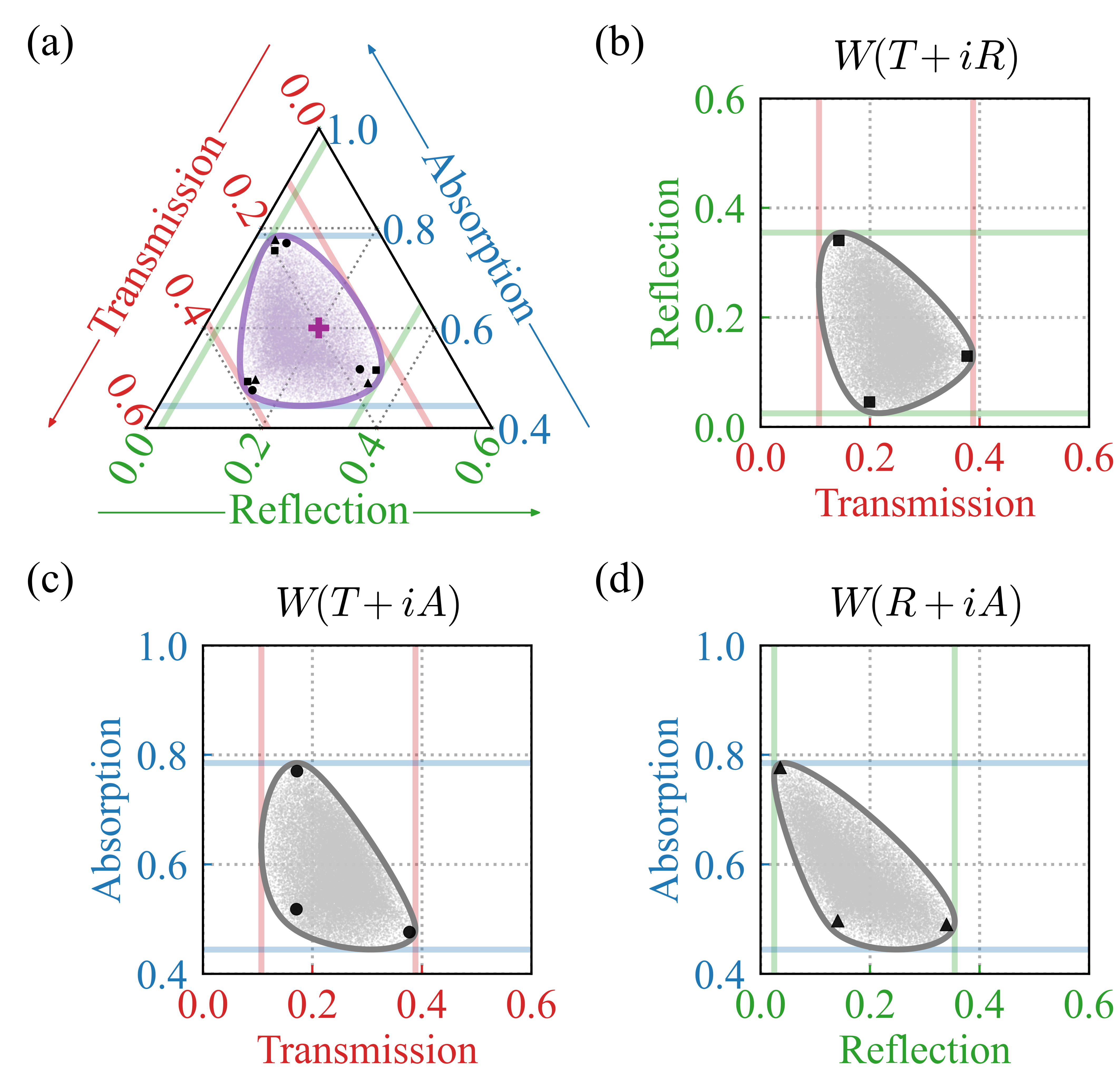}
    \caption{Attainable responses for a three-mode ($n=3$) disordered waveguide. (a) Ternary plot showing $\Omega$, which forms an ovular shape bounded by a smooth curve (purple). Purple dots show numerical results from random input states. Lines indicate bounds as in Fig.~\ref{fig:example-2modes}(a). Purple cross indicates the assigned goal in Eq.~(\ref{eq:goal_tuple}). (b-d) Projections of $\Omega$ onto the $(\tau,\rho)$, $(\tau,\alpha)$, and $(\rho,\alpha)$ planes coincide with the numerical ranges of $T+iR$, $T+iA$, and $R+iA$. Each projection is ovular and contains the three eigenvalues of the corresponding matrix (marked with distinct symbols).}
    \label{fig:example-3modes}
\end{figure}

This example illustrates our answer to Question~1:
\begin{theorem}\label{theorem:main}
For a passive linear time-invariant system with $t$ and $r$ matrices, the set of attainable tuples $(\tau, \rho, \alpha)$ under joint coherent control is given by 
\begin{align}
    &\Omega = \{ (\tau, \rho , 1- \tau - \rho) \in \mathbb{R}^{3} \mid \tau + i \rho \in W(T + i R) \} \label{eq:main_result-1}\\
    &= \{ (\tau, 1- \tau - \alpha, \alpha) \in \mathbb{R}^{3} \mid \tau + i \alpha \in W(T + i A)\} \label{eq:main_result-2}\\
    &= \{ (1- \rho - \alpha, \rho , \alpha) \in \mathbb{R}^{3} \mid \rho + i \alpha \in W(R + i A)\}, \label{eq:main_result-3}
\end{align}
where $T$, $R$, and $A$ are defined in Eq.~(\ref{eq:def-T_R_A}), and $W(M)$ denotes the numerical range of an $n\times n$ matrix $M$:
\begin{equation}
W(M) \coloneqq \{ z \in \mathbb{C}: z = \bm{a}^\dagger M \bm{a}, \bm{a} \in \mathbb{C}^{n}, \bm{a}^{\dagger} \bm{a} = 1  \}.
\end{equation}
\end{theorem}
Theorem~\ref{theorem:main} is proved in Appendix~\ref{SI-sec:proof_main_theorem}. It has a geometric interpretation: $\Omega$ is a subset of the equilateral triangle
\begin{equation}\label{eq:set_Delta}
\Delta \coloneqq \{(\tau,\rho,\alpha) \in \mathbb{R}^3: \tau + \rho + \alpha = 1, \tau \geq 0, \rho \geq 0, \alpha \geq 0\}.
\end{equation}
The projections of $\Omega$ onto the $(\tau,\rho)$, $(\tau,\alpha)$, and $(\rho,\alpha)$ planes coincide with the numerical ranges $W(T+iR)$, $W(T+iA)$, and $W(R+iA)$, respectively.

Theorem~\ref{theorem:main} also leads to a numerical method to determine $\Omega$ from $t$ and $r$ matrices. We calculate a sequence of boundary points $(\tau_j,\rho_j)$ of $W(t^\dagger t + ir^\dagger r)$ using Johnson's algorithm~\cite{johnson1978} (see Appendix~\ref{SI-sec:boundary_numerical_range}). The points $(\tau_j,\rho_j,\alpha_j=1-\tau_j-\rho_j)$ lie on the boundary of $\Omega$. Their convex hull provides a converging inner approximation of $\Omega$ as the number of boundary points increases.

\begin{figure}[htbp]
    \centering
    \includegraphics[width=0.4\textwidth]{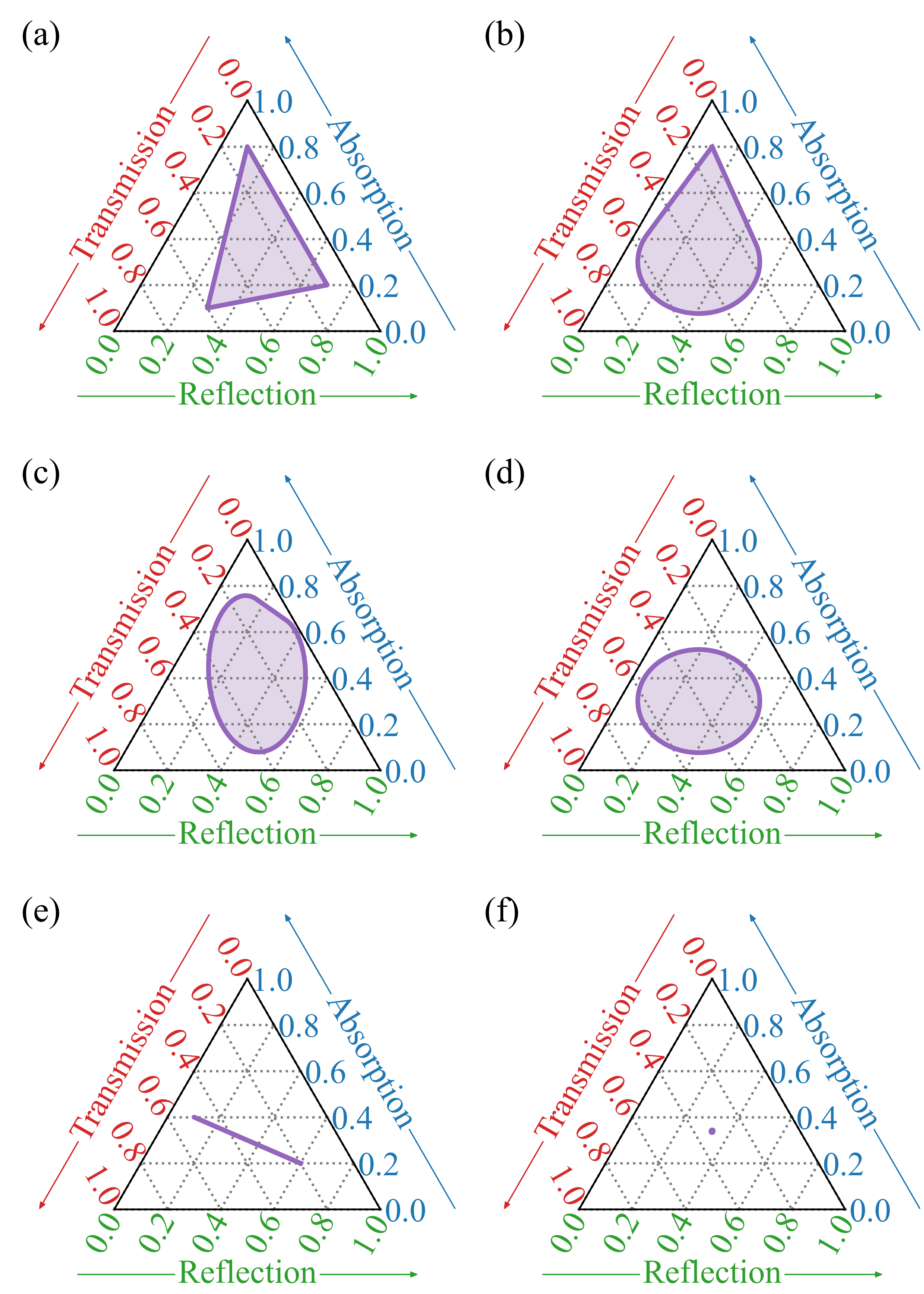}
    \caption{Possible shapes of $\Omega$ for $n=3$ beyond the ovular shape shown in Fig.~\ref{fig:example-3modes}(a). (a) A triangle. (b) The convex hull of an ellipse and an external point. (c) A two-dimensional shape with one flat boundary segment. (d) An elliptic disk. (e) A line segment. (f) A point. The $T$ and $R$ matrices used to generate each panel are listed in Appendix~\ref{SI-subsec:T_R_A_matrices-Fig4}.}
    \label{fig:shape-n=3}
\end{figure}

To illustrate Theorem~\ref{theorem:main}, we provide a second example of the disordered waveguide. We set $w=0.52~\mu\text{m}$ so that the waveguide supports $n=3$ modes. We perform similar calculations for the first example. See Appendix~\ref{SI-subsec:T_R_A_matrices-Fig3} for the numerical values of $t$, $r$, $T$, $R$, and $A$ matrices. The scatter plot in Fig.~\ref{fig:example-3modes}(a) indicates that $\Omega$ now forms an ovular disk. The projections shown in Fig.~\ref{fig:example-3modes}(b-d) also form ovular disks. The smooth boundary curves are generated using the numerical method provided above.

Now, we discuss the general properties of the set $\Omega$ for an $n$-input-port system. First, $\Omega$ is a compact (closed and bounded) and convex subset of $\Delta$ defined in Eq.~\eqref{eq:set_Delta}~\cite{horn1991}. (See Appendices~\ref{SI-subsec:proof_compactness} and~\ref{SI-subsec:proof_convexity} for proof.) The shape of $\Omega$ depends on $n$. For $n=2$, $\Omega$ must be an elliptical disk [Fig.~\ref{fig:example-2modes}(a)], a line segment, or a point~\cite{horn1991}. For $n=3$, $\Omega$ has seven possible shapes~\cite{keeler1997} [Figs.~\ref{fig:example-3modes}(a),~\ref{fig:shape-n=3}(a-f)]: (1) an ovular shape, (2) a triangle, (3) the convex hull of an ellipse and a point outside, (4) a shape with one flat boundary portion, (5) an elliptic disk, (6) a line segment, and (7) a point. For $n\geq 4$, the classification of possible shapes remains an open problem.

Second, we can bound $\Omega$ from both inside and outside. The set $\Omega$ is inscribed in the bounding hexagon
\begin{align}
\Omega_{\text{out}} \coloneqq \{(\tau,\rho,\alpha) \in \Delta: \lambda_{\min}(T) \leq \tau \leq \lambda_{\max}(T), \nonumber \\
\lambda_{\min}(R) \leq \rho \leq \lambda_{\max}(R),\lambda_{\min}(A) \leq \alpha \leq \lambda_{\max}(A) \}, \label{eq:def-hexagon}     
\end{align}
where $\lambda_{\min}(N)$ and $\lambda_{\max}(N)$ denote the minimum and maximum eigenvalues of a Hermitian matrix $N$. Additionally, $\Omega$ contains the convex hull $\Omega_{\text{in}}$ of the following $3n$ points in $\Delta$:
\begin{align}
(\tau'_k,\rho'_k,1-\tau'_k-\rho'_k)&: \tau'_k + i\rho'_k = \lambda_k(T+iR); \label{eq:point-set-1}\\
(\tau''_k,1-\tau''_k-\alpha''_k,\alpha''_k)&: \tau''_k + i\alpha''_k = \lambda_k(T+iA); \label{eq:point-set-2}\\
(1-\rho'''_k-\alpha'''_k,\rho'''_k,\alpha'''_k)&: \rho'''_k + i\alpha'''_k = \lambda_k(R+iA), \label{eq:point-set-3}
\end{align}
where $\lambda_k(M)$ denotes the $k$-th eigenvalue of a complex matrix $M$, and $k=1,2,\ldots,n$. Therefore,
\begin{equation}\label{eq:bound}
\Omega_{\text{in}} \subseteq \Omega \subseteq \Omega_{\text{out}}.
\end{equation}
See Appendices~\ref{SI-subsec:proof_outer_bound} and~\ref{SI-subsec:proof_inner_bound} for detailed proof of Eq.~(\ref{eq:bound}). These bounds are useful since computing $\Omega$ requires significant computational resources for large $n$~\cite{johnson1978}, while $\Omega_{\text{in}}$ and $\Omega_{\text{out}}$ can be determined more efficiently.

\begin{figure}
    \centering
    \includegraphics[width=0.4\textwidth]{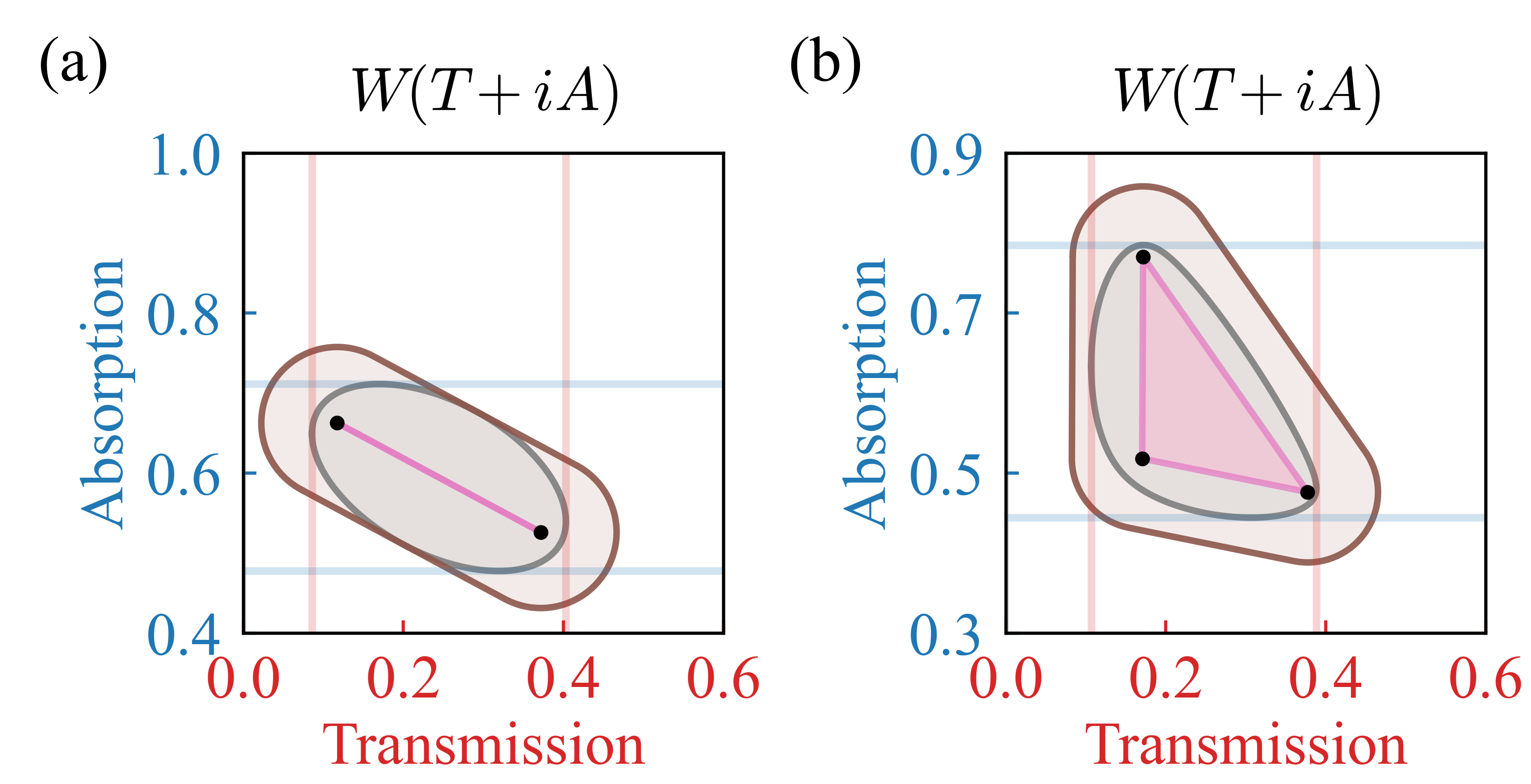}
    \caption{Non-abelian bounds on the numerical range $W(T+iA)$ (gray). The convex hull $C(T+iA)$ (pink) of eigenvalues (black dots) provides an inner bound, while the set $B(T+iA)$ (brown) gives an outer bound. Examples are shown for: (a) The $n=2$ case in Fig.~\ref{fig:example-2modes}(c). (b) The $n=3$ case in Fig.~\ref{fig:example-3modes}(c).}
    \label{fig:non-abelian-effects}
\end{figure}

Third, we reveal the non-abelian effects in joint coherent control. We examine how the non-commutativity among the $T$, $R$, and $A$ matrices affects the shape of $\Omega$. From  Eq.~\eqref{eq:energy_conservation_condition}, the commutators satisfy
\begin{equation}
    [T,R] = [R,A] = [A,T].
\end{equation}
Thus, a passive system is either abelian, for which these three matrices commute pairwise, or non-abelian, for which no pair of these matrices commutes. We provide concrete physical examples of abelian and non-abelian systems using a simple dielectric slab in Appendix~\ref{SI-sec:example_abelian_non-abelian_systems}.

Our main result for this section is that abelian systems achieve the inner bound in Eq.~\eqref{eq:bound}:
\begin{equation}
\Omega_{\text{in}} = \Omega.
\end{equation}
In contrast, for non-abelian systems, $\Omega$ can extend beyond $\Omega_{\text{in}}$, and the size of their gap is governed by the degree of non-abelianness. 

More specifically, for abelian systems, there exists a unitary matrix $U$ such that $U^\dagger TU = D_T$, $U^\dagger RU = D_R$, and $U^\dagger AU = D_A$ are diagonal matrices~\cite{horn2012}, and the columns of $U$ provide the simultaneous eigenvectors of $T$, $R$, and $A$. It follows that the three sets of points in Eqs.~\eqref{eq:point-set-1}--\eqref{eq:point-set-3} coincide, yielding at most $n$ distinct points. The set $\Omega$ equals the convex hull $\Omega_{\text{in}}$ of these points, which forms a convex polygon. (A line segment or a point is a degenerate polygon). (See Appendix~\ref{SI-subsec:abelian} for detailed proof.) Fig.~\ref{fig:shape-n=3}(a) shows an example of $\Omega$ for an abelian system with $n=3$.

For non-abelian systems, $T$, $R$, and $A$ cannot be simultaneously diagonalized by unitary similarity. The three sets of points in Eqs.~\eqref{eq:point-set-1}--\eqref{eq:point-set-3} are generally distinct, and $\Omega$ can extend beyond the convex hull $\Omega_{\text{in}}$ of these points. Figs.~\ref{fig:example-2modes}(a) and~\ref{fig:example-3modes}(a) demonstrate these behaviors for non-abelian systems with $n=2$ and $n=3$, respectively.

The above analysis suggests that the gap size between $\Omega$ and $\Omega_{\text{in}}$ depends on the degree to which the matrices fail to commute. We make this intuitive argument precise by introducing a measure of non-abelianness. Note that two Hermitian matrices such as $T$ and $R$ commute if and only if $T+ i R$ is normal~\cite{horn2012}. Thus, the degree of non-abelianness between $T$ and $R$ can be characterized by the departure from normality~\cite{henrici1962} for $T+iR$ defined as:
\begin{equation}
\operatorname{dep}(T + iR) \coloneqq \sqrt{\sum_{j}\left[\sigma_{j}^{2}(T+iR) - |\lambda_{j}(T+iR)|^{2}\right]} \geq 0  
\end{equation}
where $\sigma_{j}(M)$ denotes the $j$-th singular value of a matrix $M$. It is known that $\operatorname{dep}(T + iR) = 0$ if and only if $T+ iR$ is normal, that is, $T$ and $R$ commute. Using this measure of non-abelianness, one can prove the following bound for the numerical range $W(T+iR)$~\cite{henrici1962}:
\begin{equation} \label{eq:CWB_TR}
C(T+iR) \subseteq W(T + iR) \subseteq B(T+iR). 
\end{equation}
Here $C(T+iR)$ denotes the convex hull of all eigenvalues of $T+iR$, and 
\begin{equation}
    B(T+iR) = C(T+iR) + D(T+iR)
\end{equation}
denotes the Minkowski sum of $C(T+iR)$ and $D(T+iR)$, a circular disk centered at the origin with a radius of
\begin{equation} \label{eq:rd}
   \sqrt{(1-1/n)/2} \, \operatorname{dep}(T+iR). 
\end{equation}
(The Minkowski sum of two subsets $P$ and $Q$ of $\mathbb{R}^2$ is the subset $P+Q \coloneqq \{u+v \in \mathbb{R}^{2} \mid u\in P, v \in Q\}$~\cite{barvinok2002,boyd2004}.) The constant $\sqrt{(1-1/n)/2}$ in Eq.~\eqref{eq:rd} is optimal~\cite{henrici1962}. Similarly, we obtain
\begin{align}
C(T+iA) \subseteq W(T + iA) \subseteq B(T+iA), \label{eq:CWB_TA}\\
C(R+iA) \subseteq W(R + iA) \subseteq B(R+iA). \label{eq:CWB_RA}
\end{align}
These bounds in Eqs.~(\ref{eq:CWB_TR}), (\ref{eq:CWB_TA}), and (\ref{eq:CWB_RA}) combined with Eqs.~(\ref{eq:main_result-1})-(\ref{eq:main_result-3}) lead to the corresponding bounds for $\Omega$. As numerical illustrations, Fig.~(\ref{fig:non-abelian-effects}) demonstrates Eq.~(\ref{eq:CWB_TA}) using previous examples of Fig.~\ref{fig:example-2modes}(c) and~\ref{fig:example-3modes}(c).

An interesting converse problem is if one can determine whether the system is abelian or non-abelian from the shape of $\Omega$. It can be proven that $\Omega_{\text{in}} = \Omega$ if and only if $\Omega$ is a polygon (see Appendix~\ref{SI-subsec:polygon}). Thus, if $\Omega$ is not a polygon, then the system is non-abeliean. If $\Omega$ is a polygon, then the result depends on the number of input ports $n$: (1) when $n \leq 4$, the system is abelian; (2) when $n \geq 5$, either the system is abelian, or the system is non-abelian and $T+iR$ (or equialently, $T+iA$ or $R+iA$) is unitarily similar to a direct sum of two matrices $M_1 \oplus M_2$ where $M_1$ is normal and $W(M_2) \subseteq W(M_1)$~\cite{horn1991}. (See Appendix~\ref{SI-subsec:shape}.) We illustrate this criterion using previous examples when $n=2$ or $3$: $\Omega$ in Figs.~\ref{fig:example-2modes}(a),~\ref{fig:example-3modes}(a),~\ref{fig:shape-n=3}(b,c,d) are not polygons, and $T$ and $R$ do not commute in these cases. In contrast, $\Omega$ in Figs.~\ref{fig:shape-n=3}(a,e,f) are polygons, and $T$ and $R$ commute in these cases. 

Now we turn to Question~2. It reduces to the following inverse numerical range problem: Find a unit vector $\bm{a_0}$ such that $\bm{a_0}^{\dagger}(t^{\dagger}t +i r^{\dagger}r) \bm{a_0} = \tau_{0} + i \rho_{0}$. This problem can be solved numerically using any algorithm provided in Refs.~\cite{uhlig2008,carden2009,chorianopoulos2010,meurant2012,bebiano2014}. As an illustration, we consider the $t$ and $r$ matrices used to generate Fig.~\ref{fig:example-3modes} with their numerical values given in Eqs.~(\ref{eq:t-matrix-3modes}) and (\ref{eq:r-matrix-3modes}) in Appendix~\ref{SI-subsec:T_R_A_matrices-Fig3}. Our task is to construct an $\bm{a_0}$ with an assigned goal:
\begin{equation}\label{eq:goal_tuple}
(\tau_0, \rho_0, \alpha_0) = (0.2,0.2,0.6).
\end{equation}
First, we verify that $(\tau_{0},\rho_{0},\alpha_{0}) \in \Omega$ as indicated by the purple cross in Fig.~\ref{fig:example-3modes}(a). We apply the algorithm in Ref.~\cite{bebiano2014} and obtain a unit input vector 
\begin{equation}
\bm{a_{0}} = (0.37 - 0.50i, 0.24 - 0.66i, -0.35 + 0.08i)^T.    
\end{equation}
Importantly, this algorithm allows us to achieve all the prescribed transmittance, reflectance, and absorptance simultaneously with a single coherent input.

In conclusion, we have developed a comprehensive theory for joint coherent control of wave transmission, reflection, and absorption. We show that the numerical range provides the mathematical framework for characterizing all achievable responses simultaneously. For any multiport wave system, we determine the set of all attainable combinations of transmission, reflection, and absorption. Our theory reveals non-abelian effects in wave control - the degree of noncommutativity between transmission, reflection, and absorption matrices constrains the achievable responses in a way quantified by departure from normality. These results establish fundamental bounds for joint coherent control and provide constructive algorithms for achieving arbitrary target responses within the attainable set. The theory applies to all wave types and can be readily extended to other physical quantities. Our results lay the foundation for advanced wavefront shaping applications requiring precise control over multiple wave characteristics. 
\begin{acknowledgments}
C.G. is supported by the Jack Kilby/Texas Instruments Endowed Faculty Fellowship. S.F. is funded by the U.S. Department of Energy (Grant No. DE-FG02-07ER46426) and by a Simons Investigator in Physics grant from the Simons Foundation (Grant No. 827065).
\end{acknowledgments}

\appendix

\section{Geometry of the disordered waveguide}\label{SI-sec:geometry_waveguide}

\begin{figure}[htbp]
    \centering
    \includegraphics[width=0.45\textwidth]{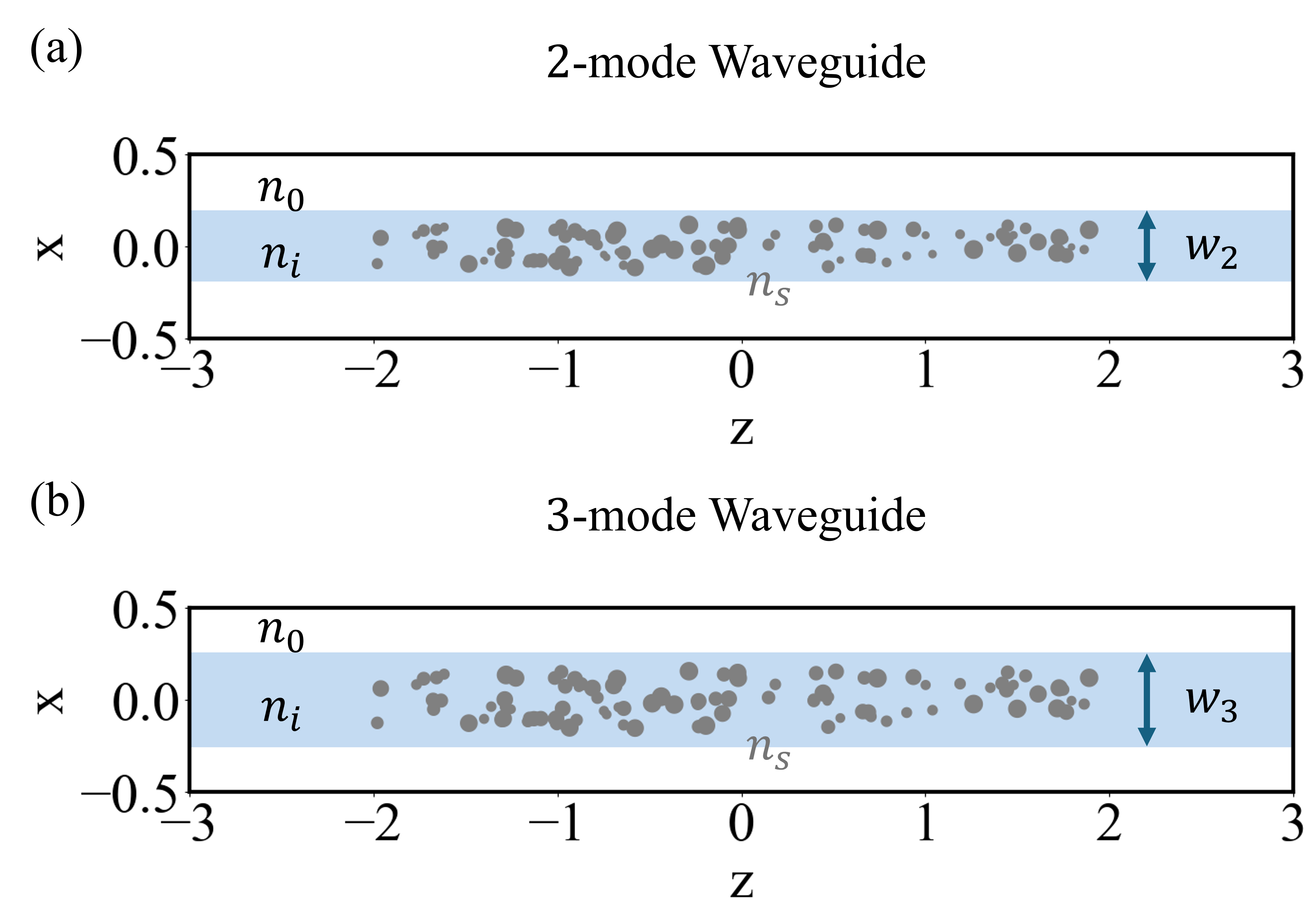}
    \caption{Structural details of the disordered waveguide. (a) Two-mode configuration with waveguide width $w_2=0.39~\mu\text{m}$, showing the distribution of silica scatterers in cross-section. (b) Three-mode configuration with increased waveguide width $w_3=0.52~\mu\text{m}$ and modified scatterer arrangement.}
    \label{fig-SI:geometry_waveguide}
\end{figure}

We present the detailed geometry of the disordered multimode waveguide shown in Fig.~\ref{fig:system}(d). The waveguide contains cylindrical scatterers arranged in a random pattern, with dielectric constants $n_0$, $n_i$, and $n_s$ as specified in the main text. Fig.~\ref{fig-SI:geometry_waveguide}(a) illustrates the two-mode waveguide with a width $w_2=0.39~\mu\text{m}$ to produce the results presented in Fig.~\ref{fig:example-2modes}. Fig.~\ref{fig-SI:geometry_waveguide}(b) illustrates the three-mode waveguide with a width $w_3=0.52~\mu\text{m}$ to produce the results presented in Fig.~\ref{fig:example-3modes}.

\section{Proof of Theorem~\ref{theorem:main}}\label{SI-sec:proof_main_theorem}
\begin{proof}
We prove Eq.~(\ref{eq:main_result-1}). Let $\Omega'$ denote its right-hand side. We first show $\Omega \subseteq \Omega'$: For any $(\tau[\bm{a}], \rho[\bm{a}], \alpha[\bm{a}]) \in \Omega$, we have
\begin{equation}
\tau[\bm{a}] + i \rho[\bm{a}] = \bm{a}^{\dagger}(T+iR) \bm{a} \in W(T + i R)
\end{equation}
and by energy conservation [Eq.~(\ref{eq:energy_conservation_condition})],
\begin{equation}
    \alpha[\bm{a}] = 1 - \tau[\bm{a}] - \rho[\bm{a}],
\end{equation}
thus $(\tau[\bm{a}], \rho[\bm{a}], \alpha[\bm{a}]) \in \Omega'$.

We then show $\Omega' \subseteq \Omega$: For any $(\tau, \rho, \alpha) \in \Omega'$, we need to find a unit vector $\bm{a}$ such that $\tau[\bm{a}] = \tau$, $\rho[\bm{a}] = \rho$, and $\alpha[\bm{a}] = \alpha$. Since $\tau + i\rho \in W(T + i R)$, there exists a unit vector $\bm{a}$ such that 
\begin{equation}
\tau + i\rho = \bm{a}^{\dagger} (T+ i R) \bm{a} = \bm{a}^{\dagger} T \bm{a} + i \bm{a}^\dagger R \bm{a}.    
\end{equation}
As $T$ and $R$ are positive semidefinite Hermitian matrices, comparing real and imaginary parts gives
\begin{equation}
\tau = \bm{a}^{\dagger} T \bm{a}, \quad   \rho = \bm{a}^{\dagger} R \bm{a}.  
\end{equation}
Therefore
\begin{equation}
\tau[\bm{a}] = \tau, \quad \rho[\bm{a}] = \rho, \quad \alpha[\bm{a}] = 1 - \tau - \rho = \alpha.
\end{equation}
This completes the proof of Eq.~(\ref{eq:main_result-1}). The proofs for Eqs.~(\ref{eq:main_result-2}) and (\ref{eq:main_result-3}) are similar. 
\end{proof}

\section{Algorithm for computing the numerical range boundary}\label{SI-sec:boundary_numerical_range}

The numerical range $W(M)$ of a matrix $M$ is convex and compact, so its boundary completely characterizes the set. Johnson's algorithm~\cite{johnson1978} computes boundary points by analyzing eigenvalues of the Hermitian part of rotated matrices, based on the following observation: For any unit vector $z \in \mathbb{C}^m$ and matrix $M \in \mathbb{C}^{m\times m}$,
\begin{equation}
z^{\dagger} M z = z^{\dagger}  \operatorname{Re}M  z + i z^{\dagger}  \operatorname{Im}M  z,
\end{equation}
where
\begin{equation}
\operatorname{Re}M = \frac{1}{2}(M+M^{\dagger}), \quad \operatorname{Im}M = \frac{1}{2i}(M-M^{\dagger}).
\end{equation}
The real part of $z^{\dagger} M z$ must lie between the largest and smallest eigenvalues of $\operatorname{Re}M$, defining a vertical strip containing $W(M)$. Since $W(e^{i\theta}M)=e^{i\theta} W(M)$, applying this to $M(\theta) = e^{i\theta} M$ for $\theta \in [0,\pi]$ generates the boundary of $W(M)$~\cite{higham2023a}. A detailed MATLAB implementation is available in the Matrix Computation Toolbox~\cite{higham2024}. More efficient algorithms have been developed by Loisel and Maxwell~\cite{loisel2018} and Uhlig~\cite{uhlig2020}.

\section{$T$, $R$, and $A$ matrices for Figs.~\ref{fig:example-2modes},~\ref{fig:example-3modes}, and~\ref{fig:shape-n=3}}\label{SI-sec:T_R_A_matrices}

\subsection{Matrices for Fig.~\ref{fig:example-2modes}}\label{SI-subsec:T_R_A_matrices-Fig2}

The two-mode disordered waveguide has transmission and reflection matrices:
\begin{align} \label{eq:t-matrix-2modes}
t &= \begin{pmatrix}
-0.24 + 0.06i & 0.15 - 0.14i \\
0.20 - 0.11i & -0.02 - 0.58i
\end{pmatrix}, \\
r &= \begin{pmatrix}
-0.39 - 0.06i & -0.05 - 0.20i \\
-0.20 + 0.05i & -0.13 - 0.25i
\end{pmatrix}.\label{eq:r-matrix-2modes}
\end{align}
We calculate $T$, $R$, and $A$ from $t$ and $r$ using Eq.~(\ref{eq:def-T_R_A}):
\begin{align}
T &= \begin{pmatrix}
0.12 & 0.01 - 0.09i \\
0.01 + 0.09i & 0.37
\end{pmatrix}
, \\
R &= \begin{pmatrix}
0.20 & 0.05 + 0.13i \\
0.05 - 0.13i & 0.12
\end{pmatrix}, \\
A &= \begin{pmatrix}
0.69 & -0.06 - 0.04i \\
-0.06 + 0.04i & 0.50
\end{pmatrix}.
\end{align}

\subsection{Matrices for Fig.~\ref{fig:example-3modes}}\label{SI-subsec:T_R_A_matrices-Fig3}

The three-mode disordered waveguide has transmission and reflection matrices:
\begin{align}\label{eq:t-matrix-3modes}
t &= \begin{pmatrix}
0.23 + 0.22i & 0.12 - 0.02i & -0.05 + 0.03i \\
0.00 - 0.17i & 0.40 + 0.44i & 0.03 - 0.01i \\
-0.07 - 0.13i & -0.07 - 0.01i & -0.43 - 0.07i
\end{pmatrix}, \\
r &= \begin{pmatrix}
0.31 + 0.35i & 0.07 - 0.08i & -0.01 + 0.25i \\
-0.03 - 0.10i & -0.37 + 0.01i & 0.03 + 0.07i \\
0.09 - 0.23i & 0.06 - 0.04i & -0.05 + 0.04i
\end{pmatrix}. \label{eq:r-matrix-3modes}
\end{align}
We calculate $T$, $R$, and $A$ from $t$ and $r$ using Eq.~(\ref{eq:def-T_R_A}):
\begin{align}
T &= \begin{pmatrix}
0.15 & -0.05 + 0.03i & 0.03 - 0.03i \\
-0.05 - 0.03i & 0.37 & 0.03 - 0.01i \\
0.03 + 0.03i & 0.03 + 0.01i & 0.20
\end{pmatrix}, \\
R &= \begin{pmatrix}
0.29 & 0.02 - 0.07i & 0.06 + 0.07i \\
0.02 + 0.07i & 0.15 & -0.03 - 0.01i \\
0.06 - 0.07i & -0.03 + 0.01i & 0.07
\end{pmatrix}, \\
A &= \begin{pmatrix}
0.56 & 0.03 + 0.04i & -0.10 - 0.05i  \\
0.03 - 0.04i & 0.47 & 0.00 + 0.02i \\
-0.10 + 0.05i & 0.00 - 0.02i & 0.73
\end{pmatrix}.
\end{align}

\subsection{Matrices for Fig.~\ref{fig:shape-n=3}}\label{SI-subsec:T_R_A_matrices-Fig4}

Here we provide the $T$ and $R$ matrices used to generate each panel in Fig.~\ref{fig:shape-n=3}. The corresponding $A$ matrices can be derived using $A = I - T - R$.  

\noindent (a) Triangular disk: 
\begin{align}
T &= \begin{pmatrix}
0.1 & 0 & 0 \\
0 & 0.6 & 0  \\
0 & 0 & 0.1
\end{pmatrix}, \\ 
R &= \begin{pmatrix}
0.7 & 0 & 0 \\
0 & 0.3 & 0 \\
0 & 0 & 0.1
\end{pmatrix}.    
\end{align}
(b) Convex hull of an ellipse and a point:
\begin{align}
T &= \begin{pmatrix}
0.2 & 0.15 + 0.05i & 0 \\
0.15 - 0.05i & 0.6 & 0  \\
0 & 0 & 0.1
\end{pmatrix}, \\
R &= \begin{pmatrix}
0.5 & 0.05 - 0.15i & 0  \\
0.05 + 0.15i & 0.1 & 0 \\
0 & 0 & 0.1
\end{pmatrix}.   
\end{align}
(c) Shape with flat boundary:
\begin{align}
T &= \begin{pmatrix}
0.2 & 0.14 + 0.06i & 0.15  \\
0.14 - 0.05i & 0.2 & 0.15  \\
0.15 & 0.15 & 0.2 
\end{pmatrix}, \\  
R &= \begin{pmatrix}
0.3 & 0.06 -0.14i & -0.15i \\
0.06 + 0.14i & 0.3 & -0.15i  \\
0.15i & 0.15i & 0.3
\end{pmatrix}.
\end{align}
(d) Elliptical disk:
\begin{align}
T &= \begin{pmatrix}
0.2 & 0.15 + 0.05i & 0 \\
0.15 - 0.05i & 0.6 & 0  \\
0 & 0 & 0.4
\end{pmatrix}, \\ 
R &= \begin{pmatrix}
0.5 & 0.05 - 0.15i & 0 \\
0.05 + 0.15i & 0.1 & 0 \\
0 & 0 & 0.3
\end{pmatrix}.    
\end{align}
(e) Line segment:
\begin{align}
T &= \begin{pmatrix}
0.2 & 0 & 0  \\
0 & 0.5 & 0 \\
0 & 0 & 0.5
\end{pmatrix}, \\
R &=\begin{pmatrix}
0.6 & 0 & 0 \\
0 & 0.1 & 0 \\
0 & 0 & 0.1
\end{pmatrix}.
\end{align}
(f) Point:
\begin{align}
T &= \begin{pmatrix}
0.33 & 0 & 0  \\
0 & 0.33 & 0 \\
0 & 0 & 0.33
\end{pmatrix}, \\
R &= \begin{pmatrix}
0.33 & 0 & 0  \\
0 & 0.33 & 0 \\
0 & 0 & 0.33
\end{pmatrix}.
\end{align}

\section{Proof of the general properties of $\Omega$}\label{SI-sec:proof_general_properties_Omega}

The set $\Omega$ inherits its properties from the numerical range $W(T+iR)$ through the affine function
\begin{equation}
f: \mathbb{R}^2 \to \mathbb{R}^3, \quad
(\tau,\rho) \mapsto (\tau,\rho,1-\tau-\rho), \label{eq:affine-function}
\end{equation}
where we identify the complex plane with $\mathbb{R}^2$. We prove four key properties of $\Omega$:

\subsection{Compactness}\label{SI-subsec:proof_compactness}
The numerical range $W(T+iR)$ is compact (see Ref.~\cite{horn1991}, p.8, Property 1.2.1.). Since $f$ is continuous and the continuous image of a compact set is compact, $\Omega$ is compact.

\subsection{Convexity} \label{SI-subsec:proof_convexity}
The numerical range $W(T+iR)$ is convex (see Ref.~\cite{horn1991}, p.8, Property 1.2.2.). Since the image of a convex set under an affine function is convex (see Ref.~\cite{boyd2004}, p.36.), $\Omega$ is convex.

\subsection{Outer bound $\Omega \subseteq \Omega_{\text{out}}$}\label{SI-subsec:proof_outer_bound}
Since $T$ is Hermitian, the quadratic form $\tau = \bm{a}^\dagger T\bm{a}$ takes all values satisfying 
\begin{equation}
\lambda_{\min}(T) \leq \tau \leq \lambda_{\max}(T)    
\end{equation}
as $\bm{a}$ ranges over complex unit vectors (see Ref.~\cite{horn1991}, p.12.). The same holds for $R$ and $A$, giving:
\begin{align}
\lambda_{\min}(R) \leq \rho &\leq \lambda_{\max}(R),\\
\lambda_{\min}(A) \leq \alpha &\leq \lambda_{\max}(A),
\end{align}
and all the bounds are attainable. Therefore, $\Omega$ is inscribed in the hexagon $\Omega_{\text{out}}$ defined in Eq.~\eqref{eq:def-hexagon}.

\subsection{Inner bound $\Omega_{\text{in}} \subseteq \Omega$} \label{SI-subsec:proof_inner_bound}
The eigenvalues of $T+iR$ lie in $W(T+iR)$ (see Ref.~\cite{horn1991}, p.10, Property 1.2.6.), giving $\tau'_k + i\rho'_k = \lambda_k(T+iR) \in W(T+iR)$. Therefore, $(\tau'_k,\rho'_k,1-\tau'_k-\rho'_k) \in \Omega$ for all $k=1,2,\ldots,n$. Similarly, the points from Eqs.~\eqref{eq:point-set-2} and~\eqref{eq:point-set-3} belong to $\Omega$. The convexity of $\Omega$ then implies that $\Omega_{\text{in}}$, the convex hull of these $3n$ points, is contained in $\Omega$.

\section{Physical examples of abelian and non-abelian systems}\label{SI-sec:example_abelian_non-abelian_systems}

\begin{figure}[htbp]
    \centering
    \includegraphics[width=0.4\textwidth]{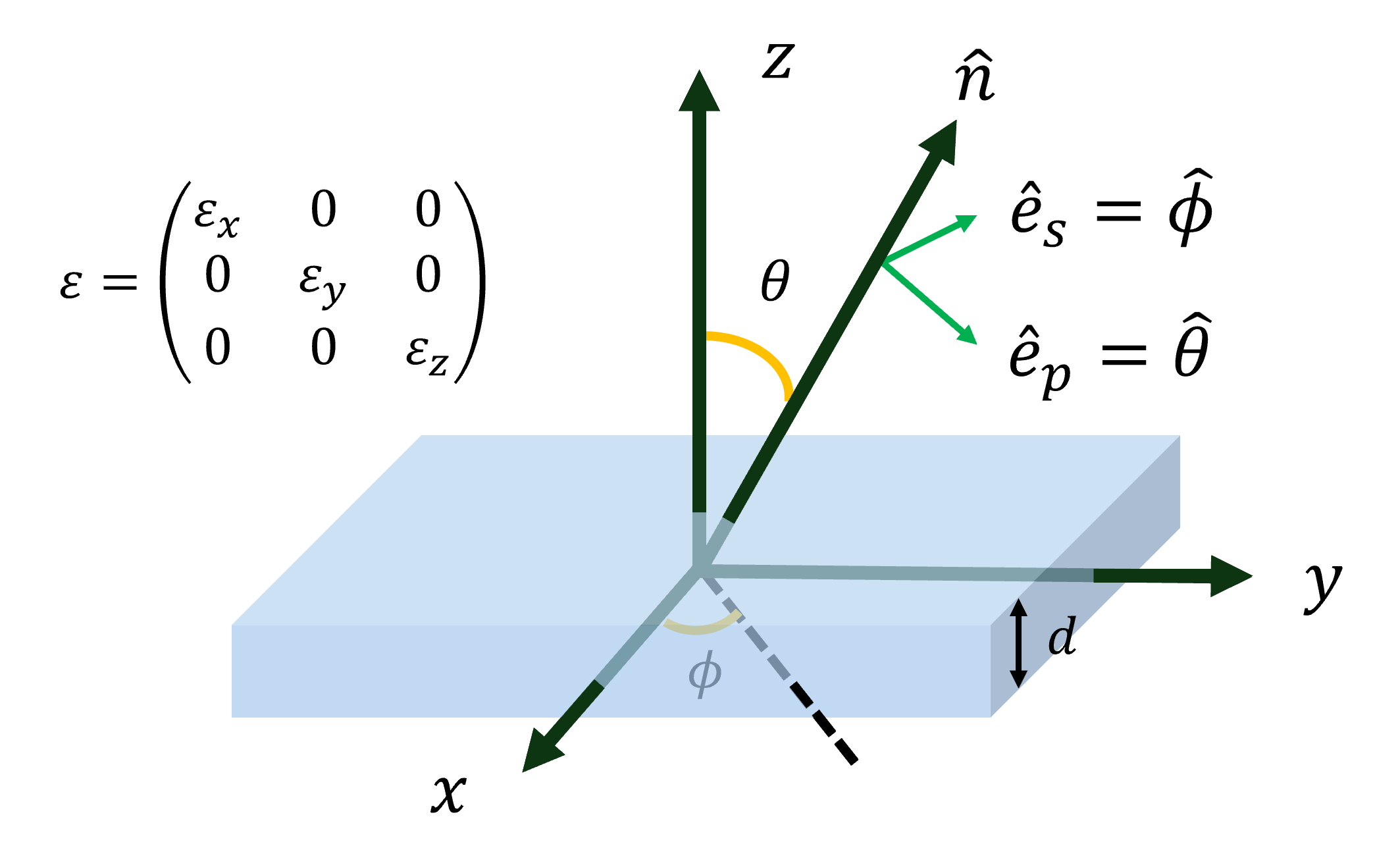}
    \caption{The geometry for the dielectric slab example. The structure is a planar slab made of a dielectric medium with a relative permittivity tensor $\varepsilon = \operatorname{diag}(\varepsilon_{x},\varepsilon_{y},\varepsilon_{z})$. The slab has a thickness of $d=1\mu m$. Light is incident from the bottom side of the slab with a wavelength of $\lambda = 1\mu m$ at a polar angle $\theta=64^\circ$ and an azimuthal angle $\phi=45^\circ$. $\hat{e}_s$ and $\hat{e}_p$ indicate the electric field directions for $s$ and $p$ polarizations.}
    \label{fig-SI:example_abelian_non-abelian_systems}
\end{figure}

Here we provide concrete physical examples of abelian and non-abelian systems. Consider a planar dielectric slab with a relative permittivity tensor $\varepsilon = \operatorname{diag}(\varepsilon_x,\varepsilon_y,\varepsilon_z)$ and thickness $d=1~\mu\text{m}$ (Fig.~\ref{fig-SI:example_abelian_non-abelian_systems}). Light of wavelength $\lambda=1~\mu\text{m}$ is incident at the polar angle $\theta=64^\circ$ and the azimuthal angle $\phi=45^\circ$. Using the transfer matrix method, we calculate the field transmission and reflection matrices $t$ and $r$ in the basis of $s$ and $p$ polarizations:
\begin{equation}
t = \begin{pmatrix} t_{ss} & t_{sp} \\ t_{ps} & t_{pp} \end{pmatrix}, \quad
r = \begin{pmatrix} r_{ss} & r_{sp} \\ r_{ps} & r_{pp} \end{pmatrix}.
\end{equation}

For an example of abelian systems, we set $\varepsilon_x = \varepsilon_y = \varepsilon_z = 9.0+0.1i$. This isotropic configuration preserves mirror symmetry with respect to the plane of incidence. Hence, the $s$ and $p$ polarization are decoupled:
\begin{align}
t &= \begin{pmatrix} 0.11-0.34i & 0 \\ 0 & 0.55-0.67i \end{pmatrix}, \\
r &= \begin{pmatrix} -0.86-0.20i & 0 \\ 0 & 0.18+0.12i \end{pmatrix}.
\end{align}
The corresponding power matrices are
\begin{align}
T &= \begin{pmatrix} 0.13 & 0 \\ 0 & 0.76 \end{pmatrix}, \\
R &= \begin{pmatrix} 0.78 & 0 \\ 0 & 0.05 \end{pmatrix}, \\
A &= \begin{pmatrix} 0.10 & 0 \\ 0 & 0.19 \end{pmatrix}.
\end{align}
These matrices commute pairwise, with all commutators vanishing:
\begin{equation}
[T,R] = [R,A] = [A,T] = O.
\end{equation}
We further confirm the abelian nature by verifying
\begin{equation}
\operatorname{dep}(T+iR) = 0.
\end{equation}

For an example of non-abelian systems, we instead set $\varepsilon_x = \varepsilon_z = 9.0+0.1i$ and $\varepsilon_y = 3.6+0.4i$. This anisotropic configuration breaks mirror symmetry with respect to the plane of incidence. Hence, the $s$ and $p$ are now coupled:
\begin{align}
t &= \begin{pmatrix} 0.07-0.30i & 0.15-0.05i \\ 0.15-0.05i & 0.23-0.60i \end{pmatrix}, \\
r &= \begin{pmatrix} -0.78-0.11i & -0.10-0.09i \\ 0.10+0.09i & 0.00+0.09i \end{pmatrix}.
\end{align}
The power matrices become
\begin{align}
T &= \begin{pmatrix} 0.12 & 0.09-0.04i \\ 0.09+0.04i & 0.44 \end{pmatrix}, \\
R &= \begin{pmatrix} 0.63 & 0.10+0.07i \\ 0.10-0.07i & 0.03 \end{pmatrix}, \\
A &= \begin{pmatrix} 0.24 & -0.19-0.03i \\ -0.19+0.03i & 0.53 \end{pmatrix}.
\end{align}
These matrices exhibit non-trivial commutators
\begin{align}
&[T,R] = [R,A] = [A,T]  \\
&= \begin{pmatrix} -0.02i & -0.09+0.00i \\ 0.09+0.00i & 0.02i \end{pmatrix} \neq O.
\end{align}
This non-abelian behaviour is quantified by a non-zero departure from normality:
\begin{equation}
\operatorname{dep}(T+iR) = 0.25 > 0.
\end{equation}

\section{Proof of non-abelian effects on $\Omega$}\label{SI-sec:proof_non-abelian_effects}

\subsection{$\Omega_{\text{in}} = \Omega$ for abelian systems}\label{SI-subsec:abelian}
For abelian systems, $T+iR$ is normal, and thus its numerical range equals the convex hull of its eigenvalues (see Ref.~\cite{horn1991}, p.11, Property 1.2.9.):
\begin{equation}
C(T+iR) = W(T+iR). \label{eq:normal-range}
\end{equation}
Here, $C(T+iR)$ denotes the convex hull of $\tau'_k + i\rho'_k = \lambda_k(T+iR)$ for $k=1,2,\ldots,n$. The set $\Omega$ is the image of $W(T+iR)$ under the affine function $f$ defined in Eq.~\eqref{eq:affine-function}. Under this mapping, a polygon transforms to a polygon with corresponding vertices. Therefore, $\Omega$ coincides with the polygon formed by the convex hull of points $(\tau'_k,\rho'_k,1-\tau'_k-\rho'_k)$ for $k=1,2,\ldots,n$ as defined in Eq.~\eqref{eq:point-set-1}. Since the three sets of points in Eqs.~\eqref{eq:point-set-1}--\eqref{eq:point-set-3} coincide in abelian systems, this polygon is identical to $\Omega_{\text{in}}$, completing the proof.

\subsection{$\Omega_{\text{in}} = \Omega$ if and only if $\Omega$ is a polygon}\label{SI-subsec:polygon}
Since $\Omega_{\text{in}}$ is the convex hull of finitely many points, it is a polygon. Thus, if $\Omega_{\text{in}}=\Omega$, then $\Omega$ must be a polygon. Conversely, suppose $\Omega$ is a polygon. Since $\Omega$ is the image of $W(T+iR)$ under the affine function $f$, and affine functions preserve polygonality, $W(T+iR)$ must also be a polygon. For numerical ranges, this occurs if and only if 
\begin{equation}
C(T+iR) = W(T+iR). \label{eq:polygon-condition}
\end{equation}
(See Ref.~\cite{horn1991}, p.~51, Corollary~1.6.4.) Applying $f$ to both sides shows that $\Omega$ equals the convex hull of the $n$ points from Eq.~\eqref{eq:point-set-1}. Similar arguments using $W(T + iA)$ or $W(R+iA)$ then establish that $\Omega$ equals the convex hull of the $n$ points from Eq.~\eqref{eq:point-set-2} or Eq.~\eqref{eq:point-set-3}. Therefore, $\Omega = \Omega_{\text{in}}$,  the convex hull of all the three sets of points.

\subsection{Determining the system type from $\Omega$'s shape}\label{SI-subsec:shape}
If $\Omega$ is not a polygon, then by the previous result in Sec.~\ref{SI-subsec:polygon}, $\Omega_{\text{in}} \neq \Omega$. The contrapositive of the result in Sec.~\ref{SI-subsec:abelian} then implies that the system is non-abelian.

If $\Omega$ is a polygon, $W(T+iR)$ must be a polygon with $C(T+iR) = W(T+iR)$. For $n \leq 4$, this equality holds if and only if $T+iR$ is normal (see Ref.~\cite{horn1991}, p.52, Corollary~1.6.9.), meaning that the system is abelian. For $n \geq 5$, this equality holds in two cases:
\begin{enumerate}
\item $T+iR$ is normal (thus the system is abelian).
\item $T+iR$ is non-normal (thus the system is non-abelian) and is unitarily similar to a matrix of the form
\begin{equation}
\begin{pmatrix}
M_1 & 0\\
0 & M_2
\end{pmatrix}, \label{eq:block-form}
\end{equation}
where $M_1$ is normal and $W(M_2) \subseteq W(M_1)$.
\end{enumerate}
The same classification applies when considering $T+iA$ or $R+iA$ instead of $T+iR$.

\bibliography{main}

\begin{thebibliography}{116}%
\makeatletter
\providecommand \@ifxundefined [1]{%
 \@ifx{#1\undefined}
}%
\providecommand \@ifnum [1]{%
 \ifnum #1\expandafter \@firstoftwo
 \else \expandafter \@secondoftwo
 \fi
}%
\providecommand \@ifx [1]{%
 \ifx #1\expandafter \@firstoftwo
 \else \expandafter \@secondoftwo
 \fi
}%
\providecommand \natexlab [1]{#1}%
\providecommand \enquote  [1]{``#1''}%
\providecommand \bibnamefont  [1]{#1}%
\providecommand \bibfnamefont [1]{#1}%
\providecommand \citenamefont [1]{#1}%
\providecommand \href@noop [0]{\@secondoftwo}%
\providecommand \href [0]{\begingroup \@sanitize@url \@href}%
\providecommand \@href[1]{\@@startlink{#1}\@@href}%
\providecommand \@@href[1]{\endgroup#1\@@endlink}%
\providecommand \@sanitize@url [0]{\catcode `\\12\catcode `\$12\catcode `\&12\catcode `\#12\catcode `\^12\catcode `\_12\catcode `\%12\relax}%
\providecommand \@@startlink[1]{}%
\providecommand \@@endlink[0]{}%
\providecommand \url  [0]{\begingroup\@sanitize@url \@url }%
\providecommand \@url [1]{\endgroup\@href {#1}{\urlprefix }}%
\providecommand \urlprefix  [0]{URL }%
\providecommand \Eprint [0]{\href }%
\providecommand \doibase [0]{https://doi.org/}%
\providecommand \selectlanguage [0]{\@gobble}%
\providecommand \bibinfo  [0]{\@secondoftwo}%
\providecommand \bibfield  [0]{\@secondoftwo}%
\providecommand \translation [1]{[#1]}%
\providecommand \BibitemOpen [0]{}%
\providecommand \bibitemStop [0]{}%
\providecommand \bibitemNoStop [0]{.\EOS\space}%
\providecommand \EOS [0]{\spacefactor3000\relax}%
\providecommand \BibitemShut  [1]{\csname bibitem#1\endcsname}%
\let\auto@bib@innerbib\@empty
\bibitem [{\citenamefont {Sebbah}(2001)}]{sebbah2001a}%
  \BibitemOpen
  \bibinfo {editor} {\bibfnamefont {P.}~\bibnamefont {Sebbah}},\ ed.,\ \href@noop {} {\emph {\bibinfo {title} {Waves and Imaging through Complex Media}}}\ (\bibinfo  {publisher} {{Kluwer Academic Publishers}},\ \bibinfo {address} {{Dordrecht ; Boston}},\ \bibinfo {year} {2001})\BibitemShut {NoStop}%
\bibitem [{\citenamefont {Ntziachristos}(2010)}]{ntziachristos2010a}%
  \BibitemOpen
  \bibfield  {author} {\bibinfo {author} {\bibfnamefont {V.}~\bibnamefont {Ntziachristos}},\ }\bibfield  {title} {\bibinfo {title} {Going deeper than microscopy: The optical imaging frontier in biology},\ }\href {https://doi.org/10.1038/nmeth.1483} {\bibfield  {journal} {\bibinfo  {journal} {Nature Methods}\ }\textbf {\bibinfo {volume} {7}},\ \bibinfo {pages} {603} (\bibinfo {year} {2010})}\BibitemShut {NoStop}%
\bibitem [{\citenamefont {{\v C}i{\v z}m{\'a}r}\ and\ \citenamefont {Dholakia}(2012)}]{cizmar2012}%
  \BibitemOpen
  \bibfield  {author} {\bibinfo {author} {\bibfnamefont {T.}~\bibnamefont {{\v C}i{\v z}m{\'a}r}}\ and\ \bibinfo {author} {\bibfnamefont {K.}~\bibnamefont {Dholakia}},\ }\bibfield  {title} {\bibinfo {title} {Exploiting multimode waveguides for pure fibre-based imaging},\ }\href {https://doi.org/10.1038/ncomms2024} {\bibfield  {journal} {\bibinfo  {journal} {Nature Communications}\ }\textbf {\bibinfo {volume} {3}},\ \bibinfo {pages} {1027} (\bibinfo {year} {2012})}\BibitemShut {NoStop}%
\bibitem [{\citenamefont {Kang}\ \emph {et~al.}(2015)\citenamefont {Kang}, \citenamefont {Jeong}, \citenamefont {Choi}, \citenamefont {Ko}, \citenamefont {Yang}, \citenamefont {Joo}, \citenamefont {Lee}, \citenamefont {Lim}, \citenamefont {Park},\ and\ \citenamefont {Choi}}]{kang2015}%
  \BibitemOpen
  \bibfield  {author} {\bibinfo {author} {\bibfnamefont {S.}~\bibnamefont {Kang}}, \bibinfo {author} {\bibfnamefont {S.}~\bibnamefont {Jeong}}, \bibinfo {author} {\bibfnamefont {W.}~\bibnamefont {Choi}}, \bibinfo {author} {\bibfnamefont {H.}~\bibnamefont {Ko}}, \bibinfo {author} {\bibfnamefont {T.~D.}\ \bibnamefont {Yang}}, \bibinfo {author} {\bibfnamefont {J.~H.}\ \bibnamefont {Joo}}, \bibinfo {author} {\bibfnamefont {J.-S.}\ \bibnamefont {Lee}}, \bibinfo {author} {\bibfnamefont {Y.-S.}\ \bibnamefont {Lim}}, \bibinfo {author} {\bibfnamefont {Q.-H.}\ \bibnamefont {Park}},\ and\ \bibinfo {author} {\bibfnamefont {W.}~\bibnamefont {Choi}},\ }\bibfield  {title} {\bibinfo {title} {Imaging deep within a scattering medium using collective accumulation of single-scattered waves},\ }\href {https://doi.org/10.1038/nphoton.2015.24} {\bibfield  {journal} {\bibinfo  {journal} {Nature Photonics}\ }\textbf {\bibinfo {volume} {9}},\ \bibinfo {pages} {253} (\bibinfo {year} {2015})}\BibitemShut {NoStop}%
\bibitem [{\citenamefont {Guo}\ \emph {et~al.}(2018{\natexlab{a}})\citenamefont {Guo}, \citenamefont {Xiao}, \citenamefont {Minkov}, \citenamefont {Shi},\ and\ \citenamefont {Fan}}]{guo2018}%
  \BibitemOpen
  \bibfield  {author} {\bibinfo {author} {\bibfnamefont {C.}~\bibnamefont {Guo}}, \bibinfo {author} {\bibfnamefont {M.}~\bibnamefont {Xiao}}, \bibinfo {author} {\bibfnamefont {M.}~\bibnamefont {Minkov}}, \bibinfo {author} {\bibfnamefont {Y.}~\bibnamefont {Shi}},\ and\ \bibinfo {author} {\bibfnamefont {S.}~\bibnamefont {Fan}},\ }\bibfield  {title} {\bibinfo {title} {Photonic crystal slab {{Laplace}} operator for image differentiation},\ }\href {https://doi.org/10.1364/OPTICA.5.000251} {\bibfield  {journal} {\bibinfo  {journal} {Optica}\ }\textbf {\bibinfo {volume} {5}},\ \bibinfo {pages} {251} (\bibinfo {year} {2018}{\natexlab{a}})}\BibitemShut {NoStop}%
\bibitem [{\citenamefont {Guo}\ \emph {et~al.}(2018{\natexlab{b}})\citenamefont {Guo}, \citenamefont {Xiao}, \citenamefont {Minkov}, \citenamefont {Shi},\ and\ \citenamefont {Fan}}]{guo2018a}%
  \BibitemOpen
  \bibfield  {author} {\bibinfo {author} {\bibfnamefont {C.}~\bibnamefont {Guo}}, \bibinfo {author} {\bibfnamefont {M.}~\bibnamefont {Xiao}}, \bibinfo {author} {\bibfnamefont {M.}~\bibnamefont {Minkov}}, \bibinfo {author} {\bibfnamefont {Y.}~\bibnamefont {Shi}},\ and\ \bibinfo {author} {\bibfnamefont {S.}~\bibnamefont {Fan}},\ }\bibfield  {title} {\bibinfo {title} {Isotropic wavevector domain image filters by a photonic crystal slab device},\ }\href {https://doi.org/10.1364/JOSAA.35.001685} {\bibfield  {journal} {\bibinfo  {journal} {JOSA A}\ }\textbf {\bibinfo {volume} {35}},\ \bibinfo {pages} {1685} (\bibinfo {year} {2018}{\natexlab{b}})}\BibitemShut {NoStop}%
\bibitem [{\citenamefont {Yoon}\ \emph {et~al.}(2020)\citenamefont {Yoon}, \citenamefont {Kim}, \citenamefont {Jang}, \citenamefont {Choi}, \citenamefont {Choi}, \citenamefont {Kang},\ and\ \citenamefont {Choi}}]{yoon2020b}%
  \BibitemOpen
  \bibfield  {author} {\bibinfo {author} {\bibfnamefont {S.}~\bibnamefont {Yoon}}, \bibinfo {author} {\bibfnamefont {M.}~\bibnamefont {Kim}}, \bibinfo {author} {\bibfnamefont {M.}~\bibnamefont {Jang}}, \bibinfo {author} {\bibfnamefont {Y.}~\bibnamefont {Choi}}, \bibinfo {author} {\bibfnamefont {W.}~\bibnamefont {Choi}}, \bibinfo {author} {\bibfnamefont {S.}~\bibnamefont {Kang}},\ and\ \bibinfo {author} {\bibfnamefont {W.}~\bibnamefont {Choi}},\ }\bibfield  {title} {\bibinfo {title} {Deep optical imaging within complex scattering media},\ }\href {https://doi.org/10.1038/s42254-019-0143-2} {\bibfield  {journal} {\bibinfo  {journal} {Nature Reviews Physics}\ }\textbf {\bibinfo {volume} {2}},\ \bibinfo {pages} {141} (\bibinfo {year} {2020})}\BibitemShut {NoStop}%
\bibitem [{\citenamefont {Wang}\ \emph {et~al.}(2020)\citenamefont {Wang}, \citenamefont {Guo}, \citenamefont {Zhao},\ and\ \citenamefont {Fan}}]{wang2020p}%
  \BibitemOpen
  \bibfield  {author} {\bibinfo {author} {\bibfnamefont {H.}~\bibnamefont {Wang}}, \bibinfo {author} {\bibfnamefont {C.}~\bibnamefont {Guo}}, \bibinfo {author} {\bibfnamefont {Z.}~\bibnamefont {Zhao}},\ and\ \bibinfo {author} {\bibfnamefont {S.}~\bibnamefont {Fan}},\ }\bibfield  {title} {\bibinfo {title} {Compact {{Incoherent Image Differentiation}} with {{Nanophotonic Structures}}},\ }\href {https://doi.org/10.1021/acsphotonics.9b01465} {\bibfield  {journal} {\bibinfo  {journal} {ACS Photonics}\ }\textbf {\bibinfo {volume} {7}},\ \bibinfo {pages} {338} (\bibinfo {year} {2020})}\BibitemShut {NoStop}%
\bibitem [{\citenamefont {Long}\ \emph {et~al.}(2021)\citenamefont {Long}, \citenamefont {Guo}, \citenamefont {Wang},\ and\ \citenamefont {Fan}}]{long2021}%
  \BibitemOpen
  \bibfield  {author} {\bibinfo {author} {\bibfnamefont {O.~Y.}\ \bibnamefont {Long}}, \bibinfo {author} {\bibfnamefont {C.}~\bibnamefont {Guo}}, \bibinfo {author} {\bibfnamefont {H.}~\bibnamefont {Wang}},\ and\ \bibinfo {author} {\bibfnamefont {S.}~\bibnamefont {Fan}},\ }\bibfield  {title} {\bibinfo {title} {Isotropic topological second-order spatial differentiator operating in transmission mode},\ }\href {https://doi.org/10.1364/OL.430699} {\bibfield  {journal} {\bibinfo  {journal} {Optics Letters}\ }\textbf {\bibinfo {volume} {46}},\ \bibinfo {pages} {3247} (\bibinfo {year} {2021})}\BibitemShut {NoStop}%
\bibitem [{\citenamefont {Bertolotti}\ and\ \citenamefont {Katz}(2022)}]{bertolotti2022}%
  \BibitemOpen
  \bibfield  {author} {\bibinfo {author} {\bibfnamefont {J.}~\bibnamefont {Bertolotti}}\ and\ \bibinfo {author} {\bibfnamefont {O.}~\bibnamefont {Katz}},\ }\bibfield  {title} {\bibinfo {title} {Imaging in complex media},\ }\href {https://doi.org/10.1038/s41567-022-01723-8} {\bibfield  {journal} {\bibinfo  {journal} {Nature Physics}\ }\textbf {\bibinfo {volume} {18}},\ \bibinfo {pages} {1008} (\bibinfo {year} {2022})}\BibitemShut {NoStop}%
\bibitem [{\citenamefont {Wang}\ \emph {et~al.}(2022)\citenamefont {Wang}, \citenamefont {Jin}, \citenamefont {Guo}, \citenamefont {Zhao}, \citenamefont {Rodrigues},\ and\ \citenamefont {Fan}}]{wang2022}%
  \BibitemOpen
  \bibfield  {author} {\bibinfo {author} {\bibfnamefont {H.}~\bibnamefont {Wang}}, \bibinfo {author} {\bibfnamefont {W.}~\bibnamefont {Jin}}, \bibinfo {author} {\bibfnamefont {C.}~\bibnamefont {Guo}}, \bibinfo {author} {\bibfnamefont {N.}~\bibnamefont {Zhao}}, \bibinfo {author} {\bibfnamefont {S.~P.}\ \bibnamefont {Rodrigues}},\ and\ \bibinfo {author} {\bibfnamefont {S.}~\bibnamefont {Fan}},\ }\bibfield  {title} {\bibinfo {title} {Design of {{Compact Meta-Crystal Slab}} for {{General Optical Convolution}}},\ }\href {https://doi.org/10.1021/acsphotonics.1c02005} {\bibfield  {journal} {\bibinfo  {journal} {ACS Photonics}\ }\textbf {\bibinfo {volume} {9}},\ \bibinfo {pages} {1358} (\bibinfo {year} {2022})}\BibitemShut {NoStop}%
\bibitem [{\citenamefont {Long}\ \emph {et~al.}(2022)\citenamefont {Long}, \citenamefont {Guo}, \citenamefont {Jin},\ and\ \citenamefont {Fan}}]{long2022}%
  \BibitemOpen
  \bibfield  {author} {\bibinfo {author} {\bibfnamefont {O.~Y.}\ \bibnamefont {Long}}, \bibinfo {author} {\bibfnamefont {C.}~\bibnamefont {Guo}}, \bibinfo {author} {\bibfnamefont {W.}~\bibnamefont {Jin}},\ and\ \bibinfo {author} {\bibfnamefont {S.}~\bibnamefont {Fan}},\ }\bibfield  {title} {\bibinfo {title} {Polarization-{{Independent Isotropic Nonlocal Metasurfaces}} with {{Wavelength-Controlled Functionality}}},\ }\href {https://doi.org/10.1103/PhysRevApplied.17.024029} {\bibfield  {journal} {\bibinfo  {journal} {Physical Review Applied}\ }\textbf {\bibinfo {volume} {17}},\ \bibinfo {pages} {024029} (\bibinfo {year} {2022})}\BibitemShut {NoStop}%
\bibitem [{\citenamefont {Aulbach}\ \emph {et~al.}(2011)\citenamefont {Aulbach}, \citenamefont {Gjonaj}, \citenamefont {Johnson}, \citenamefont {Mosk},\ and\ \citenamefont {Lagendijk}}]{aulbach2011}%
  \BibitemOpen
  \bibfield  {author} {\bibinfo {author} {\bibfnamefont {J.}~\bibnamefont {Aulbach}}, \bibinfo {author} {\bibfnamefont {B.}~\bibnamefont {Gjonaj}}, \bibinfo {author} {\bibfnamefont {P.~M.}\ \bibnamefont {Johnson}}, \bibinfo {author} {\bibfnamefont {A.~P.}\ \bibnamefont {Mosk}},\ and\ \bibinfo {author} {\bibfnamefont {A.}~\bibnamefont {Lagendijk}},\ }\bibfield  {title} {\bibinfo {title} {Control of {{Light Transmission}} through {{Opaque Scattering Media}} in {{Space}} and {{Time}}},\ }\href {https://doi.org/10.1103/physrevlett.106.103901} {\bibfield  {journal} {\bibinfo  {journal} {Physical Review Letters}\ }\textbf {\bibinfo {volume} {106}},\ \bibinfo {pages} {103901} (\bibinfo {year} {2011})}\BibitemShut {NoStop}%
\bibitem [{\citenamefont {Sarma}\ \emph {et~al.}(2015)\citenamefont {Sarma}, \citenamefont {Yamilov}, \citenamefont {Liew}, \citenamefont {Guy},\ and\ \citenamefont {Cao}}]{sarma2015}%
  \BibitemOpen
  \bibfield  {author} {\bibinfo {author} {\bibfnamefont {R.}~\bibnamefont {Sarma}}, \bibinfo {author} {\bibfnamefont {A.}~\bibnamefont {Yamilov}}, \bibinfo {author} {\bibfnamefont {S.~F.}\ \bibnamefont {Liew}}, \bibinfo {author} {\bibfnamefont {M.}~\bibnamefont {Guy}},\ and\ \bibinfo {author} {\bibfnamefont {H.}~\bibnamefont {Cao}},\ }\bibfield  {title} {\bibinfo {title} {Control of mesoscopic transport by modifying transmission channels in opaque media},\ }\href {https://doi.org/10.1103/PhysRevB.92.214206} {\bibfield  {journal} {\bibinfo  {journal} {Physical Review B}\ }\textbf {\bibinfo {volume} {92}},\ \bibinfo {pages} {214206} (\bibinfo {year} {2015})}\BibitemShut {NoStop}%
\bibitem [{\citenamefont {Mounaix}\ \emph {et~al.}(2016{\natexlab{a}})\citenamefont {Mounaix}, \citenamefont {Andreoli}, \citenamefont {Defienne}, \citenamefont {Volpe}, \citenamefont {Katz}, \citenamefont {Gr{\'e}sillon},\ and\ \citenamefont {Gigan}}]{mounaix2016}%
  \BibitemOpen
  \bibfield  {author} {\bibinfo {author} {\bibfnamefont {M.}~\bibnamefont {Mounaix}}, \bibinfo {author} {\bibfnamefont {D.}~\bibnamefont {Andreoli}}, \bibinfo {author} {\bibfnamefont {H.}~\bibnamefont {Defienne}}, \bibinfo {author} {\bibfnamefont {G.}~\bibnamefont {Volpe}}, \bibinfo {author} {\bibfnamefont {O.}~\bibnamefont {Katz}}, \bibinfo {author} {\bibfnamefont {S.}~\bibnamefont {Gr{\'e}sillon}},\ and\ \bibinfo {author} {\bibfnamefont {S.}~\bibnamefont {Gigan}},\ }\bibfield  {title} {\bibinfo {title} {Spatiotemporal {{Coherent Control}} of {{Light}} through a {{Multiple Scattering Medium}} with the {{Multispectral Transmission Matrix}}},\ }\href {https://doi.org/10.1103/physrevlett.116.253901} {\bibfield  {journal} {\bibinfo  {journal} {Physical Review Letters}\ }\textbf {\bibinfo {volume} {116}},\ \bibinfo {pages} {253901} (\bibinfo {year} {2016}{\natexlab{a}})}\BibitemShut {NoStop}%
\bibitem [{\citenamefont {Jeong}\ \emph {et~al.}(2018)\citenamefont {Jeong}, \citenamefont {Lee}, \citenamefont {Choi}, \citenamefont {Kang}, \citenamefont {Hong}, \citenamefont {Park}, \citenamefont {Lim}, \citenamefont {Park},\ and\ \citenamefont {Choi}}]{jeong2018a}%
  \BibitemOpen
  \bibfield  {author} {\bibinfo {author} {\bibfnamefont {S.}~\bibnamefont {Jeong}}, \bibinfo {author} {\bibfnamefont {Y.-R.}\ \bibnamefont {Lee}}, \bibinfo {author} {\bibfnamefont {W.}~\bibnamefont {Choi}}, \bibinfo {author} {\bibfnamefont {S.}~\bibnamefont {Kang}}, \bibinfo {author} {\bibfnamefont {J.~H.}\ \bibnamefont {Hong}}, \bibinfo {author} {\bibfnamefont {J.-S.}\ \bibnamefont {Park}}, \bibinfo {author} {\bibfnamefont {Y.-S.}\ \bibnamefont {Lim}}, \bibinfo {author} {\bibfnamefont {H.-G.}\ \bibnamefont {Park}},\ and\ \bibinfo {author} {\bibfnamefont {W.}~\bibnamefont {Choi}},\ }\bibfield  {title} {\bibinfo {title} {Focusing of light energy inside a scattering medium by controlling the time-gated multiple light scattering},\ }\href {https://doi.org/10.1038/s41566-018-0120-9} {\bibfield  {journal} {\bibinfo  {journal} {Nature Photonics}\ }\textbf {\bibinfo {volume} {12}},\ \bibinfo {pages} {277} (\bibinfo {year} {2018})}\BibitemShut {NoStop}%
\bibitem [{\citenamefont {Liu}\ and\ \citenamefont {Fiore}(2020)}]{liu2020s}%
  \BibitemOpen
  \bibfield  {author} {\bibinfo {author} {\bibfnamefont {T.}~\bibnamefont {Liu}}\ and\ \bibinfo {author} {\bibfnamefont {A.}~\bibnamefont {Fiore}},\ }\bibfield  {title} {\bibinfo {title} {Designing open channels in random scattering media for on-chip spectrometers},\ }\href {https://doi.org/10.1364/optica.391612} {\bibfield  {journal} {\bibinfo  {journal} {Optica}\ }\textbf {\bibinfo {volume} {7}},\ \bibinfo {pages} {934} (\bibinfo {year} {2020})}\BibitemShut {NoStop}%
\bibitem [{\citenamefont {Miller}(2013)}]{miller2013c}%
  \BibitemOpen
  \bibfield  {author} {\bibinfo {author} {\bibfnamefont {D.~A.~B.}\ \bibnamefont {Miller}},\ }\bibfield  {title} {\bibinfo {title} {Establishing {{Optimal Wave Communication Channels Automatically}}},\ }\href {https://doi.org/10.1109/JLT.2013.2278809} {\bibfield  {journal} {\bibinfo  {journal} {Journal of Lightwave Technology}\ }\textbf {\bibinfo {volume} {31}},\ \bibinfo {pages} {3987} (\bibinfo {year} {2013})}\BibitemShut {NoStop}%
\bibitem [{\citenamefont {Miller}(2019)}]{miller2019}%
  \BibitemOpen
  \bibfield  {author} {\bibinfo {author} {\bibfnamefont {D.~A.~B.}\ \bibnamefont {Miller}},\ }\bibfield  {title} {\bibinfo {title} {Waves, modes, communications, and optics: A tutorial},\ }\href {https://doi.org/10.1364/AOP.11.000679} {\bibfield  {journal} {\bibinfo  {journal} {Advances in Optics and Photonics}\ }\textbf {\bibinfo {volume} {11}},\ \bibinfo {pages} {679} (\bibinfo {year} {2019})}\BibitemShut {NoStop}%
\bibitem [{\citenamefont {SeyedinNavadeh}\ \emph {et~al.}(2024)\citenamefont {SeyedinNavadeh}, \citenamefont {Milanizadeh}, \citenamefont {Zanetto}, \citenamefont {Ferrari}, \citenamefont {Sampietro}, \citenamefont {Sorel}, \citenamefont {Miller}, \citenamefont {Melloni},\ and\ \citenamefont {Morichetti}}]{seyedinnavadeh2024}%
  \BibitemOpen
  \bibfield  {author} {\bibinfo {author} {\bibfnamefont {S.}~\bibnamefont {SeyedinNavadeh}}, \bibinfo {author} {\bibfnamefont {M.}~\bibnamefont {Milanizadeh}}, \bibinfo {author} {\bibfnamefont {F.}~\bibnamefont {Zanetto}}, \bibinfo {author} {\bibfnamefont {G.}~\bibnamefont {Ferrari}}, \bibinfo {author} {\bibfnamefont {M.}~\bibnamefont {Sampietro}}, \bibinfo {author} {\bibfnamefont {M.}~\bibnamefont {Sorel}}, \bibinfo {author} {\bibfnamefont {D.~A.~B.}\ \bibnamefont {Miller}}, \bibinfo {author} {\bibfnamefont {A.}~\bibnamefont {Melloni}},\ and\ \bibinfo {author} {\bibfnamefont {F.}~\bibnamefont {Morichetti}},\ }\bibfield  {title} {\bibinfo {title} {Determining the optimal communication channels of arbitrary optical systems using integrated photonic processors},\ }\href {https://doi.org/10.1038/s41566-023-01330-w} {\bibfield  {journal} {\bibinfo  {journal} {Nature Photonics}\ }\textbf {\bibinfo {volume} {18}},\ \bibinfo {pages} {149} (\bibinfo {year} {2024})}\BibitemShut {NoStop}%
\bibitem [{\citenamefont {Ottens}\ \emph {et~al.}(2011)\citenamefont {Ottens}, \citenamefont {Quetschke}, \citenamefont {Wise}, \citenamefont {Alemi}, \citenamefont {Lundock}, \citenamefont {Mueller}, \citenamefont {Reitze}, \citenamefont {Tanner},\ and\ \citenamefont {Whiting}}]{ottens2011}%
  \BibitemOpen
  \bibfield  {author} {\bibinfo {author} {\bibfnamefont {R.~S.}\ \bibnamefont {Ottens}}, \bibinfo {author} {\bibfnamefont {V.}~\bibnamefont {Quetschke}}, \bibinfo {author} {\bibfnamefont {S.}~\bibnamefont {Wise}}, \bibinfo {author} {\bibfnamefont {A.~A.}\ \bibnamefont {Alemi}}, \bibinfo {author} {\bibfnamefont {R.}~\bibnamefont {Lundock}}, \bibinfo {author} {\bibfnamefont {G.}~\bibnamefont {Mueller}}, \bibinfo {author} {\bibfnamefont {D.~H.}\ \bibnamefont {Reitze}}, \bibinfo {author} {\bibfnamefont {D.~B.}\ \bibnamefont {Tanner}},\ and\ \bibinfo {author} {\bibfnamefont {B.~F.}\ \bibnamefont {Whiting}},\ }\bibfield  {title} {\bibinfo {title} {Near-{{Field Radiative Heat Transfer}} between {{Macroscopic Planar Surfaces}}},\ }\href@noop {} {\bibfield  {journal} {\bibinfo  {journal} {Physical Review Letters}\ }\textbf {\bibinfo {volume} {107}},\ \bibinfo {pages} {014301} (\bibinfo {year} {2011})}\BibitemShut {NoStop}%
\bibitem [{\citenamefont {Guha}\ \emph {et~al.}(2012)\citenamefont {Guha}, \citenamefont {Otey}, \citenamefont {Poitras}, \citenamefont {Fan},\ and\ \citenamefont {Lipson}}]{guha2012}%
  \BibitemOpen
  \bibfield  {author} {\bibinfo {author} {\bibfnamefont {B.}~\bibnamefont {Guha}}, \bibinfo {author} {\bibfnamefont {C.}~\bibnamefont {Otey}}, \bibinfo {author} {\bibfnamefont {C.~B.}\ \bibnamefont {Poitras}}, \bibinfo {author} {\bibfnamefont {S.}~\bibnamefont {Fan}},\ and\ \bibinfo {author} {\bibfnamefont {M.}~\bibnamefont {Lipson}},\ }\bibfield  {title} {\bibinfo {title} {Near-{{Field Radiative Cooling}} of {{Nanostructures}}},\ }\href@noop {} {\bibfield  {journal} {\bibinfo  {journal} {Nano Letters}\ }\textbf {\bibinfo {volume} {12}},\ \bibinfo {pages} {4546} (\bibinfo {year} {2012})}\BibitemShut {NoStop}%
\bibitem [{\citenamefont {Babuty}\ \emph {et~al.}(2013)\citenamefont {Babuty}, \citenamefont {Joulain}, \citenamefont {Chapuis}, \citenamefont {Greffet},\ and\ \citenamefont {De~Wilde}}]{babuty2013}%
  \BibitemOpen
  \bibfield  {author} {\bibinfo {author} {\bibfnamefont {A.}~\bibnamefont {Babuty}}, \bibinfo {author} {\bibfnamefont {K.}~\bibnamefont {Joulain}}, \bibinfo {author} {\bibfnamefont {P.-O.}\ \bibnamefont {Chapuis}}, \bibinfo {author} {\bibfnamefont {J.-J.}\ \bibnamefont {Greffet}},\ and\ \bibinfo {author} {\bibfnamefont {Y.}~\bibnamefont {De~Wilde}},\ }\bibfield  {title} {\bibinfo {title} {Blackbody {{Spectrum Revisited}} in the {{Near Field}}},\ }\href@noop {} {\bibfield  {journal} {\bibinfo  {journal} {Physical Review Letters}\ }\textbf {\bibinfo {volume} {110}},\ \bibinfo {pages} {146103} (\bibinfo {year} {2013})}\BibitemShut {NoStop}%
\bibitem [{\citenamefont {Rephaeli}\ \emph {et~al.}(2013)\citenamefont {Rephaeli}, \citenamefont {Raman},\ and\ \citenamefont {Fan}}]{rephaeli2013}%
  \BibitemOpen
  \bibfield  {author} {\bibinfo {author} {\bibfnamefont {E.}~\bibnamefont {Rephaeli}}, \bibinfo {author} {\bibfnamefont {A.}~\bibnamefont {Raman}},\ and\ \bibinfo {author} {\bibfnamefont {S.}~\bibnamefont {Fan}},\ }\bibfield  {title} {\bibinfo {title} {Ultrabroadband {{Photonic Structures To Achieve High-Performance Daytime Radiative Cooling}}},\ }\href {https://doi.org/10.1021/nl4004283} {\bibfield  {journal} {\bibinfo  {journal} {Nano Letters}\ }\textbf {\bibinfo {volume} {13}},\ \bibinfo {pages} {1457} (\bibinfo {year} {2013})}\BibitemShut {NoStop}%
\bibitem [{\citenamefont {Raman}\ \emph {et~al.}(2014)\citenamefont {Raman}, \citenamefont {Anoma}, \citenamefont {Zhu}, \citenamefont {Rephaeli},\ and\ \citenamefont {Fan}}]{raman2014}%
  \BibitemOpen
  \bibfield  {author} {\bibinfo {author} {\bibfnamefont {A.~P.}\ \bibnamefont {Raman}}, \bibinfo {author} {\bibfnamefont {M.~A.}\ \bibnamefont {Anoma}}, \bibinfo {author} {\bibfnamefont {L.}~\bibnamefont {Zhu}}, \bibinfo {author} {\bibfnamefont {E.}~\bibnamefont {Rephaeli}},\ and\ \bibinfo {author} {\bibfnamefont {S.}~\bibnamefont {Fan}},\ }\bibfield  {title} {\bibinfo {title} {Passive radiative cooling below ambient air temperature under direct sunlight},\ }\href@noop {} {\bibfield  {journal} {\bibinfo  {journal} {Nature}\ }\textbf {\bibinfo {volume} {515}},\ \bibinfo {pages} {540} (\bibinfo {year} {2014})}\BibitemShut {NoStop}%
\bibitem [{\citenamefont {Boriskina}\ \emph {et~al.}(2016)\citenamefont {Boriskina}, \citenamefont {Green}, \citenamefont {Catchpole}, \citenamefont {Yablonovitch}, \citenamefont {Beard}, \citenamefont {Okada}, \citenamefont {Lany}, \citenamefont {Gershon}, \citenamefont {Zakutayev}, \citenamefont {Tahersima}, \citenamefont {Sorger}, \citenamefont {Naughton}, \citenamefont {Kempa}, \citenamefont {Dagenais}, \citenamefont {Yao}, \citenamefont {Xu}, \citenamefont {Sheng}, \citenamefont {Bronstein}, \citenamefont {Rogers}, \citenamefont {Alivisatos}, \citenamefont {Nuzzo}, \citenamefont {Gordon}, \citenamefont {Wu}, \citenamefont {Wisser}, \citenamefont {Salleo}, \citenamefont {Dionne}, \citenamefont {Bermel}, \citenamefont {Greffet}, \citenamefont {Celanovic}, \citenamefont {Soljacic}, \citenamefont {Manor}, \citenamefont {Rotschild}, \citenamefont {Raman}, \citenamefont {Zhu}, \citenamefont {Fan},\ and\ \citenamefont {Chen}}]{boriskina2016}%
  \BibitemOpen
  \bibfield  {author} {\bibinfo {author} {\bibfnamefont {S.~V.}\ \bibnamefont {Boriskina}}, \bibinfo {author} {\bibfnamefont {M.~A.}\ \bibnamefont {Green}}, \bibinfo {author} {\bibfnamefont {K.}~\bibnamefont {Catchpole}}, \bibinfo {author} {\bibfnamefont {E.}~\bibnamefont {Yablonovitch}}, \bibinfo {author} {\bibfnamefont {M.~C.}\ \bibnamefont {Beard}}, \bibinfo {author} {\bibfnamefont {Y.}~\bibnamefont {Okada}}, \bibinfo {author} {\bibfnamefont {S.}~\bibnamefont {Lany}}, \bibinfo {author} {\bibfnamefont {T.}~\bibnamefont {Gershon}}, \bibinfo {author} {\bibfnamefont {A.}~\bibnamefont {Zakutayev}}, \bibinfo {author} {\bibfnamefont {M.~H.}\ \bibnamefont {Tahersima}}, \bibinfo {author} {\bibfnamefont {V.~J.}\ \bibnamefont {Sorger}}, \bibinfo {author} {\bibfnamefont {M.~J.}\ \bibnamefont {Naughton}}, \bibinfo {author} {\bibfnamefont {K.}~\bibnamefont {Kempa}}, \bibinfo {author} {\bibfnamefont {M.}~\bibnamefont {Dagenais}}, \bibinfo {author} {\bibfnamefont {Y.}~\bibnamefont {Yao}}, \bibinfo {author} {\bibfnamefont
  {L.}~\bibnamefont {Xu}}, \bibinfo {author} {\bibfnamefont {X.}~\bibnamefont {Sheng}}, \bibinfo {author} {\bibfnamefont {N.~D.}\ \bibnamefont {Bronstein}}, \bibinfo {author} {\bibfnamefont {J.~A.}\ \bibnamefont {Rogers}}, \bibinfo {author} {\bibfnamefont {A.~P.}\ \bibnamefont {Alivisatos}}, \bibinfo {author} {\bibfnamefont {R.~G.}\ \bibnamefont {Nuzzo}}, \bibinfo {author} {\bibfnamefont {J.~M.}\ \bibnamefont {Gordon}}, \bibinfo {author} {\bibfnamefont {D.~M.}\ \bibnamefont {Wu}}, \bibinfo {author} {\bibfnamefont {M.~D.}\ \bibnamefont {Wisser}}, \bibinfo {author} {\bibfnamefont {A.}~\bibnamefont {Salleo}}, \bibinfo {author} {\bibfnamefont {J.}~\bibnamefont {Dionne}}, \bibinfo {author} {\bibfnamefont {P.}~\bibnamefont {Bermel}}, \bibinfo {author} {\bibfnamefont {J.-J.}\ \bibnamefont {Greffet}}, \bibinfo {author} {\bibfnamefont {I.}~\bibnamefont {Celanovic}}, \bibinfo {author} {\bibfnamefont {M.}~\bibnamefont {Soljacic}}, \bibinfo {author} {\bibfnamefont {A.}~\bibnamefont {Manor}}, \bibinfo {author}
  {\bibfnamefont {C.}~\bibnamefont {Rotschild}}, \bibinfo {author} {\bibfnamefont {A.}~\bibnamefont {Raman}}, \bibinfo {author} {\bibfnamefont {L.}~\bibnamefont {Zhu}}, \bibinfo {author} {\bibfnamefont {S.}~\bibnamefont {Fan}},\ and\ \bibinfo {author} {\bibfnamefont {G.}~\bibnamefont {Chen}},\ }\bibfield  {title} {\bibinfo {title} {Roadmap on optical energy conversion},\ }\href@noop {} {\bibfield  {journal} {\bibinfo  {journal} {Journal of Optics}\ }\textbf {\bibinfo {volume} {18}},\ \bibinfo {pages} {073004} (\bibinfo {year} {2016})}\BibitemShut {NoStop}%
\bibitem [{\citenamefont {Raj}\ \emph {et~al.}(2017)\citenamefont {Raj}, \citenamefont {van~de Voorde},\ and\ \citenamefont {Mahajan}}]{raj2017a}%
  \BibitemOpen
  \bibinfo {editor} {\bibfnamefont {B.}~\bibnamefont {Raj}}, \bibinfo {editor} {\bibfnamefont {M.~H.}\ \bibnamefont {van~de Voorde}},\ and\ \bibinfo {editor} {\bibfnamefont {Y.~R.}\ \bibnamefont {Mahajan}},\ eds.,\ \href@noop {} {\emph {\bibinfo {title} {Nanotechnology for Energy Sustainability. {{Volume}} 3}}}\ (\bibinfo  {publisher} {{Wiley-VCH}},\ \bibinfo {address} {{Weinheim}},\ \bibinfo {year} {2017})\BibitemShut {NoStop}%
\bibitem [{\citenamefont {Fiorino}\ \emph {et~al.}(2018)\citenamefont {Fiorino}, \citenamefont {Zhu}, \citenamefont {Thompson}, \citenamefont {Mittapally}, \citenamefont {Reddy},\ and\ \citenamefont {Meyhofer}}]{fiorino2018a}%
  \BibitemOpen
  \bibfield  {author} {\bibinfo {author} {\bibfnamefont {A.}~\bibnamefont {Fiorino}}, \bibinfo {author} {\bibfnamefont {L.}~\bibnamefont {Zhu}}, \bibinfo {author} {\bibfnamefont {D.}~\bibnamefont {Thompson}}, \bibinfo {author} {\bibfnamefont {R.}~\bibnamefont {Mittapally}}, \bibinfo {author} {\bibfnamefont {P.}~\bibnamefont {Reddy}},\ and\ \bibinfo {author} {\bibfnamefont {E.}~\bibnamefont {Meyhofer}},\ }\bibfield  {title} {\bibinfo {title} {Nanogap near-field thermophotovoltaics},\ }\href@noop {} {\bibfield  {journal} {\bibinfo  {journal} {Nature Nanotechnology}\ }\textbf {\bibinfo {volume} {13}},\ \bibinfo {pages} {806} (\bibinfo {year} {2018})}\BibitemShut {NoStop}%
\bibitem [{\citenamefont {Park}\ \emph {et~al.}(2021)\citenamefont {Park}, \citenamefont {Asadchy}, \citenamefont {Zhao}, \citenamefont {Guo}, \citenamefont {Wang},\ and\ \citenamefont {Fan}}]{park2021}%
  \BibitemOpen
  \bibfield  {author} {\bibinfo {author} {\bibfnamefont {Y.}~\bibnamefont {Park}}, \bibinfo {author} {\bibfnamefont {V.~S.}\ \bibnamefont {Asadchy}}, \bibinfo {author} {\bibfnamefont {B.}~\bibnamefont {Zhao}}, \bibinfo {author} {\bibfnamefont {C.}~\bibnamefont {Guo}}, \bibinfo {author} {\bibfnamefont {J.}~\bibnamefont {Wang}},\ and\ \bibinfo {author} {\bibfnamefont {S.}~\bibnamefont {Fan}},\ }\bibfield  {title} {\bibinfo {title} {Violating {{Kirchhoff}}'s {{Law}} of {{Thermal Radiation}} in {{Semitransparent Structures}}},\ }\href {https://doi.org/10.1021/acsphotonics.1c00612} {\bibfield  {journal} {\bibinfo  {journal} {ACS Photonics}\ }\textbf {\bibinfo {volume} {8}},\ \bibinfo {pages} {2417} (\bibinfo {year} {2021})}\BibitemShut {NoStop}%
\bibitem [{\citenamefont {Zhu}\ \emph {et~al.}(2019)\citenamefont {Zhu}, \citenamefont {Fiorino}, \citenamefont {Thompson}, \citenamefont {Mittapally}, \citenamefont {Meyhofer},\ and\ \citenamefont {Reddy}}]{zhu2019c}%
  \BibitemOpen
  \bibfield  {author} {\bibinfo {author} {\bibfnamefont {L.}~\bibnamefont {Zhu}}, \bibinfo {author} {\bibfnamefont {A.}~\bibnamefont {Fiorino}}, \bibinfo {author} {\bibfnamefont {D.}~\bibnamefont {Thompson}}, \bibinfo {author} {\bibfnamefont {R.}~\bibnamefont {Mittapally}}, \bibinfo {author} {\bibfnamefont {E.}~\bibnamefont {Meyhofer}},\ and\ \bibinfo {author} {\bibfnamefont {P.}~\bibnamefont {Reddy}},\ }\bibfield  {title} {\bibinfo {title} {Near-field photonic cooling through control of the chemical potential of photons},\ }\href@noop {} {\bibfield  {journal} {\bibinfo  {journal} {Nature}\ }\textbf {\bibinfo {volume} {566}},\ \bibinfo {pages} {239} (\bibinfo {year} {2019})}\BibitemShut {NoStop}%
\bibitem [{\citenamefont {Li}\ \emph {et~al.}(2019)\citenamefont {Li}, \citenamefont {Zhai}, \citenamefont {He}, \citenamefont {Gan}, \citenamefont {Wei}, \citenamefont {Heidarinejad}, \citenamefont {Dalgo}, \citenamefont {Mi}, \citenamefont {Zhao}, \citenamefont {Song}, \citenamefont {Dai}, \citenamefont {Chen}, \citenamefont {Aili}, \citenamefont {Vellore}, \citenamefont {Martini}, \citenamefont {Yang}, \citenamefont {Srebric}, \citenamefont {Yin},\ and\ \citenamefont {Hu}}]{li2019g}%
  \BibitemOpen
  \bibfield  {author} {\bibinfo {author} {\bibfnamefont {T.}~\bibnamefont {Li}}, \bibinfo {author} {\bibfnamefont {Y.}~\bibnamefont {Zhai}}, \bibinfo {author} {\bibfnamefont {S.}~\bibnamefont {He}}, \bibinfo {author} {\bibfnamefont {W.}~\bibnamefont {Gan}}, \bibinfo {author} {\bibfnamefont {Z.}~\bibnamefont {Wei}}, \bibinfo {author} {\bibfnamefont {M.}~\bibnamefont {Heidarinejad}}, \bibinfo {author} {\bibfnamefont {D.}~\bibnamefont {Dalgo}}, \bibinfo {author} {\bibfnamefont {R.}~\bibnamefont {Mi}}, \bibinfo {author} {\bibfnamefont {X.}~\bibnamefont {Zhao}}, \bibinfo {author} {\bibfnamefont {J.}~\bibnamefont {Song}}, \bibinfo {author} {\bibfnamefont {J.}~\bibnamefont {Dai}}, \bibinfo {author} {\bibfnamefont {C.}~\bibnamefont {Chen}}, \bibinfo {author} {\bibfnamefont {A.}~\bibnamefont {Aili}}, \bibinfo {author} {\bibfnamefont {A.}~\bibnamefont {Vellore}}, \bibinfo {author} {\bibfnamefont {A.}~\bibnamefont {Martini}}, \bibinfo {author} {\bibfnamefont {R.}~\bibnamefont {Yang}}, \bibinfo {author} {\bibfnamefont
  {J.}~\bibnamefont {Srebric}}, \bibinfo {author} {\bibfnamefont {X.}~\bibnamefont {Yin}},\ and\ \bibinfo {author} {\bibfnamefont {L.}~\bibnamefont {Hu}},\ }\bibfield  {title} {\bibinfo {title} {A radiative cooling structural material},\ }\href@noop {} {\bibfield  {journal} {\bibinfo  {journal} {Science}\ }\textbf {\bibinfo {volume} {364}},\ \bibinfo {pages} {760} (\bibinfo {year} {2019})}\BibitemShut {NoStop}%
\bibitem [{\citenamefont {Popoff}\ \emph {et~al.}(2014)\citenamefont {Popoff}, \citenamefont {Goetschy}, \citenamefont {Liew}, \citenamefont {Stone},\ and\ \citenamefont {Cao}}]{popoff2014}%
  \BibitemOpen
  \bibfield  {author} {\bibinfo {author} {\bibfnamefont {S.~M.}\ \bibnamefont {Popoff}}, \bibinfo {author} {\bibfnamefont {A.}~\bibnamefont {Goetschy}}, \bibinfo {author} {\bibfnamefont {S.~F.}\ \bibnamefont {Liew}}, \bibinfo {author} {\bibfnamefont {A.~D.}\ \bibnamefont {Stone}},\ and\ \bibinfo {author} {\bibfnamefont {H.}~\bibnamefont {Cao}},\ }\bibfield  {title} {\bibinfo {title} {Coherent {{Control}} of {{Total Transmission}} of {{Light}} through {{Disordered Media}}},\ }\href {https://doi.org/10.1103/physrevlett.112.133903} {\bibfield  {journal} {\bibinfo  {journal} {Physical Review Letters}\ }\textbf {\bibinfo {volume} {112}},\ \bibinfo {pages} {133903} (\bibinfo {year} {2014})}\BibitemShut {NoStop}%
\bibitem [{\citenamefont {Liew}\ \emph {et~al.}(2016)\citenamefont {Liew}, \citenamefont {Popoff}, \citenamefont {Sheehan}, \citenamefont {Goetschy}, \citenamefont {Schmuttenmaer}, \citenamefont {Stone},\ and\ \citenamefont {Cao}}]{liew2016}%
  \BibitemOpen
  \bibfield  {author} {\bibinfo {author} {\bibfnamefont {S.~F.}\ \bibnamefont {Liew}}, \bibinfo {author} {\bibfnamefont {S.~M.}\ \bibnamefont {Popoff}}, \bibinfo {author} {\bibfnamefont {S.~W.}\ \bibnamefont {Sheehan}}, \bibinfo {author} {\bibfnamefont {A.}~\bibnamefont {Goetschy}}, \bibinfo {author} {\bibfnamefont {C.~A.}\ \bibnamefont {Schmuttenmaer}}, \bibinfo {author} {\bibfnamefont {A.~D.}\ \bibnamefont {Stone}},\ and\ \bibinfo {author} {\bibfnamefont {H.}~\bibnamefont {Cao}},\ }\bibfield  {title} {\bibinfo {title} {Coherent {{Control}} of {{Photocurrent}} in a {{Strongly Scattering Photoelectrochemical System}}},\ }\href {https://doi.org/10.1021/acsphotonics.5b00642} {\bibfield  {journal} {\bibinfo  {journal} {ACS Photonics}\ }\textbf {\bibinfo {volume} {3}},\ \bibinfo {pages} {449} (\bibinfo {year} {2016})}\BibitemShut {NoStop}%
\bibitem [{\citenamefont {Vellekoop}\ and\ \citenamefont {Mosk}(2007)}]{vellekoop2007}%
  \BibitemOpen
  \bibfield  {author} {\bibinfo {author} {\bibfnamefont {I.~M.}\ \bibnamefont {Vellekoop}}\ and\ \bibinfo {author} {\bibfnamefont {A.~P.}\ \bibnamefont {Mosk}},\ }\bibfield  {title} {\bibinfo {title} {Focusing coherent light through opaque strongly scattering media},\ }\href {https://doi.org/10.1364/OL.32.002309} {\bibfield  {journal} {\bibinfo  {journal} {Optics Letters}\ }\textbf {\bibinfo {volume} {32}},\ \bibinfo {pages} {2309} (\bibinfo {year} {2007})}\BibitemShut {NoStop}%
\bibitem [{\citenamefont {Yu}\ \emph {et~al.}(2017)\citenamefont {Yu}, \citenamefont {Lee},\ and\ \citenamefont {Park}}]{yu2017e}%
  \BibitemOpen
  \bibfield  {author} {\bibinfo {author} {\bibfnamefont {H.}~\bibnamefont {Yu}}, \bibinfo {author} {\bibfnamefont {K.}~\bibnamefont {Lee}},\ and\ \bibinfo {author} {\bibfnamefont {Y.}~\bibnamefont {Park}},\ }\bibfield  {title} {\bibinfo {title} {Ultrahigh enhancement of light focusing through disordered media controlled by mega-pixel modes},\ }\href {https://doi.org/10.1364/oe.25.008036} {\bibfield  {journal} {\bibinfo  {journal} {Optics Express}\ }\textbf {\bibinfo {volume} {25}},\ \bibinfo {pages} {8036} (\bibinfo {year} {2017})}\BibitemShut {NoStop}%
\bibitem [{\citenamefont {Guan}\ \emph {et~al.}(2012)\citenamefont {Guan}, \citenamefont {Katz}, \citenamefont {Small}, \citenamefont {Zhou},\ and\ \citenamefont {Silberberg}}]{guan2012}%
  \BibitemOpen
  \bibfield  {author} {\bibinfo {author} {\bibfnamefont {Y.}~\bibnamefont {Guan}}, \bibinfo {author} {\bibfnamefont {O.}~\bibnamefont {Katz}}, \bibinfo {author} {\bibfnamefont {E.}~\bibnamefont {Small}}, \bibinfo {author} {\bibfnamefont {J.}~\bibnamefont {Zhou}},\ and\ \bibinfo {author} {\bibfnamefont {Y.}~\bibnamefont {Silberberg}},\ }\bibfield  {title} {\bibinfo {title} {Polarization control of multiply scattered light through random media by wavefront shaping},\ }\href {https://doi.org/10.1364/ol.37.004663} {\bibfield  {journal} {\bibinfo  {journal} {Optics Letters}\ }\textbf {\bibinfo {volume} {37}},\ \bibinfo {pages} {4663} (\bibinfo {year} {2012})}\BibitemShut {NoStop}%
\bibitem [{\citenamefont {Katz}\ \emph {et~al.}(2012)\citenamefont {Katz}, \citenamefont {Small},\ and\ \citenamefont {Silberberg}}]{katz2012}%
  \BibitemOpen
  \bibfield  {author} {\bibinfo {author} {\bibfnamefont {O.}~\bibnamefont {Katz}}, \bibinfo {author} {\bibfnamefont {E.}~\bibnamefont {Small}},\ and\ \bibinfo {author} {\bibfnamefont {Y.}~\bibnamefont {Silberberg}},\ }\bibfield  {title} {\bibinfo {title} {Looking around corners and through thin turbid layers in real time with scattered incoherent light},\ }\href {https://doi.org/10.1038/nphoton.2012.150} {\bibfield  {journal} {\bibinfo  {journal} {Nature Photonics}\ }\textbf {\bibinfo {volume} {6}},\ \bibinfo {pages} {549} (\bibinfo {year} {2012})}\BibitemShut {NoStop}%
\bibitem [{\citenamefont {Horstmeyer}\ \emph {et~al.}(2015)\citenamefont {Horstmeyer}, \citenamefont {Ruan},\ and\ \citenamefont {Yang}}]{horstmeyer2015}%
  \BibitemOpen
  \bibfield  {author} {\bibinfo {author} {\bibfnamefont {R.}~\bibnamefont {Horstmeyer}}, \bibinfo {author} {\bibfnamefont {H.}~\bibnamefont {Ruan}},\ and\ \bibinfo {author} {\bibfnamefont {C.}~\bibnamefont {Yang}},\ }\bibfield  {title} {\bibinfo {title} {Guidestar-assisted wavefront-shaping methods for focusing light into biological tissue},\ }\href {https://doi.org/10.1038/nphoton.2015.140} {\bibfield  {journal} {\bibinfo  {journal} {Nature Photonics}\ }\textbf {\bibinfo {volume} {9}},\ \bibinfo {pages} {563} (\bibinfo {year} {2015})}\BibitemShut {NoStop}%
\bibitem [{\citenamefont {Vellekoop}(2015)}]{vellekoop2015}%
  \BibitemOpen
  \bibfield  {author} {\bibinfo {author} {\bibfnamefont {I.~M.}\ \bibnamefont {Vellekoop}},\ }\bibfield  {title} {\bibinfo {title} {Feedback-based wavefront shaping},\ }\href {https://doi.org/10.1364/oe.23.012189} {\bibfield  {journal} {\bibinfo  {journal} {Optics Express}\ }\textbf {\bibinfo {volume} {23}},\ \bibinfo {pages} {12189} (\bibinfo {year} {2015})}\BibitemShut {NoStop}%
\bibitem [{\citenamefont {Yu}\ \emph {et~al.}(2015)\citenamefont {Yu}, \citenamefont {Park}, \citenamefont {Lee}, \citenamefont {Yoon}, \citenamefont {Kim}, \citenamefont {Lee},\ and\ \citenamefont {Park}}]{yu2015d}%
  \BibitemOpen
  \bibfield  {author} {\bibinfo {author} {\bibfnamefont {H.}~\bibnamefont {Yu}}, \bibinfo {author} {\bibfnamefont {J.}~\bibnamefont {Park}}, \bibinfo {author} {\bibfnamefont {K.}~\bibnamefont {Lee}}, \bibinfo {author} {\bibfnamefont {J.}~\bibnamefont {Yoon}}, \bibinfo {author} {\bibfnamefont {K.}~\bibnamefont {Kim}}, \bibinfo {author} {\bibfnamefont {S.}~\bibnamefont {Lee}},\ and\ \bibinfo {author} {\bibfnamefont {Y.}~\bibnamefont {Park}},\ }\bibfield  {title} {\bibinfo {title} {Recent advances in wavefront shaping techniques for biomedical applications},\ }\href {https://doi.org/10.1016/j.cap.2015.02.015} {\bibfield  {journal} {\bibinfo  {journal} {Current Applied Physics}\ }\textbf {\bibinfo {volume} {15}},\ \bibinfo {pages} {632} (\bibinfo {year} {2015})}\BibitemShut {NoStop}%
\bibitem [{\citenamefont {Pai}\ \emph {et~al.}(2021)\citenamefont {Pai}, \citenamefont {Bosch}, \citenamefont {K{\"u}hmayer}, \citenamefont {Rotter},\ and\ \citenamefont {Mosk}}]{pai2021b}%
  \BibitemOpen
  \bibfield  {author} {\bibinfo {author} {\bibfnamefont {P.}~\bibnamefont {Pai}}, \bibinfo {author} {\bibfnamefont {J.}~\bibnamefont {Bosch}}, \bibinfo {author} {\bibfnamefont {M.}~\bibnamefont {K{\"u}hmayer}}, \bibinfo {author} {\bibfnamefont {S.}~\bibnamefont {Rotter}},\ and\ \bibinfo {author} {\bibfnamefont {A.~P.}\ \bibnamefont {Mosk}},\ }\bibfield  {title} {\bibinfo {title} {Scattering invariant modes of light in complex media},\ }\href {https://doi.org/10.1038/s41566-021-00789-9} {\bibfield  {journal} {\bibinfo  {journal} {Nature Photonics}\ }\textbf {\bibinfo {volume} {15}},\ \bibinfo {pages} {431} (\bibinfo {year} {2021})}\BibitemShut {NoStop}%
\bibitem [{\citenamefont {Rotter}\ and\ \citenamefont {Gigan}(2017)}]{rotter2017}%
  \BibitemOpen
  \bibfield  {author} {\bibinfo {author} {\bibfnamefont {S.}~\bibnamefont {Rotter}}\ and\ \bibinfo {author} {\bibfnamefont {S.}~\bibnamefont {Gigan}},\ }\bibfield  {title} {\bibinfo {title} {Light fields in complex media: {{Mesoscopic}} scattering meets wave control},\ }\href {https://doi.org/10.1103/RevModPhys.89.015005} {\bibfield  {journal} {\bibinfo  {journal} {Reviews of Modern Physics}\ }\textbf {\bibinfo {volume} {89}},\ \bibinfo {pages} {015005} (\bibinfo {year} {2017})}\BibitemShut {NoStop}%
\bibitem [{\citenamefont {Cao}\ \emph {et~al.}(2022)\citenamefont {Cao}, \citenamefont {Mosk},\ and\ \citenamefont {Rotter}}]{cao2022a}%
  \BibitemOpen
  \bibfield  {author} {\bibinfo {author} {\bibfnamefont {H.}~\bibnamefont {Cao}}, \bibinfo {author} {\bibfnamefont {A.~P.}\ \bibnamefont {Mosk}},\ and\ \bibinfo {author} {\bibfnamefont {S.}~\bibnamefont {Rotter}},\ }\bibfield  {title} {\bibinfo {title} {Shaping the propagation of light in complex media},\ }\href {https://doi.org/10.1038/s41567-022-01677-x} {\bibfield  {journal} {\bibinfo  {journal} {Nature Physics}\ }\textbf {\bibinfo {volume} {18}},\ \bibinfo {pages} {994} (\bibinfo {year} {2022})}\BibitemShut {NoStop}%
\bibitem [{\citenamefont {Yaqoob}\ \emph {et~al.}(2008)\citenamefont {Yaqoob}, \citenamefont {Psaltis}, \citenamefont {Feld},\ and\ \citenamefont {Yang}}]{yaqoob2008}%
  \BibitemOpen
  \bibfield  {author} {\bibinfo {author} {\bibfnamefont {Z.}~\bibnamefont {Yaqoob}}, \bibinfo {author} {\bibfnamefont {D.}~\bibnamefont {Psaltis}}, \bibinfo {author} {\bibfnamefont {M.~S.}\ \bibnamefont {Feld}},\ and\ \bibinfo {author} {\bibfnamefont {C.}~\bibnamefont {Yang}},\ }\bibfield  {title} {\bibinfo {title} {Optical phase conjugation for turbidity suppression in biological samples},\ }\href {https://doi.org/10.1038/nphoton.2007.297} {\bibfield  {journal} {\bibinfo  {journal} {Nature Photonics}\ }\textbf {\bibinfo {volume} {2}},\ \bibinfo {pages} {110} (\bibinfo {year} {2008})}\BibitemShut {NoStop}%
\bibitem [{\citenamefont {Fan}\ and\ \citenamefont {Kahn}(2005)}]{fan2005}%
  \BibitemOpen
  \bibfield  {author} {\bibinfo {author} {\bibfnamefont {S.}~\bibnamefont {Fan}}\ and\ \bibinfo {author} {\bibfnamefont {J.~M.}\ \bibnamefont {Kahn}},\ }\bibfield  {title} {\bibinfo {title} {Principal modes in multimode waveguides},\ }\href {https://doi.org/10.1364/OL.30.000135} {\bibfield  {journal} {\bibinfo  {journal} {Optics Letters}\ }\textbf {\bibinfo {volume} {30}},\ \bibinfo {pages} {135} (\bibinfo {year} {2005})}\BibitemShut {NoStop}%
\bibitem [{\citenamefont {{\v C}i{\v z}m{\'a}r}\ and\ \citenamefont {Dholakia}(2011)}]{cizmar2011}%
  \BibitemOpen
  \bibfield  {author} {\bibinfo {author} {\bibfnamefont {T.}~\bibnamefont {{\v C}i{\v z}m{\'a}r}}\ and\ \bibinfo {author} {\bibfnamefont {K.}~\bibnamefont {Dholakia}},\ }\bibfield  {title} {\bibinfo {title} {Shaping the light transmission through a multimode optical fibre: Complex transformation analysis and applications in biophotonics},\ }\href {https://doi.org/10.1364/oe.19.018871} {\bibfield  {journal} {\bibinfo  {journal} {Optics Express}\ }\textbf {\bibinfo {volume} {19}},\ \bibinfo {pages} {18871} (\bibinfo {year} {2011})}\BibitemShut {NoStop}%
\bibitem [{\citenamefont {Papadopoulos}\ \emph {et~al.}(2012)\citenamefont {Papadopoulos}, \citenamefont {Farahi}, \citenamefont {Moser},\ and\ \citenamefont {Psaltis}}]{papadopoulos2012}%
  \BibitemOpen
  \bibfield  {author} {\bibinfo {author} {\bibfnamefont {I.~N.}\ \bibnamefont {Papadopoulos}}, \bibinfo {author} {\bibfnamefont {S.}~\bibnamefont {Farahi}}, \bibinfo {author} {\bibfnamefont {C.}~\bibnamefont {Moser}},\ and\ \bibinfo {author} {\bibfnamefont {D.}~\bibnamefont {Psaltis}},\ }\bibfield  {title} {\bibinfo {title} {Focusing and scanning light through a multimode optical fiber using digital phase conjugation},\ }\href {https://doi.org/10.1364/oe.20.010583} {\bibfield  {journal} {\bibinfo  {journal} {Optics Express}\ }\textbf {\bibinfo {volume} {20}},\ \bibinfo {pages} {10583} (\bibinfo {year} {2012})}\BibitemShut {NoStop}%
\bibitem [{\citenamefont {Carpenter}\ \emph {et~al.}(2015)\citenamefont {Carpenter}, \citenamefont {Eggleton},\ and\ \citenamefont {Schr{\"o}der}}]{carpenter2015}%
  \BibitemOpen
  \bibfield  {author} {\bibinfo {author} {\bibfnamefont {J.}~\bibnamefont {Carpenter}}, \bibinfo {author} {\bibfnamefont {B.~J.}\ \bibnamefont {Eggleton}},\ and\ \bibinfo {author} {\bibfnamefont {J.}~\bibnamefont {Schr{\"o}der}},\ }\bibfield  {title} {\bibinfo {title} {Observation of {{Eisenbud}}\textendash{{Wigner}}\textendash{{Smith}} states as principal modes in multimode fibre},\ }\href {https://doi.org/10.1038/nphoton.2015.188} {\bibfield  {journal} {\bibinfo  {journal} {Nature Photonics}\ }\textbf {\bibinfo {volume} {9}},\ \bibinfo {pages} {751} (\bibinfo {year} {2015})}\BibitemShut {NoStop}%
\bibitem [{\citenamefont {Xiong}\ \emph {et~al.}(2016)\citenamefont {Xiong}, \citenamefont {Ambichl}, \citenamefont {Bromberg}, \citenamefont {Redding}, \citenamefont {Rotter},\ and\ \citenamefont {Cao}}]{xiong2016}%
  \BibitemOpen
  \bibfield  {author} {\bibinfo {author} {\bibfnamefont {W.}~\bibnamefont {Xiong}}, \bibinfo {author} {\bibfnamefont {P.}~\bibnamefont {Ambichl}}, \bibinfo {author} {\bibfnamefont {Y.}~\bibnamefont {Bromberg}}, \bibinfo {author} {\bibfnamefont {B.}~\bibnamefont {Redding}}, \bibinfo {author} {\bibfnamefont {S.}~\bibnamefont {Rotter}},\ and\ \bibinfo {author} {\bibfnamefont {H.}~\bibnamefont {Cao}},\ }\bibfield  {title} {\bibinfo {title} {Spatiotemporal {{Control}} of {{Light Transmission}} through a {{Multimode Fiber}} with {{Strong Mode Coupling}}},\ }\href {https://doi.org/10.1103/physrevlett.117.053901} {\bibfield  {journal} {\bibinfo  {journal} {Physical Review Letters}\ }\textbf {\bibinfo {volume} {117}},\ \bibinfo {pages} {053901} (\bibinfo {year} {2016})}\BibitemShut {NoStop}%
\bibitem [{\citenamefont {Lerosey}\ \emph {et~al.}(2007)\citenamefont {Lerosey}, \citenamefont {De~Rosny}, \citenamefont {Tourin},\ and\ \citenamefont {Fink}}]{lerosey2007}%
  \BibitemOpen
  \bibfield  {author} {\bibinfo {author} {\bibfnamefont {G.}~\bibnamefont {Lerosey}}, \bibinfo {author} {\bibfnamefont {J.}~\bibnamefont {De~Rosny}}, \bibinfo {author} {\bibfnamefont {A.}~\bibnamefont {Tourin}},\ and\ \bibinfo {author} {\bibfnamefont {M.}~\bibnamefont {Fink}},\ }\bibfield  {title} {\bibinfo {title} {Focusing {{Beyond}} the {{Diffraction Limit}} with {{Far-Field Time Reversal}}},\ }\href {https://doi.org/10.1126/science.1134824} {\bibfield  {journal} {\bibinfo  {journal} {Science}\ }\textbf {\bibinfo {volume} {315}},\ \bibinfo {pages} {1120} (\bibinfo {year} {2007})}\BibitemShut {NoStop}%
\bibitem [{\citenamefont {Vellekoop}\ and\ \citenamefont {Mosk}(2008{\natexlab{a}})}]{vellekoop2008b}%
  \BibitemOpen
  \bibfield  {author} {\bibinfo {author} {\bibfnamefont {I.}~\bibnamefont {Vellekoop}}\ and\ \bibinfo {author} {\bibfnamefont {A.}~\bibnamefont {Mosk}},\ }\bibfield  {title} {\bibinfo {title} {Phase control algorithms for focusing light through turbid media},\ }\href {https://doi.org/10.1016/j.optcom.2008.02.022} {\bibfield  {journal} {\bibinfo  {journal} {Optics Communications}\ }\textbf {\bibinfo {volume} {281}},\ \bibinfo {pages} {3071} (\bibinfo {year} {2008}{\natexlab{a}})}\BibitemShut {NoStop}%
\bibitem [{\citenamefont {Katz}\ \emph {et~al.}(2011)\citenamefont {Katz}, \citenamefont {Small}, \citenamefont {Bromberg},\ and\ \citenamefont {Silberberg}}]{katz2011}%
  \BibitemOpen
  \bibfield  {author} {\bibinfo {author} {\bibfnamefont {O.}~\bibnamefont {Katz}}, \bibinfo {author} {\bibfnamefont {E.}~\bibnamefont {Small}}, \bibinfo {author} {\bibfnamefont {Y.}~\bibnamefont {Bromberg}},\ and\ \bibinfo {author} {\bibfnamefont {Y.}~\bibnamefont {Silberberg}},\ }\bibfield  {title} {\bibinfo {title} {Focusing and compression of ultrashort pulses through scattering media},\ }\href {https://doi.org/10.1038/nphoton.2011.72} {\bibfield  {journal} {\bibinfo  {journal} {Nature Photonics}\ }\textbf {\bibinfo {volume} {5}},\ \bibinfo {pages} {372} (\bibinfo {year} {2011})}\BibitemShut {NoStop}%
\bibitem [{\citenamefont {McCabe}\ \emph {et~al.}(2011)\citenamefont {McCabe}, \citenamefont {Tajalli}, \citenamefont {Austin}, \citenamefont {Bondareff}, \citenamefont {Walmsley}, \citenamefont {Gigan},\ and\ \citenamefont {Chatel}}]{mccabe2011}%
  \BibitemOpen
  \bibfield  {author} {\bibinfo {author} {\bibfnamefont {D.~J.}\ \bibnamefont {McCabe}}, \bibinfo {author} {\bibfnamefont {A.}~\bibnamefont {Tajalli}}, \bibinfo {author} {\bibfnamefont {D.~R.}\ \bibnamefont {Austin}}, \bibinfo {author} {\bibfnamefont {P.}~\bibnamefont {Bondareff}}, \bibinfo {author} {\bibfnamefont {I.~A.}\ \bibnamefont {Walmsley}}, \bibinfo {author} {\bibfnamefont {S.}~\bibnamefont {Gigan}},\ and\ \bibinfo {author} {\bibfnamefont {B.}~\bibnamefont {Chatel}},\ }\bibfield  {title} {\bibinfo {title} {Spatio-temporal focusing of an ultrafast pulse through a multiply scattering medium},\ }\href {https://doi.org/10.1038/ncomms1434} {\bibfield  {journal} {\bibinfo  {journal} {Nature Communications}\ }\textbf {\bibinfo {volume} {2}},\ \bibinfo {pages} {447} (\bibinfo {year} {2011})}\BibitemShut {NoStop}%
\bibitem [{\citenamefont {Xu}\ \emph {et~al.}(2011)\citenamefont {Xu}, \citenamefont {Liu},\ and\ \citenamefont {Wang}}]{xu2011a}%
  \BibitemOpen
  \bibfield  {author} {\bibinfo {author} {\bibfnamefont {X.}~\bibnamefont {Xu}}, \bibinfo {author} {\bibfnamefont {H.}~\bibnamefont {Liu}},\ and\ \bibinfo {author} {\bibfnamefont {L.~V.}\ \bibnamefont {Wang}},\ }\bibfield  {title} {\bibinfo {title} {Time-reversed ultrasonically encoded optical focusing into scattering media},\ }\href {https://doi.org/10.1038/nphoton.2010.306} {\bibfield  {journal} {\bibinfo  {journal} {Nature Photonics}\ }\textbf {\bibinfo {volume} {5}},\ \bibinfo {pages} {154} (\bibinfo {year} {2011})}\BibitemShut {NoStop}%
\bibitem [{\citenamefont {Park}\ \emph {et~al.}(2013)\citenamefont {Park}, \citenamefont {Park}, \citenamefont {Yu}, \citenamefont {Park}, \citenamefont {Han}, \citenamefont {Shin}, \citenamefont {Ko}, \citenamefont {Nam}, \citenamefont {Cho},\ and\ \citenamefont {Park}}]{park2013a}%
  \BibitemOpen
  \bibfield  {author} {\bibinfo {author} {\bibfnamefont {J.-H.}\ \bibnamefont {Park}}, \bibinfo {author} {\bibfnamefont {C.}~\bibnamefont {Park}}, \bibinfo {author} {\bibfnamefont {H.}~\bibnamefont {Yu}}, \bibinfo {author} {\bibfnamefont {J.}~\bibnamefont {Park}}, \bibinfo {author} {\bibfnamefont {S.}~\bibnamefont {Han}}, \bibinfo {author} {\bibfnamefont {J.}~\bibnamefont {Shin}}, \bibinfo {author} {\bibfnamefont {S.~H.}\ \bibnamefont {Ko}}, \bibinfo {author} {\bibfnamefont {K.~T.}\ \bibnamefont {Nam}}, \bibinfo {author} {\bibfnamefont {Y.-H.}\ \bibnamefont {Cho}},\ and\ \bibinfo {author} {\bibfnamefont {Y.}~\bibnamefont {Park}},\ }\bibfield  {title} {\bibinfo {title} {Subwavelength light focusing using random nanoparticles},\ }\href {https://doi.org/10.1038/nphoton.2013.95} {\bibfield  {journal} {\bibinfo  {journal} {Nature Photonics}\ }\textbf {\bibinfo {volume} {7}},\ \bibinfo {pages} {454} (\bibinfo {year} {2013})}\BibitemShut {NoStop}%
\bibitem [{\citenamefont {Blochet}\ \emph {et~al.}(2017)\citenamefont {Blochet}, \citenamefont {Bourdieu},\ and\ \citenamefont {Gigan}}]{blochet2017}%
  \BibitemOpen
  \bibfield  {author} {\bibinfo {author} {\bibfnamefont {B.}~\bibnamefont {Blochet}}, \bibinfo {author} {\bibfnamefont {L.}~\bibnamefont {Bourdieu}},\ and\ \bibinfo {author} {\bibfnamefont {S.}~\bibnamefont {Gigan}},\ }\bibfield  {title} {\bibinfo {title} {Focusing light through dynamical samples using fast continuous wavefront optimization},\ }\href {https://doi.org/10.1364/ol.42.004994} {\bibfield  {journal} {\bibinfo  {journal} {Optics Letters}\ }\textbf {\bibinfo {volume} {42}},\ \bibinfo {pages} {4994} (\bibinfo {year} {2017})}\BibitemShut {NoStop}%
\bibitem [{\citenamefont {Vellekoop}\ and\ \citenamefont {Mosk}(2008{\natexlab{b}})}]{vellekoop2008}%
  \BibitemOpen
  \bibfield  {author} {\bibinfo {author} {\bibfnamefont {I.~M.}\ \bibnamefont {Vellekoop}}\ and\ \bibinfo {author} {\bibfnamefont {A.~P.}\ \bibnamefont {Mosk}},\ }\bibfield  {title} {\bibinfo {title} {Universal {{Optimal Transmission}} of {{Light Through Disordered Materials}}},\ }\href {https://doi.org/10.1103/physrevlett.101.120601} {\bibfield  {journal} {\bibinfo  {journal} {Physical Review Letters}\ }\textbf {\bibinfo {volume} {101}},\ \bibinfo {pages} {120601} (\bibinfo {year} {2008}{\natexlab{b}})}\BibitemShut {NoStop}%
\bibitem [{\citenamefont {Popoff}\ \emph {et~al.}(2010)\citenamefont {Popoff}, \citenamefont {Lerosey}, \citenamefont {Carminati}, \citenamefont {Fink}, \citenamefont {Boccara},\ and\ \citenamefont {Gigan}}]{popoff2010}%
  \BibitemOpen
  \bibfield  {author} {\bibinfo {author} {\bibfnamefont {S.~M.}\ \bibnamefont {Popoff}}, \bibinfo {author} {\bibfnamefont {G.}~\bibnamefont {Lerosey}}, \bibinfo {author} {\bibfnamefont {R.}~\bibnamefont {Carminati}}, \bibinfo {author} {\bibfnamefont {M.}~\bibnamefont {Fink}}, \bibinfo {author} {\bibfnamefont {A.~C.}\ \bibnamefont {Boccara}},\ and\ \bibinfo {author} {\bibfnamefont {S.}~\bibnamefont {Gigan}},\ }\bibfield  {title} {\bibinfo {title} {Measuring the {{Transmission Matrix}} in {{Optics}}: {{An Approach}} to the {{Study}} and {{Control}} of {{Light Propagation}} in {{Disordered Media}}},\ }\href {https://doi.org/10.1103/physrevlett.104.100601} {\bibfield  {journal} {\bibinfo  {journal} {Physical Review Letters}\ }\textbf {\bibinfo {volume} {104}},\ \bibinfo {pages} {100601} (\bibinfo {year} {2010})}\BibitemShut {NoStop}%
\bibitem [{\citenamefont {Kim}\ \emph {et~al.}(2012)\citenamefont {Kim}, \citenamefont {Choi}, \citenamefont {Yoon}, \citenamefont {Choi}, \citenamefont {Kim}, \citenamefont {Park},\ and\ \citenamefont {Choi}}]{kim2012a}%
  \BibitemOpen
  \bibfield  {author} {\bibinfo {author} {\bibfnamefont {M.}~\bibnamefont {Kim}}, \bibinfo {author} {\bibfnamefont {Y.}~\bibnamefont {Choi}}, \bibinfo {author} {\bibfnamefont {C.}~\bibnamefont {Yoon}}, \bibinfo {author} {\bibfnamefont {W.}~\bibnamefont {Choi}}, \bibinfo {author} {\bibfnamefont {J.}~\bibnamefont {Kim}}, \bibinfo {author} {\bibfnamefont {Q.-H.}\ \bibnamefont {Park}},\ and\ \bibinfo {author} {\bibfnamefont {W.}~\bibnamefont {Choi}},\ }\bibfield  {title} {\bibinfo {title} {Maximal energy transport through disordered media with the implementation of transmission eigenchannels},\ }\href {https://doi.org/10.1038/nphoton.2012.159} {\bibfield  {journal} {\bibinfo  {journal} {Nature Photonics}\ }\textbf {\bibinfo {volume} {6}},\ \bibinfo {pages} {581} (\bibinfo {year} {2012})}\BibitemShut {NoStop}%
\bibitem [{\citenamefont {Shi}\ and\ \citenamefont {Genack}(2012)}]{shi2012}%
  \BibitemOpen
  \bibfield  {author} {\bibinfo {author} {\bibfnamefont {Z.}~\bibnamefont {Shi}}\ and\ \bibinfo {author} {\bibfnamefont {A.~Z.}\ \bibnamefont {Genack}},\ }\bibfield  {title} {\bibinfo {title} {Transmission {{Eigenvalues}} and the {{Bare Conductance}} in the {{Crossover}} to {{Anderson Localization}}},\ }\href {https://doi.org/10.1103/physrevlett.108.043901} {\bibfield  {journal} {\bibinfo  {journal} {Physical Review Letters}\ }\textbf {\bibinfo {volume} {108}},\ \bibinfo {pages} {043901} (\bibinfo {year} {2012})}\BibitemShut {NoStop}%
\bibitem [{\citenamefont {Yu}\ \emph {et~al.}(2013)\citenamefont {Yu}, \citenamefont {Hillman}, \citenamefont {Choi}, \citenamefont {Lee}, \citenamefont {Feld}, \citenamefont {Dasari},\ and\ \citenamefont {Park}}]{yu2013a}%
  \BibitemOpen
  \bibfield  {author} {\bibinfo {author} {\bibfnamefont {H.}~\bibnamefont {Yu}}, \bibinfo {author} {\bibfnamefont {T.~R.}\ \bibnamefont {Hillman}}, \bibinfo {author} {\bibfnamefont {W.}~\bibnamefont {Choi}}, \bibinfo {author} {\bibfnamefont {J.~O.}\ \bibnamefont {Lee}}, \bibinfo {author} {\bibfnamefont {M.~S.}\ \bibnamefont {Feld}}, \bibinfo {author} {\bibfnamefont {R.~R.}\ \bibnamefont {Dasari}},\ and\ \bibinfo {author} {\bibfnamefont {Y.}~\bibnamefont {Park}},\ }\bibfield  {title} {\bibinfo {title} {Measuring {{Large Optical Transmission Matrices}} of {{Disordered Media}}},\ }\href {https://doi.org/10.1103/physrevlett.111.153902} {\bibfield  {journal} {\bibinfo  {journal} {Physical Review Letters}\ }\textbf {\bibinfo {volume} {111}},\ \bibinfo {pages} {153902} (\bibinfo {year} {2013})}\BibitemShut {NoStop}%
\bibitem [{\citenamefont {Yan}\ \emph {et~al.}(2014)\citenamefont {Yan}, \citenamefont {Cui}, \citenamefont {Gu}, \citenamefont {Tian}, \citenamefont {Fu},\ and\ \citenamefont {Wu}}]{yan2014a}%
  \BibitemOpen
  \bibfield  {author} {\bibinfo {author} {\bibfnamefont {X.-B.}\ \bibnamefont {Yan}}, \bibinfo {author} {\bibfnamefont {C.-L.}\ \bibnamefont {Cui}}, \bibinfo {author} {\bibfnamefont {K.-H.}\ \bibnamefont {Gu}}, \bibinfo {author} {\bibfnamefont {X.-D.}\ \bibnamefont {Tian}}, \bibinfo {author} {\bibfnamefont {C.-B.}\ \bibnamefont {Fu}},\ and\ \bibinfo {author} {\bibfnamefont {J.-H.}\ \bibnamefont {Wu}},\ }\bibfield  {title} {\bibinfo {title} {Coherent perfect absorption, transmission, and synthesis in a double-cavity optomechanical system},\ }\href {https://doi.org/10.1364/OE.22.004886} {\bibfield  {journal} {\bibinfo  {journal} {Optics Express}\ }\textbf {\bibinfo {volume} {22}},\ \bibinfo {pages} {4886} (\bibinfo {year} {2014})}\BibitemShut {NoStop}%
\bibitem [{\citenamefont {G{\'e}rardin}\ \emph {et~al.}(2014)\citenamefont {G{\'e}rardin}, \citenamefont {Laurent}, \citenamefont {Derode}, \citenamefont {Prada},\ and\ \citenamefont {Aubry}}]{gerardin2014}%
  \BibitemOpen
  \bibfield  {author} {\bibinfo {author} {\bibfnamefont {B.}~\bibnamefont {G{\'e}rardin}}, \bibinfo {author} {\bibfnamefont {J.}~\bibnamefont {Laurent}}, \bibinfo {author} {\bibfnamefont {A.}~\bibnamefont {Derode}}, \bibinfo {author} {\bibfnamefont {C.}~\bibnamefont {Prada}},\ and\ \bibinfo {author} {\bibfnamefont {A.}~\bibnamefont {Aubry}},\ }\bibfield  {title} {\bibinfo {title} {Full {{Transmission}} and {{Reflection}} of {{Waves Propagating}} through a {{Maze}} of {{Disorder}}},\ }\href {https://doi.org/10.1103/physrevlett.113.173901} {\bibfield  {journal} {\bibinfo  {journal} {Physical Review Letters}\ }\textbf {\bibinfo {volume} {113}},\ \bibinfo {pages} {173901} (\bibinfo {year} {2014})}\BibitemShut {NoStop}%
\bibitem [{\citenamefont {Pe{\~n}a}\ \emph {et~al.}(2014)\citenamefont {Pe{\~n}a}, \citenamefont {Girschik}, \citenamefont {Libisch}, \citenamefont {Rotter},\ and\ \citenamefont {Chabanov}}]{pena2014a}%
  \BibitemOpen
  \bibfield  {author} {\bibinfo {author} {\bibfnamefont {A.}~\bibnamefont {Pe{\~n}a}}, \bibinfo {author} {\bibfnamefont {A.}~\bibnamefont {Girschik}}, \bibinfo {author} {\bibfnamefont {F.}~\bibnamefont {Libisch}}, \bibinfo {author} {\bibfnamefont {S.}~\bibnamefont {Rotter}},\ and\ \bibinfo {author} {\bibfnamefont {A.~A.}\ \bibnamefont {Chabanov}},\ }\bibfield  {title} {\bibinfo {title} {The single-channel regime of transport through random media},\ }\href {https://doi.org/10.1038/ncomms4488} {\bibfield  {journal} {\bibinfo  {journal} {Nature Communications}\ }\textbf {\bibinfo {volume} {5}},\ \bibinfo {pages} {3488} (\bibinfo {year} {2014})}\BibitemShut {NoStop}%
\bibitem [{\citenamefont {Davy}\ \emph {et~al.}(2015)\citenamefont {Davy}, \citenamefont {Shi}, \citenamefont {Wang}, \citenamefont {Cheng},\ and\ \citenamefont {Genack}}]{davy2015}%
  \BibitemOpen
  \bibfield  {author} {\bibinfo {author} {\bibfnamefont {M.}~\bibnamefont {Davy}}, \bibinfo {author} {\bibfnamefont {Z.}~\bibnamefont {Shi}}, \bibinfo {author} {\bibfnamefont {J.}~\bibnamefont {Wang}}, \bibinfo {author} {\bibfnamefont {X.}~\bibnamefont {Cheng}},\ and\ \bibinfo {author} {\bibfnamefont {A.~Z.}\ \bibnamefont {Genack}},\ }\bibfield  {title} {\bibinfo {title} {Transmission {{Eigenchannels}} and the {{Densities}} of {{States}} of {{Random Media}}},\ }\href {https://doi.org/10.1103/physrevlett.114.033901} {\bibfield  {journal} {\bibinfo  {journal} {Physical Review Letters}\ }\textbf {\bibinfo {volume} {114}},\ \bibinfo {pages} {033901} (\bibinfo {year} {2015})}\BibitemShut {NoStop}%
\bibitem [{\citenamefont {Shi}\ \emph {et~al.}(2015)\citenamefont {Shi}, \citenamefont {Davy},\ and\ \citenamefont {Genack}}]{shi2015a}%
  \BibitemOpen
  \bibfield  {author} {\bibinfo {author} {\bibfnamefont {Z.}~\bibnamefont {Shi}}, \bibinfo {author} {\bibfnamefont {M.}~\bibnamefont {Davy}},\ and\ \bibinfo {author} {\bibfnamefont {A.~Z.}\ \bibnamefont {Genack}},\ }\bibfield  {title} {\bibinfo {title} {Statistics and control of waves in disordered media},\ }\href {https://doi.org/10.1364/oe.23.012293} {\bibfield  {journal} {\bibinfo  {journal} {Optics Express}\ }\textbf {\bibinfo {volume} {23}},\ \bibinfo {pages} {12293} (\bibinfo {year} {2015})}\BibitemShut {NoStop}%
\bibitem [{\citenamefont {Bender}\ \emph {et~al.}(2020)\citenamefont {Bender}, \citenamefont {Yamilov}, \citenamefont {Y{\i}lmaz},\ and\ \citenamefont {Cao}}]{bender2020c}%
  \BibitemOpen
  \bibfield  {author} {\bibinfo {author} {\bibfnamefont {N.}~\bibnamefont {Bender}}, \bibinfo {author} {\bibfnamefont {A.}~\bibnamefont {Yamilov}}, \bibinfo {author} {\bibfnamefont {H.}~\bibnamefont {Y{\i}lmaz}},\ and\ \bibinfo {author} {\bibfnamefont {H.}~\bibnamefont {Cao}},\ }\bibfield  {title} {\bibinfo {title} {Fluctuations and {{Correlations}} of {{Transmission Eigenchannels}} in {{Diffusive Media}}},\ }\href {https://doi.org/10.1103/physrevlett.125.165901} {\bibfield  {journal} {\bibinfo  {journal} {Physical Review Letters}\ }\textbf {\bibinfo {volume} {125}},\ \bibinfo {pages} {165901} (\bibinfo {year} {2020})}\BibitemShut {NoStop}%
\bibitem [{\citenamefont {Guo}\ and\ \citenamefont {Fan}(2024{\natexlab{a}})}]{guo2024c}%
  \BibitemOpen
  \bibfield  {author} {\bibinfo {author} {\bibfnamefont {C.}~\bibnamefont {Guo}}\ and\ \bibinfo {author} {\bibfnamefont {S.}~\bibnamefont {Fan}},\ }\bibfield  {title} {\bibinfo {title} {Unitary control of partially coherent waves. {{II}}. {{Transmission}} or reflection},\ }\href {https://doi.org/10.1103/PhysRevB.110.035431} {\bibfield  {journal} {\bibinfo  {journal} {Physical Review B}\ }\textbf {\bibinfo {volume} {110}},\ \bibinfo {pages} {035431} (\bibinfo {year} {2024}{\natexlab{a}})}\BibitemShut {NoStop}%
\bibitem [{\citenamefont {Chong}\ \emph {et~al.}(2010)\citenamefont {Chong}, \citenamefont {Ge}, \citenamefont {Cao},\ and\ \citenamefont {Stone}}]{chong2010a}%
  \BibitemOpen
  \bibfield  {author} {\bibinfo {author} {\bibfnamefont {Y.~D.}\ \bibnamefont {Chong}}, \bibinfo {author} {\bibfnamefont {L.}~\bibnamefont {Ge}}, \bibinfo {author} {\bibfnamefont {H.}~\bibnamefont {Cao}},\ and\ \bibinfo {author} {\bibfnamefont {A.~D.}\ \bibnamefont {Stone}},\ }\bibfield  {title} {\bibinfo {title} {Coherent {{Perfect Absorbers}}: {{Time-Reversed Lasers}}},\ }\href {https://doi.org/10.1103/PhysRevLett.105.053901} {\bibfield  {journal} {\bibinfo  {journal} {Physical Review Letters}\ }\textbf {\bibinfo {volume} {105}},\ \bibinfo {pages} {053901} (\bibinfo {year} {2010})}\BibitemShut {NoStop}%
\bibitem [{\citenamefont {Wan}\ \emph {et~al.}(2011)\citenamefont {Wan}, \citenamefont {Chong}, \citenamefont {Ge}, \citenamefont {Noh}, \citenamefont {Stone},\ and\ \citenamefont {Cao}}]{wan2011}%
  \BibitemOpen
  \bibfield  {author} {\bibinfo {author} {\bibfnamefont {W.}~\bibnamefont {Wan}}, \bibinfo {author} {\bibfnamefont {Y.}~\bibnamefont {Chong}}, \bibinfo {author} {\bibfnamefont {L.}~\bibnamefont {Ge}}, \bibinfo {author} {\bibfnamefont {H.}~\bibnamefont {Noh}}, \bibinfo {author} {\bibfnamefont {A.~D.}\ \bibnamefont {Stone}},\ and\ \bibinfo {author} {\bibfnamefont {H.}~\bibnamefont {Cao}},\ }\bibfield  {title} {\bibinfo {title} {Time-reversed lasing and interferometric control of absorption.},\ }\href {https://doi.org/10.1126/science.1200735} {\bibfield  {journal} {\bibinfo  {journal} {Science}\ }\textbf {\bibinfo {volume} {331}},\ \bibinfo {pages} {889} (\bibinfo {year} {2011})}\BibitemShut {NoStop}%
\bibitem [{\citenamefont {Sun}\ \emph {et~al.}(2014)\citenamefont {Sun}, \citenamefont {Tan}, \citenamefont {Li}, \citenamefont {Li},\ and\ \citenamefont {Chen}}]{sun2014}%
  \BibitemOpen
  \bibfield  {author} {\bibinfo {author} {\bibfnamefont {Y.}~\bibnamefont {Sun}}, \bibinfo {author} {\bibfnamefont {W.}~\bibnamefont {Tan}}, \bibinfo {author} {\bibfnamefont {H.-q.}\ \bibnamefont {Li}}, \bibinfo {author} {\bibfnamefont {J.}~\bibnamefont {Li}},\ and\ \bibinfo {author} {\bibfnamefont {H.}~\bibnamefont {Chen}},\ }\bibfield  {title} {\bibinfo {title} {Experimental {{Demonstration}} of a {{Coherent Perfect Absorber}} with {{PT Phase Transition}}},\ }\href {https://doi.org/10.1103/PhysRevLett.112.143903} {\bibfield  {journal} {\bibinfo  {journal} {Physical Review Letters}\ }\textbf {\bibinfo {volume} {112}},\ \bibinfo {pages} {143903} (\bibinfo {year} {2014})}\BibitemShut {NoStop}%
\bibitem [{\citenamefont {Baranov}\ \emph {et~al.}(2017)\citenamefont {Baranov}, \citenamefont {Krasnok}, \citenamefont {Shegai}, \citenamefont {Al{\`u}},\ and\ \citenamefont {Chong}}]{baranov2017}%
  \BibitemOpen
  \bibfield  {author} {\bibinfo {author} {\bibfnamefont {D.~G.}\ \bibnamefont {Baranov}}, \bibinfo {author} {\bibfnamefont {A.}~\bibnamefont {Krasnok}}, \bibinfo {author} {\bibfnamefont {T.}~\bibnamefont {Shegai}}, \bibinfo {author} {\bibfnamefont {A.}~\bibnamefont {Al{\`u}}},\ and\ \bibinfo {author} {\bibfnamefont {Y.}~\bibnamefont {Chong}},\ }\bibfield  {title} {\bibinfo {title} {Coherent perfect absorbers: Linear control of light with light},\ }\href {https://doi.org/10.1038/natrevmats.2017.64} {\bibfield  {journal} {\bibinfo  {journal} {Nature Reviews Materials}\ }\textbf {\bibinfo {volume} {2}},\ \bibinfo {pages} {1} (\bibinfo {year} {2017})}\BibitemShut {NoStop}%
\bibitem [{\citenamefont {M{\"u}llers}\ \emph {et~al.}(2018)\citenamefont {M{\"u}llers}, \citenamefont {Santra}, \citenamefont {Baals}, \citenamefont {Baals}, \citenamefont {Jiang}, \citenamefont {Benary}, \citenamefont {Labouvie}, \citenamefont {Zezyulin}, \citenamefont {Konotop},\ and\ \citenamefont {Ott}}]{mullers2018a}%
  \BibitemOpen
  \bibfield  {author} {\bibinfo {author} {\bibfnamefont {A.}~\bibnamefont {M{\"u}llers}}, \bibinfo {author} {\bibfnamefont {B.}~\bibnamefont {Santra}}, \bibinfo {author} {\bibfnamefont {C.}~\bibnamefont {Baals}}, \bibinfo {author} {\bibfnamefont {C.}~\bibnamefont {Baals}}, \bibinfo {author} {\bibfnamefont {J.}~\bibnamefont {Jiang}}, \bibinfo {author} {\bibfnamefont {J.}~\bibnamefont {Benary}}, \bibinfo {author} {\bibfnamefont {R.}~\bibnamefont {Labouvie}}, \bibinfo {author} {\bibfnamefont {D.~A.}\ \bibnamefont {Zezyulin}}, \bibinfo {author} {\bibfnamefont {V.~V.}\ \bibnamefont {Konotop}},\ and\ \bibinfo {author} {\bibfnamefont {H.}~\bibnamefont {Ott}},\ }\bibfield  {title} {\bibinfo {title} {Coherent perfect absorption of nonlinear matter waves},\ }\href@noop {} {\bibfield  {journal} {\bibinfo  {journal} {Science Advances}\ }\textbf {\bibinfo {volume} {4}} (\bibinfo {year} {2018})}\BibitemShut {NoStop}%
\bibitem [{\citenamefont {Pichler}\ \emph {et~al.}(2019)\citenamefont {Pichler}, \citenamefont {K{\"u}hmayer}, \citenamefont {B{\"o}hm}, \citenamefont {Brandst{\"o}tter}, \citenamefont {Ambichl}, \citenamefont {Kuhl},\ and\ \citenamefont {Rotter}}]{pichler2019a}%
  \BibitemOpen
  \bibfield  {author} {\bibinfo {author} {\bibfnamefont {K.}~\bibnamefont {Pichler}}, \bibinfo {author} {\bibfnamefont {M.}~\bibnamefont {K{\"u}hmayer}}, \bibinfo {author} {\bibfnamefont {J.}~\bibnamefont {B{\"o}hm}}, \bibinfo {author} {\bibfnamefont {A.}~\bibnamefont {Brandst{\"o}tter}}, \bibinfo {author} {\bibfnamefont {P.}~\bibnamefont {Ambichl}}, \bibinfo {author} {\bibfnamefont {U.}~\bibnamefont {Kuhl}},\ and\ \bibinfo {author} {\bibfnamefont {S.}~\bibnamefont {Rotter}},\ }\bibfield  {title} {\bibinfo {title} {Random anti-lasing through coherent perfect absorption in a disordered medium},\ }\href {https://doi.org/10.1038/s41586-019-0971-3} {\bibfield  {journal} {\bibinfo  {journal} {Nature}\ }\textbf {\bibinfo {volume} {567}},\ \bibinfo {pages} {351} (\bibinfo {year} {2019})}\BibitemShut {NoStop}%
\bibitem [{\citenamefont {Sweeney}\ \emph {et~al.}(2019)\citenamefont {Sweeney}, \citenamefont {Hsu}, \citenamefont {Rotter},\ and\ \citenamefont {Stone}}]{sweeney2019a}%
  \BibitemOpen
  \bibfield  {author} {\bibinfo {author} {\bibfnamefont {W.~R.}\ \bibnamefont {Sweeney}}, \bibinfo {author} {\bibfnamefont {C.~W.}\ \bibnamefont {Hsu}}, \bibinfo {author} {\bibfnamefont {S.}~\bibnamefont {Rotter}},\ and\ \bibinfo {author} {\bibfnamefont {A.~D.}\ \bibnamefont {Stone}},\ }\bibfield  {title} {\bibinfo {title} {Perfectly {{Absorbing Exceptional Points}} and {{Chiral Absorbers}}},\ }\href {https://doi.org/10.1103/PhysRevLett.122.093901} {\bibfield  {journal} {\bibinfo  {journal} {Physical Review Letters}\ }\textbf {\bibinfo {volume} {122}},\ \bibinfo {pages} {093901} (\bibinfo {year} {2019})}\BibitemShut {NoStop}%
\bibitem [{\citenamefont {Chen}\ \emph {et~al.}(2020)\citenamefont {Chen}, \citenamefont {Kottos},\ and\ \citenamefont {Anlage}}]{chen2020ac}%
  \BibitemOpen
  \bibfield  {author} {\bibinfo {author} {\bibfnamefont {L.}~\bibnamefont {Chen}}, \bibinfo {author} {\bibfnamefont {T.}~\bibnamefont {Kottos}},\ and\ \bibinfo {author} {\bibfnamefont {S.~M.}\ \bibnamefont {Anlage}},\ }\bibfield  {title} {\bibinfo {title} {Perfect absorption in complex scattering systems with or without hidden symmetries},\ }\href {https://doi.org/10.1038/s41467-020-19645-5} {\bibfield  {journal} {\bibinfo  {journal} {Nature Communications}\ }\textbf {\bibinfo {volume} {11}},\ \bibinfo {pages} {5826} (\bibinfo {year} {2020})}\BibitemShut {NoStop}%
\bibitem [{\citenamefont {Wang}\ \emph {et~al.}(2021)\citenamefont {Wang}, \citenamefont {Sweeney}, \citenamefont {Stone},\ and\ \citenamefont {Yang}}]{wang2021h}%
  \BibitemOpen
  \bibfield  {author} {\bibinfo {author} {\bibfnamefont {C.}~\bibnamefont {Wang}}, \bibinfo {author} {\bibfnamefont {W.~R.}\ \bibnamefont {Sweeney}}, \bibinfo {author} {\bibfnamefont {A.~D.}\ \bibnamefont {Stone}},\ and\ \bibinfo {author} {\bibfnamefont {L.}~\bibnamefont {Yang}},\ }\bibfield  {title} {\bibinfo {title} {Coherent perfect absorption at an exceptional point},\ }\href {https://doi.org/10.1126/science.abj1028} {\bibfield  {journal} {\bibinfo  {journal} {Science}\ }\textbf {\bibinfo {volume} {373}},\ \bibinfo {pages} {1261} (\bibinfo {year} {2021})}\BibitemShut {NoStop}%
\bibitem [{\citenamefont {Slobodkin}\ \emph {et~al.}(2022)\citenamefont {Slobodkin}, \citenamefont {Weinberg}, \citenamefont {H{\"o}rner}, \citenamefont {Pichler}, \citenamefont {Rotter},\ and\ \citenamefont {Katz}}]{slobodkin2022}%
  \BibitemOpen
  \bibfield  {author} {\bibinfo {author} {\bibfnamefont {Y.}~\bibnamefont {Slobodkin}}, \bibinfo {author} {\bibfnamefont {G.}~\bibnamefont {Weinberg}}, \bibinfo {author} {\bibfnamefont {H.}~\bibnamefont {H{\"o}rner}}, \bibinfo {author} {\bibfnamefont {K.}~\bibnamefont {Pichler}}, \bibinfo {author} {\bibfnamefont {S.}~\bibnamefont {Rotter}},\ and\ \bibinfo {author} {\bibfnamefont {O.}~\bibnamefont {Katz}},\ }\bibfield  {title} {\bibinfo {title} {Massively degenerate coherent perfect absorber for arbitrary wavefronts},\ }\href {https://doi.org/10.1126/science.abq8103} {\bibfield  {journal} {\bibinfo  {journal} {Science}\ }\textbf {\bibinfo {volume} {377}},\ \bibinfo {pages} {995} (\bibinfo {year} {2022})}\BibitemShut {NoStop}%
\bibitem [{\citenamefont {Guo}\ and\ \citenamefont {Fan}(2024{\natexlab{b}})}]{guo2024b}%
  \BibitemOpen
  \bibfield  {author} {\bibinfo {author} {\bibfnamefont {C.}~\bibnamefont {Guo}}\ and\ \bibinfo {author} {\bibfnamefont {S.}~\bibnamefont {Fan}},\ }\bibfield  {title} {\bibinfo {title} {Unitary control of partially coherent waves. {{I}}. {{Absorption}}},\ }\href {https://doi.org/10.1103/PhysRevB.110.035430} {\bibfield  {journal} {\bibinfo  {journal} {Physical Review B}\ }\textbf {\bibinfo {volume} {110}},\ \bibinfo {pages} {035430} (\bibinfo {year} {2024}{\natexlab{b}})}\BibitemShut {NoStop}%
\bibitem [{\citenamefont {Sweeney}\ \emph {et~al.}(2020)\citenamefont {Sweeney}, \citenamefont {Hsu},\ and\ \citenamefont {Stone}}]{sweeney2020a}%
  \BibitemOpen
  \bibfield  {author} {\bibinfo {author} {\bibfnamefont {W.~R.}\ \bibnamefont {Sweeney}}, \bibinfo {author} {\bibfnamefont {C.~W.}\ \bibnamefont {Hsu}},\ and\ \bibinfo {author} {\bibfnamefont {A.~D.}\ \bibnamefont {Stone}},\ }\bibfield  {title} {\bibinfo {title} {Theory of reflectionless scattering modes},\ }\href {https://doi.org/10.1103/PhysRevA.102.063511} {\bibfield  {journal} {\bibinfo  {journal} {Physical Review A}\ }\textbf {\bibinfo {volume} {102}},\ \bibinfo {pages} {063511} (\bibinfo {year} {2020})}\BibitemShut {NoStop}%
\bibitem [{\citenamefont {Stone}\ \emph {et~al.}(2021)\citenamefont {Stone}, \citenamefont {Sweeney}, \citenamefont {Hsu}, \citenamefont {Wisal},\ and\ \citenamefont {Wang}}]{stone2021}%
  \BibitemOpen
  \bibfield  {author} {\bibinfo {author} {\bibfnamefont {A.~D.}\ \bibnamefont {Stone}}, \bibinfo {author} {\bibfnamefont {W.~R.}\ \bibnamefont {Sweeney}}, \bibinfo {author} {\bibfnamefont {C.~W.}\ \bibnamefont {Hsu}}, \bibinfo {author} {\bibfnamefont {K.}~\bibnamefont {Wisal}},\ and\ \bibinfo {author} {\bibfnamefont {Z.}~\bibnamefont {Wang}},\ }\bibfield  {title} {\bibinfo {title} {Reflectionless excitation of arbitrary photonic structures: A general theory},\ }\href {https://doi.org/10.1515/nanoph-2020-0403} {\bibfield  {journal} {\bibinfo  {journal} {Nanophotonics}\ }\textbf {\bibinfo {volume} {10}},\ \bibinfo {pages} {343} (\bibinfo {year} {2021})}\BibitemShut {NoStop}%
\bibitem [{\citenamefont {Horodynski}\ \emph {et~al.}(2022)\citenamefont {Horodynski}, \citenamefont {K{\"u}hmayer}, \citenamefont {Ferise}, \citenamefont {Rotter},\ and\ \citenamefont {Davy}}]{horodynski2022}%
  \BibitemOpen
  \bibfield  {author} {\bibinfo {author} {\bibfnamefont {M.}~\bibnamefont {Horodynski}}, \bibinfo {author} {\bibfnamefont {M.}~\bibnamefont {K{\"u}hmayer}}, \bibinfo {author} {\bibfnamefont {C.}~\bibnamefont {Ferise}}, \bibinfo {author} {\bibfnamefont {S.}~\bibnamefont {Rotter}},\ and\ \bibinfo {author} {\bibfnamefont {M.}~\bibnamefont {Davy}},\ }\bibfield  {title} {\bibinfo {title} {Anti-reflection structure for perfect transmission through complex media},\ }\href {https://doi.org/10.1038/s41586-022-04843-6} {\bibfield  {journal} {\bibinfo  {journal} {Nature}\ }\textbf {\bibinfo {volume} {607}},\ \bibinfo {pages} {281} (\bibinfo {year} {2022})}\BibitemShut {NoStop}%
\bibitem [{\citenamefont {{\v C}i{\v z}m{\'a}r}\ \emph {et~al.}(2010)\citenamefont {{\v C}i{\v z}m{\'a}r}, \citenamefont {Mazilu},\ and\ \citenamefont {Dholakia}}]{cizmar2010a}%
  \BibitemOpen
  \bibfield  {author} {\bibinfo {author} {\bibfnamefont {T.}~\bibnamefont {{\v C}i{\v z}m{\'a}r}}, \bibinfo {author} {\bibfnamefont {M.}~\bibnamefont {Mazilu}},\ and\ \bibinfo {author} {\bibfnamefont {K.}~\bibnamefont {Dholakia}},\ }\bibfield  {title} {\bibinfo {title} {In situ wavefront correction and its application to micromanipulation},\ }\href {https://doi.org/10.1038/nphoton.2010.85} {\bibfield  {journal} {\bibinfo  {journal} {Nature Photonics}\ }\textbf {\bibinfo {volume} {4}},\ \bibinfo {pages} {388} (\bibinfo {year} {2010})}\BibitemShut {NoStop}%
\bibitem [{\citenamefont {{Gonzalez-Ballestero}}\ \emph {et~al.}(2021)\citenamefont {{Gonzalez-Ballestero}}, \citenamefont {Aspelmeyer}, \citenamefont {Novotny}, \citenamefont {Quidant},\ and\ \citenamefont {{Romero-Isart}}}]{gonzalez-ballestero2021}%
  \BibitemOpen
  \bibfield  {author} {\bibinfo {author} {\bibfnamefont {C.}~\bibnamefont {{Gonzalez-Ballestero}}}, \bibinfo {author} {\bibfnamefont {M.}~\bibnamefont {Aspelmeyer}}, \bibinfo {author} {\bibfnamefont {L.}~\bibnamefont {Novotny}}, \bibinfo {author} {\bibfnamefont {R.}~\bibnamefont {Quidant}},\ and\ \bibinfo {author} {\bibfnamefont {O.}~\bibnamefont {{Romero-Isart}}},\ }\bibfield  {title} {\bibinfo {title} {Levitodynamics: {{Levitation}} and control of microscopic objects in vacuum},\ }\href {https://doi.org/10.1126/science.abg3027} {\bibfield  {journal} {\bibinfo  {journal} {Science}\ }\textbf {\bibinfo {volume} {374}},\ \bibinfo {pages} {eabg3027} (\bibinfo {year} {2021})}\BibitemShut {NoStop}%
\bibitem [{\citenamefont {H{\"u}pfl}\ \emph {et~al.}(2023)\citenamefont {H{\"u}pfl}, \citenamefont {Bachelard}, \citenamefont {Kaczvinszki}, \citenamefont {Horodynski}, \citenamefont {K{\"u}hmayer},\ and\ \citenamefont {Rotter}}]{hupfl2023}%
  \BibitemOpen
  \bibfield  {author} {\bibinfo {author} {\bibfnamefont {J.}~\bibnamefont {H{\"u}pfl}}, \bibinfo {author} {\bibfnamefont {N.}~\bibnamefont {Bachelard}}, \bibinfo {author} {\bibfnamefont {M.}~\bibnamefont {Kaczvinszki}}, \bibinfo {author} {\bibfnamefont {M.}~\bibnamefont {Horodynski}}, \bibinfo {author} {\bibfnamefont {M.}~\bibnamefont {K{\"u}hmayer}},\ and\ \bibinfo {author} {\bibfnamefont {S.}~\bibnamefont {Rotter}},\ }\bibfield  {title} {\bibinfo {title} {Optimal {{Cooling}} of {{Multiple Levitated Particles}} through {{Far-Field Wavefront Shaping}}},\ }\href {https://doi.org/10.1103/PhysRevLett.130.083203} {\bibfield  {journal} {\bibinfo  {journal} {Physical Review Letters}\ }\textbf {\bibinfo {volume} {130}},\ \bibinfo {pages} {083203} (\bibinfo {year} {2023})}\BibitemShut {NoStop}%
\bibitem [{\citenamefont {Shaughnessy}\ \emph {et~al.}(2024)\citenamefont {Shaughnessy}, \citenamefont {McIntosh}, \citenamefont {Goetschy}, \citenamefont {Hsu}, \citenamefont {Bender}, \citenamefont {Y{\i}lmaz}, \citenamefont {Yamilov},\ and\ \citenamefont {Cao}}]{shaughnessy2024}%
  \BibitemOpen
  \bibfield  {author} {\bibinfo {author} {\bibfnamefont {L.}~\bibnamefont {Shaughnessy}}, \bibinfo {author} {\bibfnamefont {R.~E.}\ \bibnamefont {McIntosh}}, \bibinfo {author} {\bibfnamefont {A.}~\bibnamefont {Goetschy}}, \bibinfo {author} {\bibfnamefont {C.~W.}\ \bibnamefont {Hsu}}, \bibinfo {author} {\bibfnamefont {N.}~\bibnamefont {Bender}}, \bibinfo {author} {\bibfnamefont {H.}~\bibnamefont {Y{\i}lmaz}}, \bibinfo {author} {\bibfnamefont {A.}~\bibnamefont {Yamilov}},\ and\ \bibinfo {author} {\bibfnamefont {H.}~\bibnamefont {Cao}},\ }\bibfield  {title} {\bibinfo {title} {Multiregion {{Light Control}} in {{Diffusive Media}} via {{Wavefront Shaping}}},\ }\href {https://doi.org/10.1103/PhysRevLett.133.146901} {\bibfield  {journal} {\bibinfo  {journal} {Physical Review Letters}\ }\textbf {\bibinfo {volume} {133}},\ \bibinfo {pages} {146901} (\bibinfo {year} {2024})}\BibitemShut {NoStop}%
\bibitem [{\citenamefont {Fink}(1997)}]{fink1997}%
  \BibitemOpen
  \bibfield  {author} {\bibinfo {author} {\bibfnamefont {M.}~\bibnamefont {Fink}},\ }\bibfield  {title} {\bibinfo {title} {Time {{Reversed Acoustics}}},\ }\href {https://doi.org/10.1063/1.881692} {\bibfield  {journal} {\bibinfo  {journal} {Physics Today}\ }\textbf {\bibinfo {volume} {50}},\ \bibinfo {pages} {34} (\bibinfo {year} {1997})}\BibitemShut {NoStop}%
\bibitem [{\citenamefont {Xie}\ \emph {et~al.}(2014)\citenamefont {Xie}, \citenamefont {Wang}, \citenamefont {Chen}, \citenamefont {Konneker}, \citenamefont {Popa},\ and\ \citenamefont {Cummer}}]{xie2014}%
  \BibitemOpen
  \bibfield  {author} {\bibinfo {author} {\bibfnamefont {Y.}~\bibnamefont {Xie}}, \bibinfo {author} {\bibfnamefont {W.}~\bibnamefont {Wang}}, \bibinfo {author} {\bibfnamefont {H.}~\bibnamefont {Chen}}, \bibinfo {author} {\bibfnamefont {A.}~\bibnamefont {Konneker}}, \bibinfo {author} {\bibfnamefont {B.-I.}\ \bibnamefont {Popa}},\ and\ \bibinfo {author} {\bibfnamefont {S.~A.}\ \bibnamefont {Cummer}},\ }\bibfield  {title} {\bibinfo {title} {Wavefront modulation and subwavelength diffractive acoustics with an acoustic metasurface},\ }\href {https://doi.org/10.1038/ncomms6553} {\bibfield  {journal} {\bibinfo  {journal} {Nature Communications}\ }\textbf {\bibinfo {volume} {5}},\ \bibinfo {pages} {5553} (\bibinfo {year} {2014})}\BibitemShut {NoStop}%
\bibitem [{\citenamefont {Mounaix}\ \emph {et~al.}(2016{\natexlab{b}})\citenamefont {Mounaix}, \citenamefont {Defienne},\ and\ \citenamefont {Gigan}}]{mounaix2016a}%
  \BibitemOpen
  \bibfield  {author} {\bibinfo {author} {\bibfnamefont {M.}~\bibnamefont {Mounaix}}, \bibinfo {author} {\bibfnamefont {H.}~\bibnamefont {Defienne}},\ and\ \bibinfo {author} {\bibfnamefont {S.}~\bibnamefont {Gigan}},\ }\bibfield  {title} {\bibinfo {title} {Deterministic light focusing in space and time through multiple scattering media with a time-resolved transmission matrix approach},\ }\href {https://doi.org/10.1103/PhysRevA.94.041802} {\bibfield  {journal} {\bibinfo  {journal} {Physical Review A}\ }\textbf {\bibinfo {volume} {94}},\ \bibinfo {pages} {041802} (\bibinfo {year} {2016}{\natexlab{b}})}\BibitemShut {NoStop}%
\bibitem [{\citenamefont {Katz}\ \emph {et~al.}(2019)\citenamefont {Katz}, \citenamefont {Ramaz}, \citenamefont {Gigan},\ and\ \citenamefont {Fink}}]{katz2019}%
  \BibitemOpen
  \bibfield  {author} {\bibinfo {author} {\bibfnamefont {O.}~\bibnamefont {Katz}}, \bibinfo {author} {\bibfnamefont {F.}~\bibnamefont {Ramaz}}, \bibinfo {author} {\bibfnamefont {S.}~\bibnamefont {Gigan}},\ and\ \bibinfo {author} {\bibfnamefont {M.}~\bibnamefont {Fink}},\ }\bibfield  {title} {\bibinfo {title} {Controlling light in complex media beyond the acoustic diffraction-limit using the acousto-optic transmission matrix},\ }\href {https://doi.org/10.1038/s41467-019-08583-6} {\bibfield  {journal} {\bibinfo  {journal} {Nature Communications}\ }\textbf {\bibinfo {volume} {10}},\ \bibinfo {pages} {717} (\bibinfo {year} {2019})}\BibitemShut {NoStop}%
\bibitem [{\citenamefont {Guo}\ and\ \citenamefont {Fan}(2023)}]{guo2023b}%
  \BibitemOpen
  \bibfield  {author} {\bibinfo {author} {\bibfnamefont {C.}~\bibnamefont {Guo}}\ and\ \bibinfo {author} {\bibfnamefont {S.}~\bibnamefont {Fan}},\ }\bibfield  {title} {\bibinfo {title} {Majorization {{Theory}} for {{Unitary Control}} of {{Optical Absorption}} and {{Emission}}},\ }\href {https://doi.org/10.1103/PhysRevLett.130.146202} {\bibfield  {journal} {\bibinfo  {journal} {Physical Review Letters}\ }\textbf {\bibinfo {volume} {130}},\ \bibinfo {pages} {146202} (\bibinfo {year} {2023})}\BibitemShut {NoStop}%
\bibitem [{\citenamefont {Guo}\ \emph {et~al.}(2023)\citenamefont {Guo}, \citenamefont {Li}, \citenamefont {Xiao},\ and\ \citenamefont {Fan}}]{guo2023g}%
  \BibitemOpen
  \bibfield  {author} {\bibinfo {author} {\bibfnamefont {C.}~\bibnamefont {Guo}}, \bibinfo {author} {\bibfnamefont {J.}~\bibnamefont {Li}}, \bibinfo {author} {\bibfnamefont {M.}~\bibnamefont {Xiao}},\ and\ \bibinfo {author} {\bibfnamefont {S.}~\bibnamefont {Fan}},\ }\bibfield  {title} {\bibinfo {title} {Singular topology of scattering matrices},\ }\href {https://doi.org/10.1103/PhysRevB.108.155418} {\bibfield  {journal} {\bibinfo  {journal} {Physical Review B}\ }\textbf {\bibinfo {volume} {108}},\ \bibinfo {pages} {155418} (\bibinfo {year} {2023})}\BibitemShut {NoStop}%
\bibitem [{\citenamefont {Guo}\ and\ \citenamefont {Fan}(2024{\natexlab{c}})}]{guo2024a}%
  \BibitemOpen
  \bibfield  {author} {\bibinfo {author} {\bibfnamefont {C.}~\bibnamefont {Guo}}\ and\ \bibinfo {author} {\bibfnamefont {S.}~\bibnamefont {Fan}},\ }\bibfield  {title} {\bibinfo {title} {Topological winding guaranteed coherent orthogonal scattering},\ }\href {https://doi.org/10.1103/PhysRevA.109.L061503} {\bibfield  {journal} {\bibinfo  {journal} {Physical Review A}\ }\textbf {\bibinfo {volume} {109}},\ \bibinfo {pages} {L061503} (\bibinfo {year} {2024}{\natexlab{c}})}\BibitemShut {NoStop}%
\bibitem [{\citenamefont {Guo}\ \emph {et~al.}(2024)\citenamefont {Guo}, \citenamefont {Miller},\ and\ \citenamefont {Fan}}]{guo2024e}%
  \BibitemOpen
  \bibfield  {author} {\bibinfo {author} {\bibfnamefont {C.}~\bibnamefont {Guo}}, \bibinfo {author} {\bibfnamefont {D.~A.~B.}\ \bibnamefont {Miller}},\ and\ \bibinfo {author} {\bibfnamefont {S.}~\bibnamefont {Fan}},\ }\href {https://arxiv.org/abs/2408.06386} {\bibinfo {title} {Transport measurements of majorization order for wave coherence}} (\bibinfo {year} {2024}),\ \Eprint {https://arxiv.org/abs/2408.06386} {arXiv:2408.06386 [physics.optics]} \BibitemShut {NoStop}%
\bibitem [{\citenamefont {Bonsall}\ and\ \citenamefont {Duncan}(1971)}]{bonsall1971}%
  \BibitemOpen
  \bibfield  {author} {\bibinfo {author} {\bibfnamefont {F.~F.}\ \bibnamefont {Bonsall}}\ and\ \bibinfo {author} {\bibfnamefont {J.}~\bibnamefont {Duncan}},\ }\href@noop {} {\emph {\bibinfo {title} {Numerical Ranges of Operators on Normed Spaces and of Elements of Normed Algebras}}}\ (\bibinfo  {publisher} {Cambridge University Press},\ \bibinfo {address} {London},\ \bibinfo {year} {1971})\BibitemShut {NoStop}%
\bibitem [{\citenamefont {Gustafson}\ and\ \citenamefont {Rao}(1997)}]{gustafson1997}%
  \BibitemOpen
  \bibfield  {author} {\bibinfo {author} {\bibfnamefont {K.~E.}\ \bibnamefont {Gustafson}}\ and\ \bibinfo {author} {\bibfnamefont {D.~K.~M.}\ \bibnamefont {Rao}},\ }\href@noop {} {\emph {\bibinfo {title} {Numerical Range: The Field of Values of Linear Operators and Matrices}}},\ Universitext\ (\bibinfo  {publisher} {Springer},\ \bibinfo {address} {New York},\ \bibinfo {year} {1997})\BibitemShut {NoStop}%
\bibitem [{\citenamefont {Wu}(2021)}]{wu2021a}%
  \BibitemOpen
  \bibfield  {author} {\bibinfo {author} {\bibfnamefont {P.~Y.}\ \bibnamefont {Wu}},\ }\href@noop {} {\emph {\bibinfo {title} {Numerical Ranges of {{Hilbert}} Space Operators}}},\ \bibinfo {series} {Encyclopedia of Mathematics and Its Applications}\ No.\ \bibinfo {number} {179}\ (\bibinfo  {publisher} {Cambridge University Press},\ \bibinfo {address} {Cambridge, United Kingdom ; New York, NY},\ \bibinfo {year} {2021})\BibitemShut {NoStop}%
\bibitem [{\citenamefont {Haus}(1984)}]{haus1984}%
  \BibitemOpen
  \bibfield  {author} {\bibinfo {author} {\bibfnamefont {H.~A.}\ \bibnamefont {Haus}},\ }\href@noop {} {\emph {\bibinfo {title} {Waves and Fields in Optoelectronics}}}\ (\bibinfo  {publisher} {{Prentice-Hall}},\ \bibinfo {address} {{Englewood Cliffs, NJ}},\ \bibinfo {year} {1984})\BibitemShut {NoStop}%
\bibitem [{\citenamefont {Guo}\ and\ \citenamefont {Fan}(2024{\natexlab{d}})}]{guo2024j}%
  \BibitemOpen
  \bibfield  {author} {\bibinfo {author} {\bibfnamefont {C.}~\bibnamefont {Guo}}\ and\ \bibinfo {author} {\bibfnamefont {S.}~\bibnamefont {Fan}},\ }\bibfield  {title} {\bibinfo {title} {Passivity constraints on the relations between transmission, reflection, and absorption eigenvalues},\ }\href {https://doi.org/10.1103/PhysRevB.110.205431} {\bibfield  {journal} {\bibinfo  {journal} {Physical Review B}\ }\textbf {\bibinfo {volume} {110}},\ \bibinfo {pages} {205431} (\bibinfo {year} {2024}{\natexlab{d}})}\BibitemShut {NoStop}%
\bibitem [{\citenamefont {Hughes}\ \emph {et~al.}(2021)\citenamefont {Hughes}, \citenamefont {Minkov}, \citenamefont {Liu}, \citenamefont {Yu},\ and\ \citenamefont {Fan}}]{hughes2021a}%
  \BibitemOpen
  \bibfield  {author} {\bibinfo {author} {\bibfnamefont {T.~W.}\ \bibnamefont {Hughes}}, \bibinfo {author} {\bibfnamefont {M.}~\bibnamefont {Minkov}}, \bibinfo {author} {\bibfnamefont {V.}~\bibnamefont {Liu}}, \bibinfo {author} {\bibfnamefont {Z.}~\bibnamefont {Yu}},\ and\ \bibinfo {author} {\bibfnamefont {S.}~\bibnamefont {Fan}},\ }\bibfield  {title} {\bibinfo {title} {A perspective on the pathway toward full wave simulation of large area metalenses},\ }\href {https://doi.org/10.1063/5.0071245} {\bibfield  {journal} {\bibinfo  {journal} {Applied Physics Letters}\ }\textbf {\bibinfo {volume} {119}},\ \bibinfo {pages} {150502} (\bibinfo {year} {2021})}\BibitemShut {NoStop}%
\bibitem [{\citenamefont {Johnson}(1978)}]{johnson1978}%
  \BibitemOpen
  \bibfield  {author} {\bibinfo {author} {\bibfnamefont {C.~R.}\ \bibnamefont {Johnson}},\ }\bibfield  {title} {\bibinfo {title} {Numerical {{Determination}} of the {{Field}} of {{Values}} of a {{General Complex Matrix}}},\ }\href {https://doi.org/10.1137/0715039} {\bibfield  {journal} {\bibinfo  {journal} {SIAM Journal on Numerical Analysis}\ }\textbf {\bibinfo {volume} {15}},\ \bibinfo {pages} {595} (\bibinfo {year} {1978})}\BibitemShut {NoStop}%
\bibitem [{\citenamefont {Horn}\ and\ \citenamefont {Johnson}(1991)}]{horn1991}%
  \BibitemOpen
  \bibfield  {author} {\bibinfo {author} {\bibfnamefont {R.~A.}\ \bibnamefont {Horn}}\ and\ \bibinfo {author} {\bibfnamefont {C.~R.}\ \bibnamefont {Johnson}},\ }\href@noop {} {\emph {\bibinfo {title} {Topics in Matrix Analysis}}}\ (\bibinfo  {publisher} {Cambridge University Press},\ \bibinfo {address} {Cambridge},\ \bibinfo {year} {1991})\BibitemShut {NoStop}%
\bibitem [{\citenamefont {Keeler}\ \emph {et~al.}(1997)\citenamefont {Keeler}, \citenamefont {Rodman},\ and\ \citenamefont {Spitkovsky}}]{keeler1997}%
  \BibitemOpen
  \bibfield  {author} {\bibinfo {author} {\bibfnamefont {D.~S.}\ \bibnamefont {Keeler}}, \bibinfo {author} {\bibfnamefont {L.}~\bibnamefont {Rodman}},\ and\ \bibinfo {author} {\bibfnamefont {I.~M.}\ \bibnamefont {Spitkovsky}},\ }\bibfield  {title} {\bibinfo {title} {The numerical range of 3 {\texttimes} 3 matrices},\ }\href {https://doi.org/10.1016/0024-3795(95)00674-5} {\bibfield  {journal} {\bibinfo  {journal} {Linear Algebra and its Applications}\ }\textbf {\bibinfo {volume} {252}},\ \bibinfo {pages} {115} (\bibinfo {year} {1997})}\BibitemShut {NoStop}%
\bibitem [{\citenamefont {Horn}\ and\ \citenamefont {Johnson}(2012)}]{horn2012}%
  \BibitemOpen
  \bibfield  {author} {\bibinfo {author} {\bibfnamefont {R.~A.}\ \bibnamefont {Horn}}\ and\ \bibinfo {author} {\bibfnamefont {C.~R.}\ \bibnamefont {Johnson}},\ }\href@noop {} {\emph {\bibinfo {title} {Matrix Analysis}}},\ \bibinfo {edition} {2nd}\ ed.\ (\bibinfo  {publisher} {{Cambridge University Press}},\ \bibinfo {address} {{Cambridge ; New York}},\ \bibinfo {year} {2012})\BibitemShut {NoStop}%
\bibitem [{\citenamefont {Henrici}(1962)}]{henrici1962}%
  \BibitemOpen
  \bibfield  {author} {\bibinfo {author} {\bibfnamefont {P.}~\bibnamefont {Henrici}},\ }\bibfield  {title} {\bibinfo {title} {Bounds for iterates, inverses, spectral variation and fields of values of non-normal matrices},\ }\href {https://doi.org/10.1007/BF01386294} {\bibfield  {journal} {\bibinfo  {journal} {Numerische Mathematik}\ }\textbf {\bibinfo {volume} {4}},\ \bibinfo {pages} {24} (\bibinfo {year} {1962})}\BibitemShut {NoStop}%
\bibitem [{\citenamefont {Barvinok}(2002)}]{barvinok2002}%
  \BibitemOpen
  \bibfield  {author} {\bibinfo {author} {\bibfnamefont {A.}~\bibnamefont {Barvinok}},\ }\href@noop {} {\emph {\bibinfo {title} {A Course in Convexity}}},\ \bibinfo {series} {Graduate Studies in Mathematics}\ No.\ \bibinfo {number} {v. 54}\ (\bibinfo  {publisher} {American Mathematical Society},\ \bibinfo {address} {Providence, R.I},\ \bibinfo {year} {2002})\BibitemShut {NoStop}%
\bibitem [{\citenamefont {Boyd}\ and\ \citenamefont {Vandenberghe}(2004)}]{boyd2004}%
  \BibitemOpen
  \bibfield  {author} {\bibinfo {author} {\bibfnamefont {S.~P.}\ \bibnamefont {Boyd}}\ and\ \bibinfo {author} {\bibfnamefont {L.}~\bibnamefont {Vandenberghe}},\ }\href@noop {} {\emph {\bibinfo {title} {Convex Optimization}}}\ (\bibinfo  {publisher} {Cambridge University Press},\ \bibinfo {address} {Cambridge, UK ; New York},\ \bibinfo {year} {2004})\BibitemShut {NoStop}%
\bibitem [{\citenamefont {Uhlig}(2008)}]{uhlig2008}%
  \BibitemOpen
  \bibfield  {author} {\bibinfo {author} {\bibfnamefont {F.}~\bibnamefont {Uhlig}},\ }\bibfield  {title} {\bibinfo {title} {An inverse field of values problem},\ }\href {https://doi.org/10.1088/0266-5611/24/5/055019} {\bibfield  {journal} {\bibinfo  {journal} {Inverse Problems}\ }\textbf {\bibinfo {volume} {24}},\ \bibinfo {pages} {055019} (\bibinfo {year} {2008})}\BibitemShut {NoStop}%
\bibitem [{\citenamefont {Carden}(2009)}]{carden2009}%
  \BibitemOpen
  \bibfield  {author} {\bibinfo {author} {\bibfnamefont {R.}~\bibnamefont {Carden}},\ }\bibfield  {title} {\bibinfo {title} {A simple algorithm for the inverse field of values problem},\ }\href {https://doi.org/10.1088/0266-5611/25/11/115019} {\bibfield  {journal} {\bibinfo  {journal} {Inverse Problems}\ }\textbf {\bibinfo {volume} {25}},\ \bibinfo {pages} {115019} (\bibinfo {year} {2009})}\BibitemShut {NoStop}%
\bibitem [{\citenamefont {Chorianopoulos}\ \emph {et~al.}(2010)\citenamefont {Chorianopoulos}, \citenamefont {Psarrakos},\ and\ \citenamefont {Uhlig}}]{chorianopoulos2010}%
  \BibitemOpen
  \bibfield  {author} {\bibinfo {author} {\bibfnamefont {C.}~\bibnamefont {Chorianopoulos}}, \bibinfo {author} {\bibfnamefont {P.}~\bibnamefont {Psarrakos}},\ and\ \bibinfo {author} {\bibfnamefont {F.}~\bibnamefont {Uhlig}},\ }\bibfield  {title} {\bibinfo {title} {A method for the inverse numerical range problem.},\ }\href {https://doi.org/10.13001/1081-3810.1368} {\bibfield  {journal} {\bibinfo  {journal} {ELA. The Electronic Journal of Linear Algebra [electronic only]}\ }\textbf {\bibinfo {volume} {20}},\ \bibinfo {pages} {198} (\bibinfo {year} {2010})}\BibitemShut {NoStop}%
\bibitem [{\citenamefont {Meurant}(2012)}]{meurant2012}%
  \BibitemOpen
  \bibfield  {author} {\bibinfo {author} {\bibfnamefont {G.}~\bibnamefont {Meurant}},\ }\bibfield  {title} {\bibinfo {title} {The computation of isotropic vectors},\ }\href {https://doi.org/10.1007/s11075-012-9537-2} {\bibfield  {journal} {\bibinfo  {journal} {Numerical Algorithms}\ }\textbf {\bibinfo {volume} {60}},\ \bibinfo {pages} {193} (\bibinfo {year} {2012})}\BibitemShut {NoStop}%
\bibitem [{\citenamefont {Bebiano}\ \emph {et~al.}(2014)\citenamefont {Bebiano}, \citenamefont {da~Provid{\^e}ncia}, \citenamefont {Nata},\ and\ \citenamefont {da~Provid{\^e}ncia}}]{bebiano2014}%
  \BibitemOpen
  \bibfield  {author} {\bibinfo {author} {\bibfnamefont {N.}~\bibnamefont {Bebiano}}, \bibinfo {author} {\bibfnamefont {J.}~\bibnamefont {da~Provid{\^e}ncia}}, \bibinfo {author} {\bibfnamefont {A.}~\bibnamefont {Nata}},\ and\ \bibinfo {author} {\bibfnamefont {J.~P.}\ \bibnamefont {da~Provid{\^e}ncia}},\ }\bibfield  {title} {\bibinfo {title} {Revisiting the inverse field of values problem},\ }\href@noop {} {\bibfield  {journal} {\bibinfo  {journal} {Electronic Transactions on Numerical Analysis}\ }\textbf {\bibinfo {volume} {42}},\ \bibinfo {pages} {1} (\bibinfo {year} {2014})}\BibitemShut {NoStop}%
\bibitem [{\citenamefont {Higham}(2023)}]{higham2023a}%
  \BibitemOpen
  \bibfield  {author} {\bibinfo {author} {\bibfnamefont {N.}~\bibnamefont {Higham}},\ }\href@noop {} {\bibinfo {title} {What {{Is}} the {{Numerical Range}} of a {{Matrix}}?}} (\bibinfo {year} {2023})\BibitemShut {NoStop}%
\bibitem [{\citenamefont {Higham}(2024)}]{higham2024}%
  \BibitemOpen
  \bibfield  {author} {\bibinfo {author} {\bibfnamefont {N.}~\bibnamefont {Higham}},\ }\href@noop {} {\bibinfo {title} {The {{Matrix Computation Toolbox}}}},\ \bibinfo {howpublished} {https://www.mathworks.com/matlabcentral/fileexchange/2360-the-matrix-computation-toolbox} (\bibinfo {year} {2024})\BibitemShut {NoStop}%
\bibitem [{\citenamefont {Loisel}\ and\ \citenamefont {Maxwell}(2018)}]{loisel2018}%
  \BibitemOpen
  \bibfield  {author} {\bibinfo {author} {\bibfnamefont {S.}~\bibnamefont {Loisel}}\ and\ \bibinfo {author} {\bibfnamefont {P.}~\bibnamefont {Maxwell}},\ }\bibfield  {title} {\bibinfo {title} {Path-{{Following Method}} to {{Determine}} the {{Field}} of {{Values}} of a {{Matrix}} with {{High Accuracy}}},\ }\href {https://doi.org/10.1137/17M1148608} {\bibfield  {journal} {\bibinfo  {journal} {SIAM Journal on Matrix Analysis and Applications}\ }\textbf {\bibinfo {volume} {39}},\ \bibinfo {pages} {1726} (\bibinfo {year} {2018})}\BibitemShut {NoStop}%
\bibitem [{\citenamefont {Uhlig}(2020)}]{uhlig2020}%
  \BibitemOpen
  \bibfield  {author} {\bibinfo {author} {\bibfnamefont {F.}~\bibnamefont {Uhlig}},\ }\bibfield  {title} {\bibinfo {title} {Zhang {{Neural Networks}} for fast and accurate computations of the field of values},\ }\href {https://doi.org/10.1080/03081087.2019.1648375} {\bibfield  {journal} {\bibinfo  {journal} {Linear and Multilinear Algebra}\ }\textbf {\bibinfo {volume} {68}},\ \bibinfo {pages} {1894} (\bibinfo {year} {2020})}\BibitemShut {NoStop}%
\end{thebibliography}%

\end{document}